\documentclass[twocolumn]{aastex631}

\usepackage{amsmath}
\usepackage{gensymb}
\usepackage{xcolor}
\usepackage{placeins}

\usepackage{float}

\usepackage{tabularx}
\usepackage{booktabs}

\newcommand{\grs}{GRS\,1747$-$312}
\newcommand{\chandra}{\textit{Chandra}}

\begin{document}

\title{A Possible Third Body in the X-Ray System \grs\ \\ and Models with Higher-Order Multiplicity}

\author[0009-0003-6274-657X]{Caleb Painter}
\affiliation{Center for Astrophysics $|$ Harvard \& Smithsonian, 60 Garden Street, Cambridge, MA 02138}

\author[0000-0003-0972-1376]{Rosanne Di Stefano}
\affiliation{Center for Astrophysics $|$ Harvard \& Smithsonian, 60 Garden Street, Cambridge, MA 02138}

\author[0000-0002-3869-7996]{Vinay L.\ Kashyap}
\affiliation{Center for Astrophysics $|$ Harvard \& Smithsonian, 60 Garden Street, Cambridge, MA 02138}

\author{Roberto Soria}
\affiliation{College of Astronomy and Space Sciences, University of the Chinese Academy of Sciences, Beijing 100049, China}
\affiliation{INAF-Osservatorio Astrofisico di Torino, Strada Osservatorio 20, I-10025 Pino Torinese, Italy}
\affiliation{Sydney Institute for Astronomy, School of Physics A28, The University of Sydney, Sydney, NSW 2006, Australia}

\author[0000-0002-6187-2713]{Jose Lopez-Miralles}
\affiliation{Aurora Technology for the European Space Agency, ESAC/ESA, Camino Bajo del Castillo s/n, Urb. Villafranca del Castillo, 28691 Villanueva de la Cañada, Madrid, Spain}
\affiliation{Departament d’Astronomia i Astrofísica, Universitat de València, C/ Dr. Moliner, 50, 46100, Burjassot, València, Spain}

\author{Ryan Urquhart}
\affiliation{Center for Data Intensive and Time Domain Astronomy, Department of Physics and Astronomy, Michigan State University, East
Lansing MI, USA}

\author[0000-0002-5872-6061]{James F. Steiner}
\affiliation{Center for Astrophysics $|$ Harvard \& Smithsonian, 60 Garden Street, Cambridge, MA 02138}

\author{Sara Motta}
\affiliation{INAF-Osservatorio Astronomico di Brera, Merate,
Italy.}

\author[0000-0003-1080-9770]{Darin Ragozzine}
\affiliation{Brigham Young University, Department of Physics and Astronomy, N283 ESC, Provo, UT 84602, USA}

\author{Hideyuki Mori}
\affiliation{Japan Aerospace Exploration Agency (JAXA), Scientific Ballooning Research and Operation Group, Sagamihara, Kanagawa, 252-5210, Japan}

\begin{abstract}

\grs\ is a bright 
Low-Mass X-ray Binary in the globular cluster Terzan 6, located at a distance of 9.5 kpc from the Earth. It exhibits regular outbursts approximately every 4.5 months, during which periodic eclipses  are known to occur. These eclipses have only been observed in the outburst phase, and are not clearly seen when the source is quiescent. Recent \chandra\ observations of the source were performed in June 2019 and April, June, and August of 2021. Two of these observations captured the source during its outburst, and showed clear flux decreases at the expected time of eclipse. The other two observations occurred when the source was quiescent. We present the discovery of a dip that occurred during the quiescent state. The dip is of longer duration and its time of occurrence does not fit the ephemeris of the shorter eclipses. We study the physical characteristics of the dip and determine that it has all the properties of an eclipse by an object with a well defined surface. We find that there are several possibilities for the nature of the object causing the $5.3$~ks eclipse. First, \grs\ may be an X-ray triple, with an LMXB orbited by an outer third object, which could be an M-dwarf, brown dwarf, or planet. Second, there could be two LMXBs in close proximity to each other, likely bound together. Whatever the true nature of the eclipser, its presence suggests that the \grs\ system is unique.


\end{abstract}


\section{Introduction} \label{sec:intro}

Low Mass X-ray Binaries (LMXBs) are stellar systems
consisting of a compact object, either a neutron star (NS) or black hole (BH), and a low-mass companion star. The companion star transfers matter onto the compact object through Roche-lobe overflow, and this accretion process releases gravitational energy, a large fraction of which emerges in the form of X-ray photons. 

Many LMXBs are transient X-ray sources, with epochs of lower and higher luminosity traditionally labelled as low and high states.  Many transients undergo extreme variations ranging from an inactive phase of quiescence punctuated by periods of X-ray bright outbursts. During outburst states, the accretion rate onto the compact object is much higher than the long-term average mass transfer rate from the donor star. 
Typical outburst durations are a few weeks, separated by several months or years of quiescence.

A different type of transient phenomenon in some LMXBs is the presence of Type-I X-ray bursts. Such bursts occur when mass transferred from the companion star accumulates on the surface of an accreting neutron star, and the accreted gas is compressed and heated to the point of thermonuclear ignition \citep{2003A&A...399..663K}. These bursts manifest as rapid increases in X-ray flux (over the course of a few seconds), followed by decay over tens to hundreds of seconds. The detection of Type-I bursts from an LMXB is a widely accepted diagnostic demonstrating that the compact object is a neutron star rather than a black hole.

\begin{table*}[htbp]
\begin{center}

    \caption{Summary of the \chandra/ACIS observations}
    \begin{tabular}{lcccccc}
    \hline\hline
    ObsID & Start time & Exp.\ Time  & Centroid$^a$ & $L_{\rm \rm {X,out}}$(0.3--10 keV)$^b$ & $L_{\rm \rm {X,ecl}}$(0.3--10 keV)$^c$ & Eclipse?\\
       &  (UTC)   &   (ks)   &   (RA, Dec) &(erg s$^{-1}$) &(erg s$^{-1}$) & \\
    \hline \\[-6pt]
     $21218^d$ & 2019-07-27  05:03:27 & $10.2$ & 267.69505, $-$31.27483 &   $(7.2 \pm 0.3) \times 10^{36}$ & $(1.2 \pm 0.2) \times 10^{35}$ &    YES, 2.6 ks\\
     23443 & 2021-04-20  20:11:18 & $30.0$ & 267.69554, $-$31.27462  & $(2.0 \pm 0.2) \times 10^{34}$ & $(2.4 \pm 0.8) \times 10^{33}$ &  YES, 5.3 ks\\
    $23444^d$ & 2021-06-01  11:27:35 & $30.1$ &267.69531, $-$31.27479  &  $(1.4 \pm 0.1) \times 10^{36}$ & $(2.1 \pm 0.4) \times 10^{34}$ & YES, 2.6 ks \\
     23441 & 2021-08-23  09:53:10 & $10.0$ & 267.69532, $-$31.27515 & $(3.5 \pm 0.5) \times 10^{34}$  & -- &  NO\\[4pt]
    \hline
    \multicolumn{7}{l}{$a$ Centroid of the source or center of the piled up region.} \\
    \multicolumn{7}{l}{$b$: X-ray luminosity out of the eclipse (power-law model)} \\
    \multicolumn{7}{l}{$c$: X-ray luminosity during the eclipse (power-law model)} \\
    \multicolumn{7}{l}{$d$: The source is 
    heavily piled up out of eclipse} \\
    \end{tabular}
    \label{tab:obslog}
\end{center}

\end{table*}

Globular clusters have high per-capita numbers of LMXBs, because of dynamical capture processes in dense stellar environments. The subject of this study, the LMXB GRS 1747$-$312, is  located in the inner region of the metal-rich, core-collapsed Milky Way globular cluster Terzan 6 \citep{1997A&AS..122..483B}. The distance to Terzan 6 is $9.5^{+3.3}_{\rm -2.5}$ kpc \citep{2003A&A...399..663K}. Terzan 6 has a core radius of $\approx$3\farcs3 and a tidal radius of  $\approx$316$^{\prime\prime}$ \citep{1995AJ....109..218T}. The most accurate position of GRS 1747$-$312, determined from X-ray observations with the High Resolution Camera on board the {\chandra\ X-ray Observatory}, is $(\alpha, \delta)_{\rm J2000.0} = (267.69526\degree, -31.27468\degree)$ \citep{2003A&A...406..233I}, with a 95\% uncertainty of 0\farcs4. This position places GRS 1747$-$312 about 0.9 core radii (in projection) from the center of Terzan 6. 

GRS 1747$-$312 was first detected in 1990 with the {\it{Roentgen Satellite}} ({\it{ROSAT}}) \citep{1991A&A...246L..21P} and with {\it{Granat}} \citep{1994ApJ...425..110P}. 
Its  X-ray luminosity varies from $\approx$5$ \times 10^{33}$ erg s$^{-1}$ in quiescence (\citealt{vats18}, from a 2004 {\it XMM-Newton} observations) up to several times $10^{36}$ erg s$^{-1}$ in its highest state (\citealt{2016PASJ...68S..15S} and this work). Its X-ray high states  recur at intervals of about 130--142 days, with a decay time of about 18 days \citep{2003A&A...406..233I}. Type-I bursts have been detected \citep{2003A&A...406..233I}, unambiguously identifying the compact object as a neutron star. It has no identified optical counterpart.

One reason why GRS 1747$-$312 is an interesting LMXB, studied with many X-ray satellites over the past three decades, is that it shows sharp X-ray eclipses during its outbursts. It is one of only 13 Galactic LMXBs with this property 
\citep{2023hxga.book..120B}.
{\it BeppoSAX} and {\it Rossi X-ray Timing Explorer} ({\it RXTE}) observations have determined an eclipse duration of 2596 s, with an orbital period of $P = 0.514980303$ days \citep{2003A&A...406..233I}. 
However, GRS 1747$-$312's eclipsing behavior is peculiar, and still not understood despite decades of observations. The most puzzling features reported in the literature \citep{2016PASJ...68S..15S} are: 1) eclipses are seen when the source is in a high state, but not when it is in a low state; 2) residual X-ray emission is seen during the eclipses, at a level comparable to the low-state luminosity, but variable from epoch to epoch; 3) a long duration dip (longer than standard eclipses) may have been observed in  2009 {\it{Suzaku}} observation \citep{2016PASJ...68S..15S}. 

A possible explanation for the on/off eclipse behavior, proposed by \cite{2016PASJ...68S..15S}, is that we are seeing two unresolved X-ray sources; the brighter one is a transient with the 2.6-ks eclipses. When the eclipsing source is in the off state, and during its X-ray eclipses, we are seeing the emission from an interloper source. The problem of this scenario is that GRS 1747$-$312 appears point-like even in \chandra/HRC images. Given the low density of X-ray sources in the region, the likelihood of a chance coincidence is low \citep{2021ApJ...923...88P}. 
Another possibility is that the high and low states correspond to dominant emission from the neutron star surface (and/or from a geometrically thin accretion disk), and from a vertically extended corona, respectively. Then, for a suitable range of the viewing angle, it is conceivable that the low-mass companion star eclipses the emission from inner disk and neutron star surface, but not (or only part of) the extended coronal emission. A third possibility is that higher and lower states are not physically different geometries of the accretion flow; instead, they could be the result of a mis-aligned and precessing inner accretion disk. For some phases of the precessing cycle, the disk itself might occult the surface emission, leaving only the extended coronal emission visible to us; thus, the apparently lower state would already be in permanent occultation, with no additional eclipses by the companion star. 

Another (possibly related) conundrum of this neutron star is its unusually high radio/X-ray luminosity ratio measured in one of three available radio observations \citep{2021ApJ...923...88P}, which places it along the black hole sequence; in the other two epochs, the ratio is at least an order of magnitude lower, consistent with the neutron stars population. This finding puts a strong caveat on the use of the radio/X-ray ratio to identify black hole transients in the low/hard state. On the other hand, the high ratio seen once in GRS 1747$-$312 might be explained if in that epoch, most of the intrinsic X-ray emission was occulted.  


In this paper, we have studied the X-ray properties of the source using archival but mostly unpublished data from various observatories: \chandra, the {\it Neutron Star Interior Composition Explorer} ({\it NICER}), \textit{Suzaku}, and {\it RXTE}. For all instruments, we focus mostly on the timing analysis, to understand the eclipse behavior; for the \chandra\ observations, we also present a short discussion of spectral properties, to place the light curve behavior in the context of the corresponding spectral state. More detailed spectral analysis is left to further work. 

We report the discovery of an occultation episode that is off-cycle from the expected time of the companion star eclipse, and is also of longer duration than a regular eclipse. We highlight the new observational findings, test previous models and scenarios, and propose new scenarios to explain the unusual eclipse behavior. We will first go through the observational results for each telescope in Section \ref{sec:observations}, perform a detailed analysis of the data in Section \ref{sec:detailed_lc}, and then present our interpretations in Section \ref{sec:models}.



\section{Observational results}\label{sec:observations}

\subsection{\chandra\ X-ray Observatory}



\subsubsection{Data analysis}
    The inner region of Terzan 6, including the location of GRS\,1747$-$312, has been imaged several times with the back-illuminated S3 chip of \chandra's Advanced CCD Imaging Spectrometer (ACIS). Here we use the four datasets that are currently publicly available: ObsID 21218 from 2019 July,  ObsID 23443 from 2021 April, ObsID 23444 from 2021 June, and ObsID 23441 from 2021 August (Table~\ref{tab:obslog}). We downloaded the data from the public archive and reprocessed them using the \chandra\ Interactive Analysis of Observation ({\sc ciao}) software package Version 4.14 \citep{2006SPIE.6270E..1VF}, with the Calibration Database Version 4.9.7.  We reprocessed the data with the {\sc ciao} task {\it chandra\_repro}, and applied barycenter corrections with {\it axbary}.

We inspected the images from the four observations with {\sc ds9}. In ObsIDs 21218 and 23444, the source is piled up, while in the other two epochs it was in a low state and not piled up. The piled-up observations can still be used for spectral and timing analysis, with some care in the selection of annular source extraction regions that excludes the most heavily piled-up pixels. Conversely, the low state of the source in ObsIDs 23441 and (especially) 23443 gives us the best chance to determine whether there are other, faint contaminating sources within a few arcsec of GRS\,1747$-$312, which only \chandra\ can resolve.
    


\begin{figure}[t]
    \centering
\includegraphics[width=4.2cm]{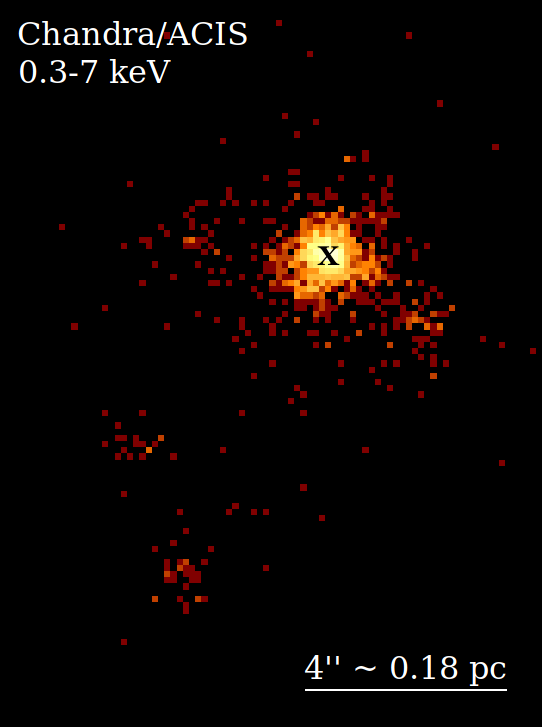}
\includegraphics[width=4.2cm]{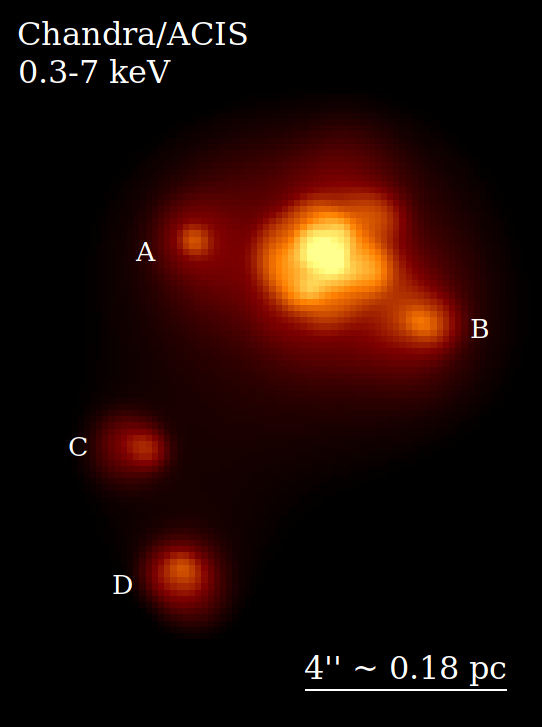}
    \caption{Left panel: \chandra/ACIS image of GRS\,1747$-$312 (marked with black ``X") and surrounding field, in the 0.3--7 keV band, from ObsID 23443. North is up and East to the left. Pixels were rebinned to 4-fold oversample the default pixel size in each dimension ({\it i.e.}, the ``effective'' pixel size shown is 0\farcs123). Right panel: same image, adaptively smoothed with the {\sc ciao} task {\it csmooth}, with minimum signal-to-noise ratio of 2. We estimate a total unabsorbed 0.3--10 keV luminosity of $(2.1 \pm 0.2) \times 10^{34}$ erg s$^{-1}$ from all the resolved and unresolved sources in this field; of this, GRS\,1747$-$312 contributes with $(1.7 \pm 0.2) \times 10^{34}$ erg s$^{-1}$. The luminosity of the four resolved sources A through D is given in Table 2. 
}
    \label{fig:smaller_sources}%
\end{figure}

For the spectral and timing analysis of GRS\,1747$-$312 from ObsID 21218, we chose an annular source region with inner radius of 3$^{\prime\prime}$ and outer radius of 10$^{\prime\prime}$, which excludes the depressed core of the point spread function. We also excluded from the source annulus its intersection with a rectangular box of width 2\farcs5, oriented at a $45\degree$ angle, to exclude the readout streak. ObsID 21218 contains an eclipse (Section 2.1.3): during the eclipse, faint residual emission from the source is still visible but not piled up. Thus, we also extracted the light curve and spectrum during the eclipse, using a circular aperture of 1\farcs5 for the source region. For ObsID 23444, the source extraction annulus had radii of 1\farcs5 and 10$^{\prime\prime}$, and the exclusion box for the readout streak was oriented at a $25\degree$ angle. ObsID 23444 also contains an eclipse (Section 2.1.3), during which residual emission is still detected but not piled up. The timing and spectral properties of the source during the eclipse interval were taken from a circular region of radius 1\farcs5.
For the other two ObsIDs when the source was in a low state (23443 and 23441), we also used a circular aperture of 1\farcs5.  We selected a background during the two high-state observations using a circular region of radius $70^{\prime\prime}$, located 250$^{\prime\prime}$ from the source at $({\rm RA},{\rm Dec})_{\rm 2000}=(267.69526,-31.27468)$. For the non-piled-up (eclipse) sub-intervals of those two observations, and for the other two (lower state) observations, we used an annular region with radii of 25$^{\prime\prime}$ and 95$^{\prime\prime}$ centered on the source.  


For our timing study, we extracted background-subtracted light curves in the 0.5--7 keV band for each \chandra\ ObsID using the {\sc ciao} {\it dmextract} tool. Subsequent light curve analysis was done with python packages such as {\it matplotlib} and {\it numpy}. We double-checked the main results with NASA's High Energy Astrophysics Software (HEASoft) Version 6.32 \citep{heasoft14}; in particular, the {\sc ftools} task {\it lcurve}.

For our spectral study, we used {\it specextract} to create source and background spectra and associate response and ancillary response files.
Whenever possible (and unless specifically indicated) regrouped the spectra to a minimum of 15 counts per bin, with the {\sc ftools} task {\it grppha}. We then modelled the spectra with {\sc xspec} Version 12.13.0 \citep{arnaud96}, using the $\chi^2$ fit statistics. For spectra with few net counts, we used the Cash fit statistics \citep{cash79}, grouping the spectra to a minimum of 1 count per bin.

\subsubsection{Source confusion near GRS\,1747$-$312}

The 30 ks observation of ObsID 23443 offers the best chance to date to inspect the contamination from nearby sources in Terzan 6. GRS\,1747$-$312 is still the brightest source in the field, even in a low state. We estimated its flux and luminosity first with a simple power-law fit, using the {\sc ciao} task {\it srcflux}, then with more complex spectral models in {\sc xspec} (Section 2.1.2). We obtain a 0.3--10 keV unabsorbed luminosity of $(1.7 \pm 0.2) \times 10^{34}$ erg s$^{-1}$. There are at least four point-like X-ray sources clearly resolved within 7$^{\prime\prime}$, labelled as A through D in Figure 1. The source region appears extended, with a bright core coming from the eclipsing X-ray binary, and probably several faint, unresolved sources within an $\approx$2$^{\prime\prime}$ radius around it (Figure 1). All the faint sources shown in Figure 1 are unresolved from GRS\,1747$-$312 in any X-ray study in the literature (including those based on {\it XMM-Newton} observations). Thus, we need to estimate the relative contribution of this contamination.

Using again {\it srcflux}, assuming a column density $N_{\rm \rm H} \approx 10^{22}$ cm$^{-2}$ (similar to the one derived for GRS\,1747$-$312, Section 2.1.2, and justified by the spectral energy distribution of the observed counts) and a power-law photon index $\Gamma \approx 1.5$ (again, similar to the one derived for GRS\,1747$-$312), we estimated the observed fluxes and unabsorbed model luminosities listed in Table 2. We also calculated the total X-ray luminosity (resolved and unresolved) within 10$^{\prime\prime}$ of GRS\,1747$-$312, and then subtracted the modelled contribution from the X-ray binary. We obtain an observed flux of $(1.1\pm0.4) \times 10^{-13}$ erg cm$^{-2}$ s$^{-1}$ in the 0.5--7 keV band, and an unabsorbed luminosity of $(4\pm1) \times 10^{33}$ erg s$^{-1}$ in the 0.3--10 keV band. This is the unavoidable contamination to the true X-ray binary contribution reported in previous X-ray studies. In particular, \cite{vats18} reported a 0.5--10 keV flux of $\approx$2 $\times 10^{-13}$ erg cm$^{-2}$ s$^{-1}$ and an unabsorbed 0.5--10 keV luminosity of $\approx$4.8 $\times 10^{33}$ erg s$^{-1}$ for the source in {\it XMM-Newton} observations from 2004 August. These values are  consistent or only slightly higher than the contaminating emission within 10$^{\prime\prime}$ estimated from our \chandra\ study. Given that this emission comes from several resolved and unresolved sources in the innermost region of Terzan 6, it is unlikely that it can decrease significantly from epoch to epoch. We conclude that the true luminosity of GRS\,1747$-$312 in the 2004 {\it XMM-Newton} observations was likely $\lesssim$10$^{33}$ erg s$^{-1}$, and the spectra modelled by \cite{vats18} are dominated by contaminating sources. 



\begin{table}
\begin{center}
    \caption{Contamination from other sources, measured from \chandra/ACIS ObsID 23443}
    \begin{tabular}{lcc}
    \hline\hline
    Source & $F$(0.5--7 keV)$^{\dagger}$ & $L$(0.3--10 keV)$^{\ddagger}$\\
           &   (erg cm$^{-2}$ s$^{-1}$)   &   (erg s$^{-1}$) \\
    \hline \\[-6pt]
     A  & $(2.5\pm0.7) \times 10^{-14}$   & $(6\pm2) \times 10^{32}$ \\
     B  &  $(4.5\pm1.0) \times 10^{-14}$ &  $(10\pm2) \times 10^{32}$\\
     C   &  $(1.2\pm0.4) \times 10^{-14}$ &  $(3\pm1) \times 10^{32}$\\
     D   & $(2.0\pm0.6) \times 10^{-14}$  &  $(5\pm2) \times 10^{32}$\\
     GRS\,1747$-$312  &  $(7.2\pm0.3) \times 10^{-13}$ &  $(1.7\pm0.2) \times 10^{34}$\\
     Total within 10$^{\prime\prime}$  &  $(8.3\pm0.3) \times 10^{-13}$ &  $(2.1\pm0.2) \times 10^{34}$\\[4pt]
    \hline
    \multicolumn{3}{l}{$\dagger$: Observed flux, estimated with {\it srcflux}} \\
    \multicolumn{3}{l}{$\ddagger$ Unabsorbed luminosity ($\Gamma = 1.5$ power-law model)} \\
    \end{tabular}
    \label{tab:contamination}
\end{center}
\end{table}

\begin{figure*}%
    \centering
{{\includegraphics[width=0.47\textwidth]{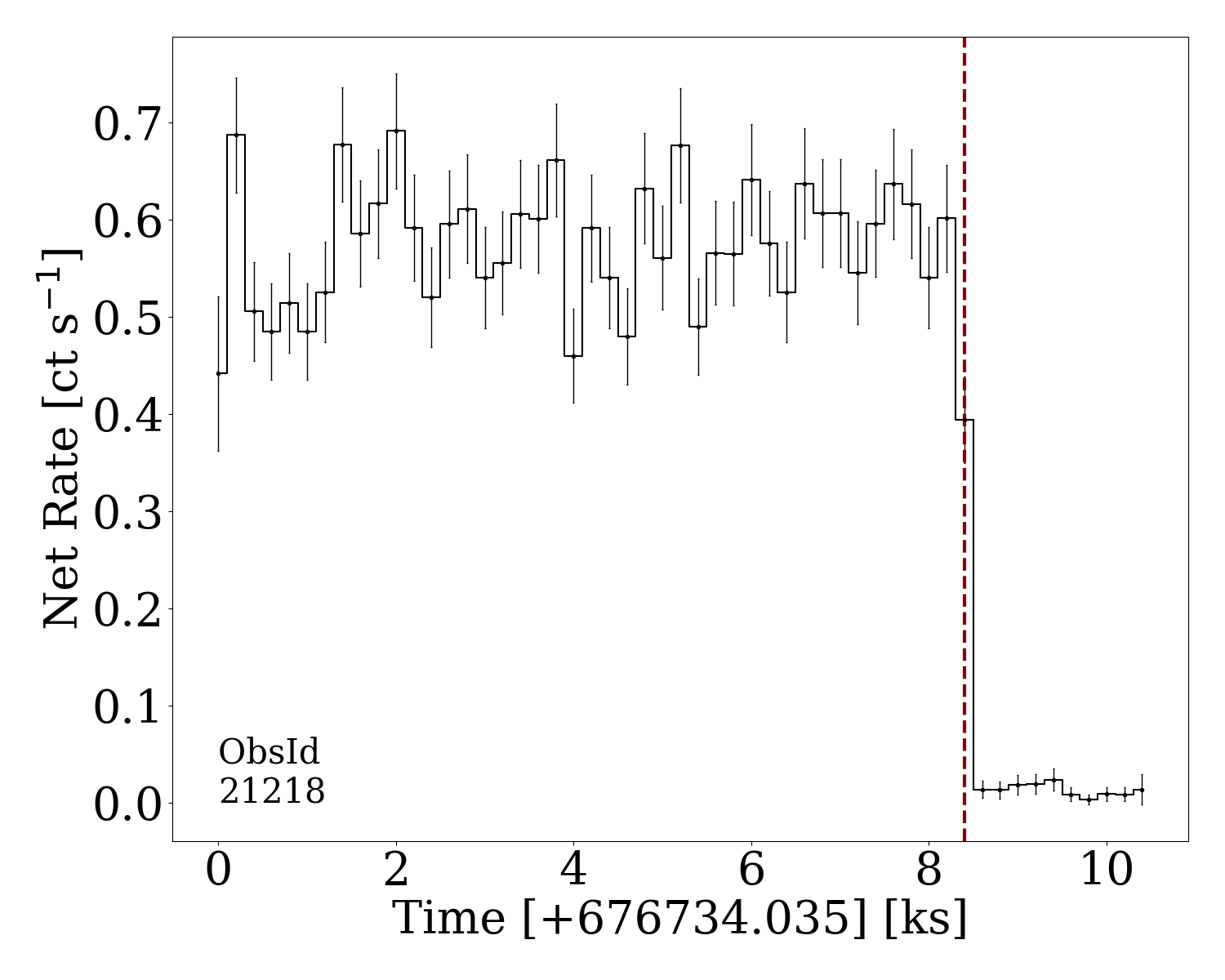} }}%
    \qquad
  {{\includegraphics[width=0.47\textwidth]{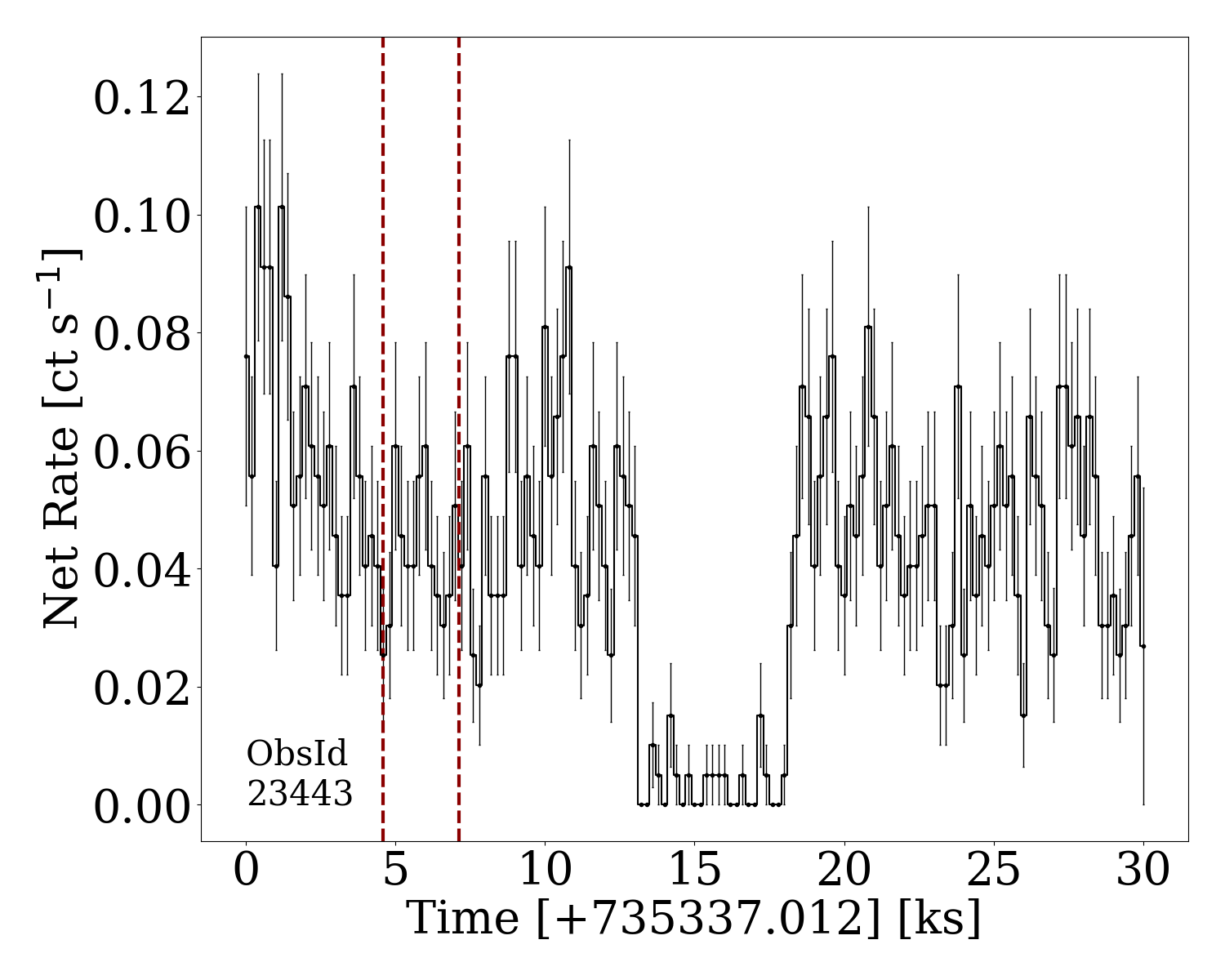} }}%
    
    {{\includegraphics[width=0.47\textwidth]{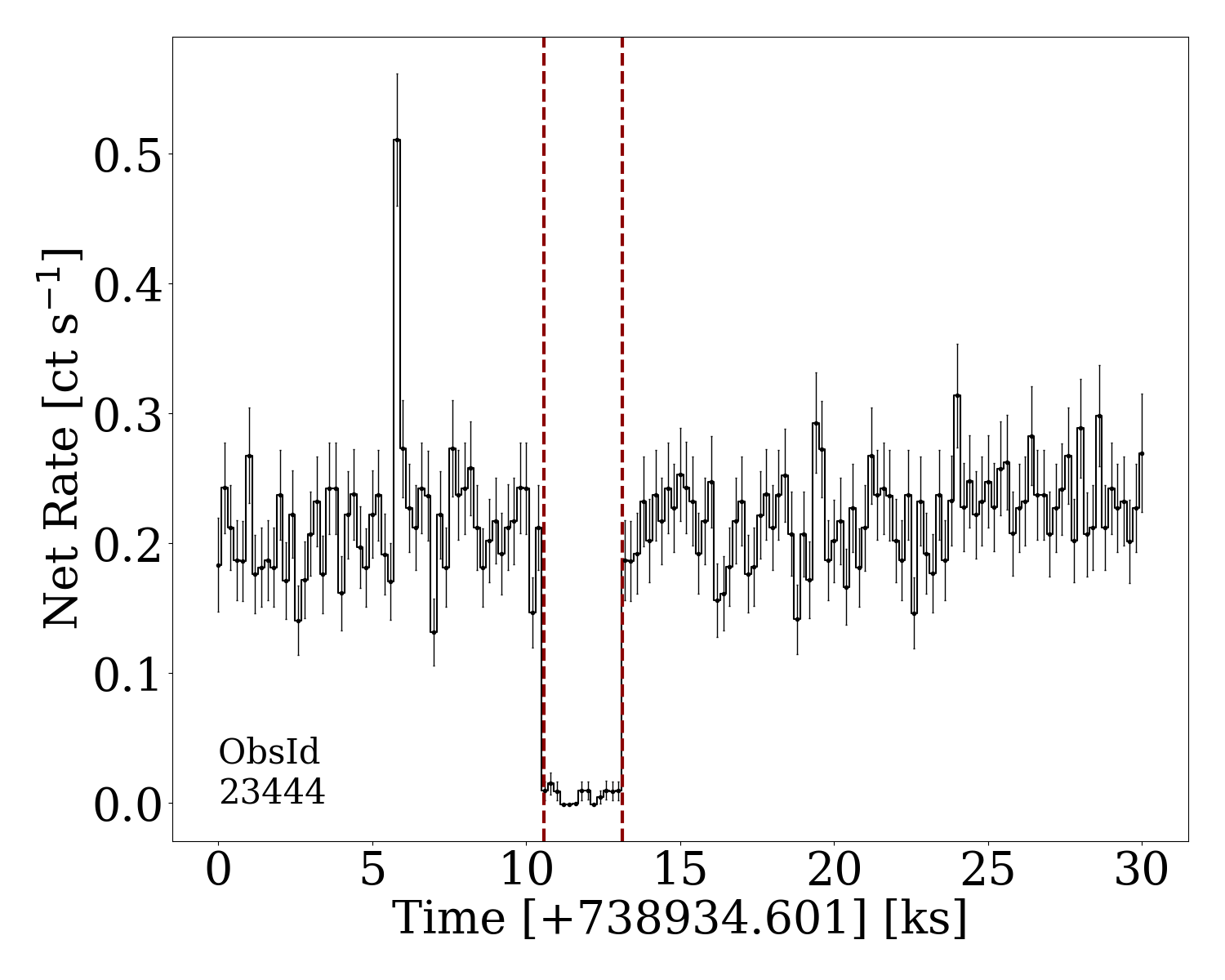} }}%
    \qquad
    {{\includegraphics[width=0.47\textwidth]{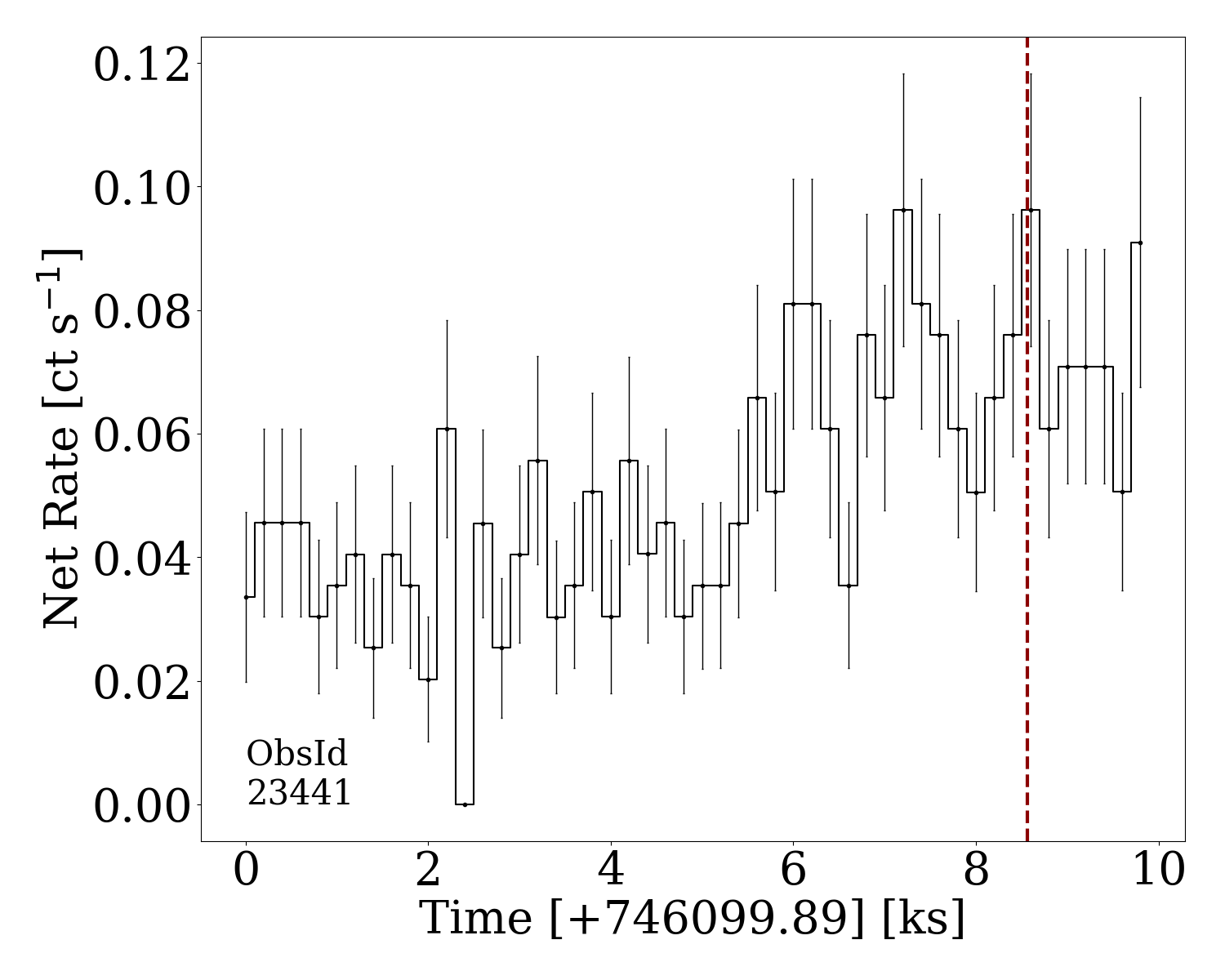} }}%

    \caption{\chandra\ light curves of GRS\,1748$-$312, binned at 200~s.  The background subracted net count rates for ObsIDs 21218 (top left), 23443 (top right), 23444 (bottom left), and 23441 (bottom right) are shown, with solid vertical bars denoting 1$\sigma$ uncertainties.  The red dashed vertical lines denote the ingress and egress points of the short $2.6$~ks eclipse.  Notice that the short eclipse is not detectable during low state (ObsIDs 23443 and 23441; see Table~\ref{tab:dips}).
   }%
    \label{fig:Chandra_lc}%
\end{figure*}

\subsection{Optical Counterparts for GRS 1747-312}

As of yet, there have been no optical counterparts to \grs\ reported in the literature. We examine HST images of Terzan 6 in the 606W and 110W filters to verify this. We check for counterparts using the radio position of \grs\ from \citet{2021ApJ...923...88P}, assuming a .15" error and using the GAIA DR2 catalogue to perform astrometric corrections to the HST image. We find no optical counterparts for the radio position in any of the filters. We estimate an upper limit on the absolute magnitude of a counterpart as $M>2.4$ in the 606W band, for which the resolution is highest. In the 110W band the field is dominated by unresolved emission, making an upper limit not very constraining. For this filter we estimate an upper limit on an optical counterpart as $M>.21$.

\subsubsection{Summary of timing results}

The two observations when GRS\,1748$-$312 was in a higher state ($L_{\rm \rm X} \sim 10^{36}$ erg s$^{-1}$) provided results consistent with previous X-ray studies (\cite{2003A&A...406..233I}, \cite{2016PASJ...68S..15S}). We recovered the previously known 2.6-ks eclipses (Figure~\ref{fig:Chandra_lc}), discussed in more detail in Section \ref{sec:short_eclipse}. The expected eclipse times are $T_{\rm \mathrm{I}}(n)=52066.259473(5)+n{\times}0.514980311(8)$~MJD and $T_{\rm \mathrm{E}}(n)=52066.289497(10)+n{\times}0.514980275(15)$~MJD for the $n^{\rm th}$ ingress and egress, respectively, based on the ephemeris of \citet{2003A&A...406..233I}. We verified the accuracy of this ephemeris: we show the predicted eclipse times as dashed vertical lines in the left-hand panels of Figure~\ref{fig:Chandra_lc}.

The two observations in the lower state ($L_{\rm \rm X} \sim 10^{34}$ erg s$^{-1}$) provided conflicting results, which make the behaviour of this X-ray binary even more puzzling. Neither observation shows eclipses at the expected times (dashed vertical lines in the right-hand panels of Figure~\ref{fig:Chandra_lc}), in line with what was known from previous X-ray monitoring \citep{2016PASJ...68S..15S}. The luminosity of the lower states is comparable to the luminosity during eclipses when the source is in a high state (Table 1). This is consistent with the previously proposed scenario \citep{2016PASJ...68S..15S} that the lower state is simply an epoch of ``permanent'' occultation of the direct neutron star emission, rather than a different accretion state. However, the most interesting finding is that one of two lower-state observations (ObsID 23443, top left panel of Figure 2) contains a different eclipse-like feature, longer (5.3 ks) than the regular 2.6-ks orbital eclipses, and starting at an orbital phase $\phi \approx 0.2$ from the expected mid-time of the (unseen) regular eclipse. The ingress and egress sharpness of this light-curve feature strongly suggest it is an eclipse by a solid object (see discussion in section \ref{sec:long_eclipse}). The luminosity of the source during the 5.3-ks eclipse (Table 1) is the lowest ever recorded ($L_{\rm \rm X} \approx (3\pm1) \times 10^{33}$ erg s$^{-1}$), an order of magnitude lower than during typical 2.6-ks eclipses. If regular donor-star eclipses (and perhaps lower-state epochs) represent an occultation of the direct emission but not of an extended corona, the 5.3-ks eclipse might represent an occultation also of most of the corona, and it might be caused by a larger occulting body.
In contrast, the lower-state light curve from ObsID 23441 shows intra-observational variability by a factor of 2 but no eclipses (Figure 2, bottom right panel). There is no feature at the expected ingress time of the 2.6-ks eclipse. The lack of a 5.3-ks eclipse in ObsID 23441 may simply be due to chance, as the system was observed only for 10 ks. However, there is an intriguing spectral difference between the emission in the two lower-state epochs.

\begin{table*}
\begin{center}
\centering
    \caption{Dips in \chandra\ Observations modeled as eclipses}
    \begin{tabular}{l c c c cc}
    \hline\hline
    ObsID & type & duration & SNR$^\Vert$ & ingress$^\dag$ & egress$^\dag$ \\
    \hfil & \hfil & [ks] & \hfil & \multicolumn{1}{c}{[ks]} & [ks]\\
    \hline
    21218 & short & $>2$  & 60.6 & $0-0.03^\Uparrow$ & \nodata \\
    23443 & short & $2.26-5.5^\Uparrow$ &  1.8 & $0-1.8^\Uparrow$ & $0-1.4^\Uparrow$ \\
    \hfil & long & $5.1-5.4^\Uparrow$ & 13.4 & $0-0.27^\Uparrow$ & $0-0.19^\Uparrow$ \\
    23444 & short & $2.6-2.68^\Uparrow$ & 38.5 & $0-0.004^\Uparrow$ & $0-0.08^\Uparrow$ \\
    23441 & short & $>1.5$  & 0.7 & $0-2.1^\Uparrow$ & \nodata \\
    \hfil & long$^?$ & $>5$ & 5.5 & \nodata & $0-3.15^\Uparrow$ \\
    \hline
    \multicolumn{6}{l}{$\dag$: Taken from in and out of eclipse fluxes in Tables \ref{tab:21218_params},\ref{tab:23444_params},\ref{tab:23443_params}.} \\
    \multicolumn{6}{l}{$\Vert$: S/N of the depth of the dip, from the baseline and dip count rates in the observations. } \\
    \multicolumn{6}{l}{$\Uparrow$: 68\% highest posterior density (HPD) interval} \\
    \multicolumn{6}{l}{$?$: Assuming that the apparent change in state at $+6$~ks is due to an eclipse egress} \\
    \end{tabular}
    \label{tab:dips}
\end{center}

\end{table*}

\subsubsection{Spectral results}
\label{sec:spectral_analysis}

\begin{figure}[t]
\hspace{-0.4cm}
\includegraphics[width=7cm, angle=270]{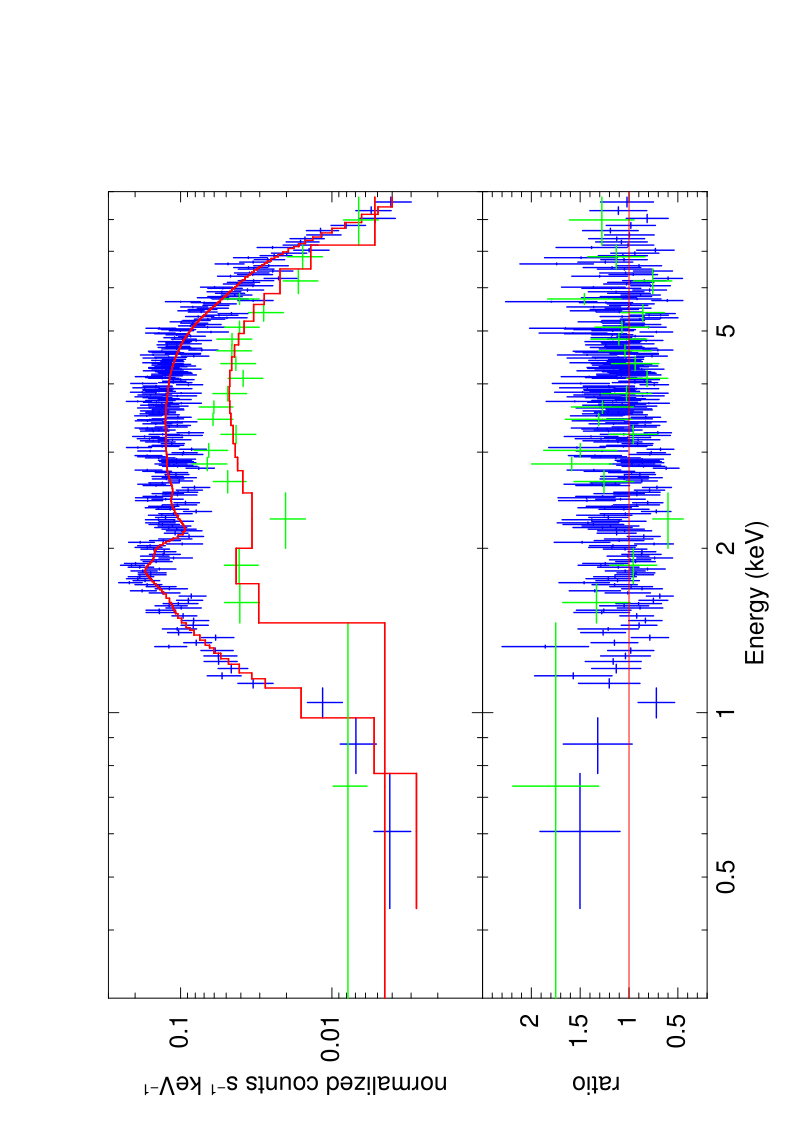}
    \caption{\chandra/ACIS spectrum from ObsID 21218, fitted with an absorbed power-law model (Table \ref{tab:21218_params} for the fit parameters). Blue datapoints are extracted from the out-of-eclipse time interval; green datapoints are from the 2.6-ks eclipse.
}
    \label{fig:21218_po_both}%
\end{figure}

\begin{figure}[t]
\hspace{-0.4cm}
\includegraphics[width=7cm, angle=270]{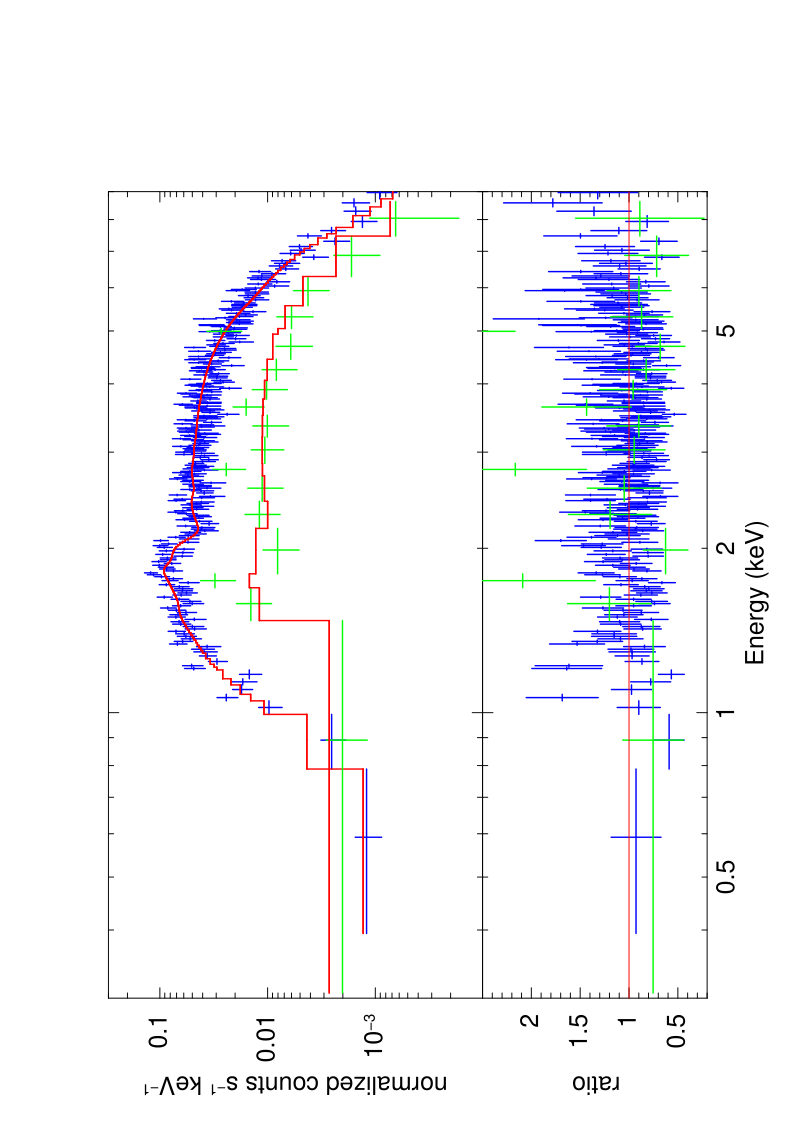}
\caption{\chandra/ACIS spectrum from ObsID 23444, fitted with an absorbed power-law model (Table \ref{tab:23444_params} for the fit parameters). Blue datapoints are extracted from the out-of-eclipse time interval; green datapoints are from the 2.6-ks eclipse.}
\label{fig:23444_po_both}
\end{figure}

\begin{figure}[t]
\hspace{-0.4cm}
\includegraphics[width=7cm, angle=270]{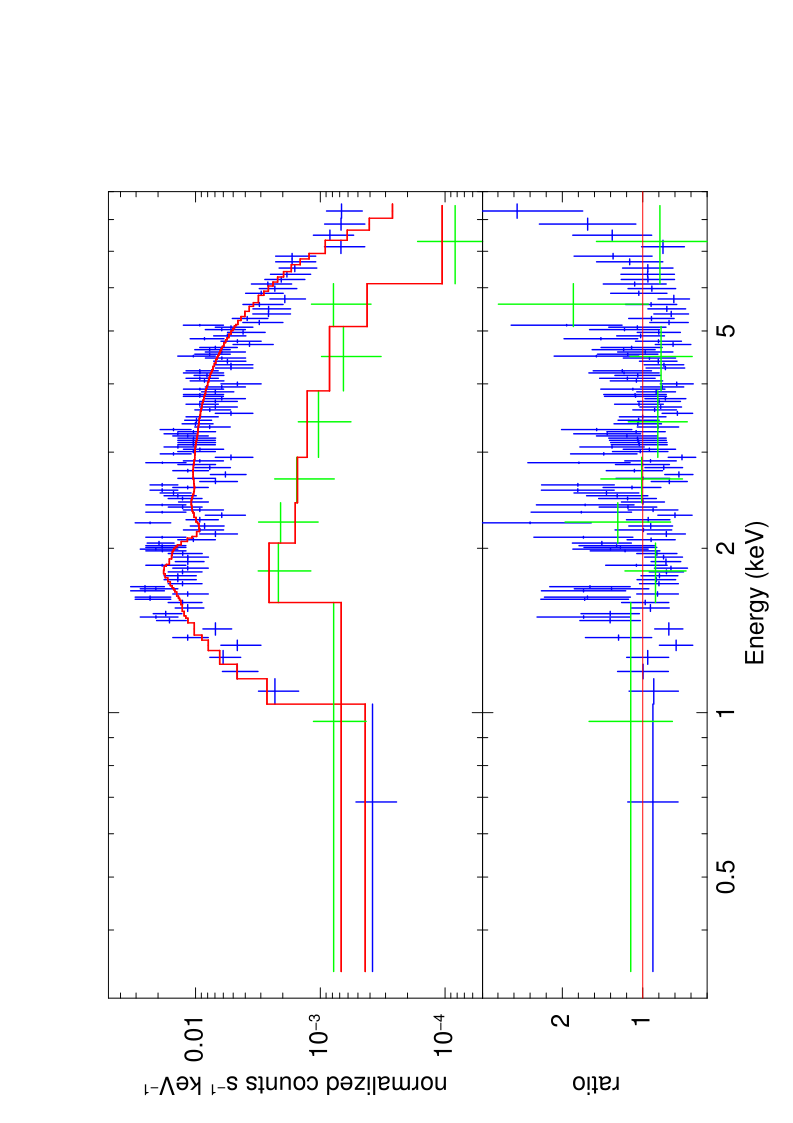}
    \caption{\chandra/ACIS spectrum from ObsID 23443, fitted with an absorbed power-law model (Table \ref{tab:23443_params} for the fit parameters). Blue datapoints are extracted from the out-of-eclipse time interval; green datapoints are from the 5.3-ks eclipse. For plotting purposes only, out-of-eclipse datapoints were rebinned to a signal-to-noise ratio $>2.8$ while eclipse datapoints were rebinned to a signal-to-noise ratio $>2.0$.
}
    \label{fig:23443_po}%
\end{figure}

\begin{figure*}[t]
\hspace{-0.4cm}
\includegraphics[width=7cm, angle=270]{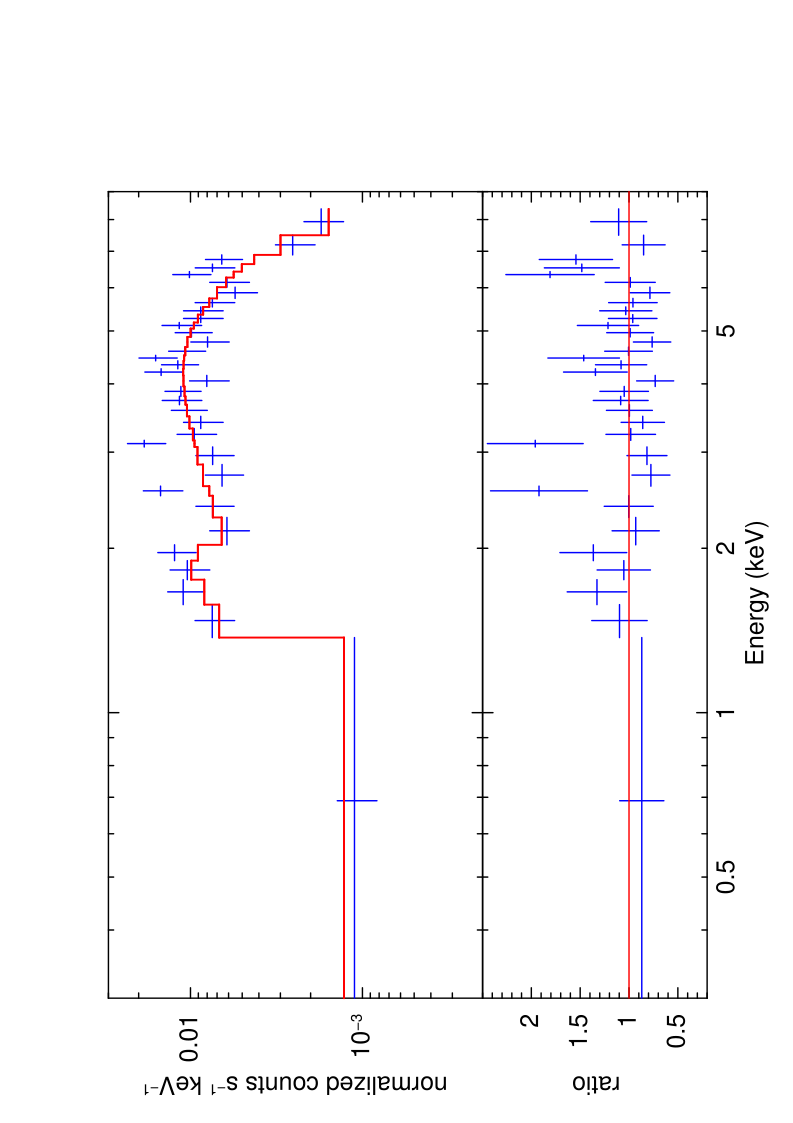}
\includegraphics[width=7cm, angle=270]{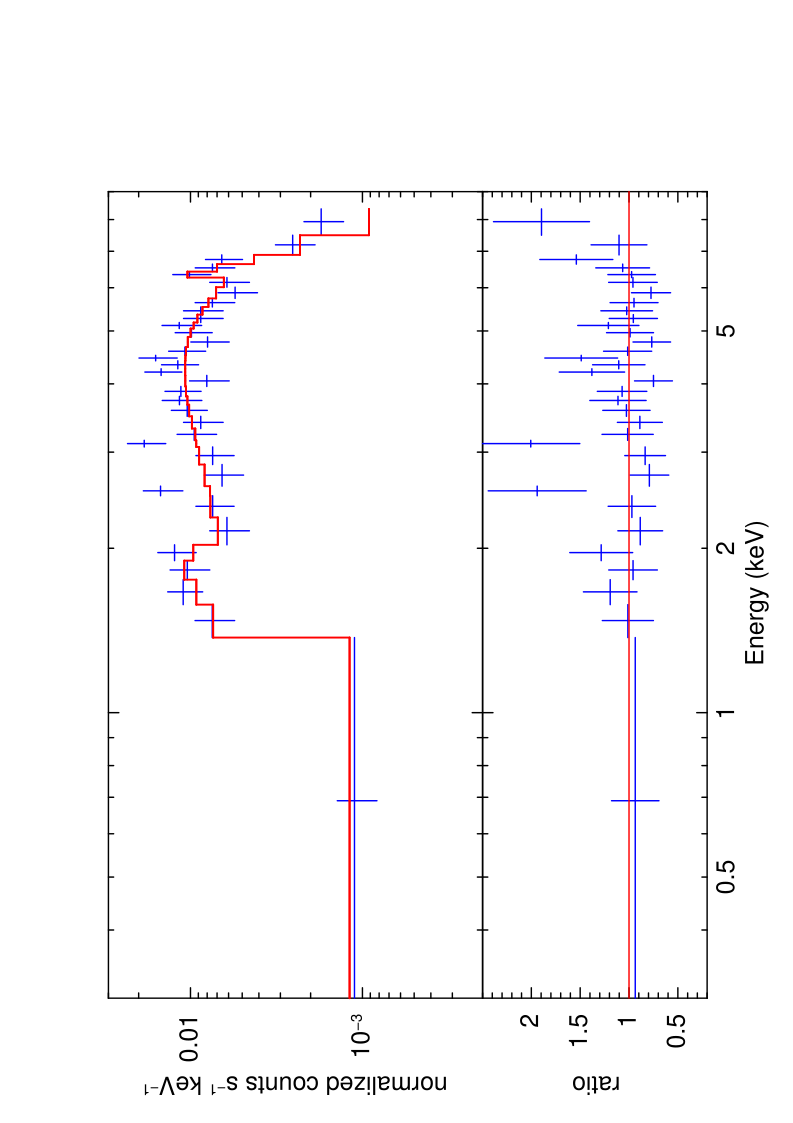}
    \caption{Left panel: \chandra/ACIS spectrum from ObsID 23441, fitted with an absorbed power-law model (Table \ref{tab:23441_params} for the fit parameters). Right panel: the same spectrum, fitted with the {\it pexrav} reflection model plus a Gaussian emission line at 6.4 keV.
}
    \label{fig:23441_po_pexrav}%
\end{figure*}

While we do not perform a comprehensive spectral study, only a few properties are necessary to understand the timing and eclipsing behaviour. In particular, we analyze the spectra during eclipse sub-intervals and in the lower luminosity states, because only {\it Chandra} can resolve the source at those low fluxes.
The in-eclipse sub-intervals during the higher luminosity epochs (ObsIDs 21218 and 23444) have significantly harder spectra than the out-of-eclipse ones (Figures \ref{fig:21218_po_both},\ref{fig:23444_po_both}, and Tables \ref{tab:21218_params},\ref{tab:23444_params}). This is even more noteworthy considering that the out-of-eclipse spectra may appear artificially hardened owing to pile-up. In contrast, the 5.3-ks eclipse spectrum during the lower luminosity ObsID 23443 is not harder than the out-of-eclipse spectrum (Figure \ref{fig:23443_po}, Table \ref{tab:23443_params}). The spectrum from ObsID 23443 is consistent with what is expected from an X-ray binary in the low/hard state, with a simple power-law, $\Gamma \approx 1.5$, and column density similar to what is found in other observations \citep{2016PASJ...68S..15S}. Instead, the spectrum from ObsID 23441 is inconsistent with typical low-state X-ray binaries. When fitted with a simple power-law (Figure \ref{fig:23441_po_pexrav} left, and Table \ref{tab:23441_params}), the photon index is unphysically low ($<-0.2$). The column density is too low compared with other observations. And, intriguingly, there is a hint of a 6.4-keV emission line (Figure \ref{fig:23441_po_pexrav} right, and Table \ref{tab:23441_params}). Taken together, these properties suggest a Compton refection spectrum. Indeed, the spectrum from ObsID 23441 is well fitted with a {\it pexrav} reflection model plus Gaussian line (other examples of reflection components in the X-ray spectra of neutron star LMXBs are in \cite{fiocchi07}). The X-ray luminosities derived from the spectral modelling and summarized in Table \ref{tab:contamination} are apparent isotropic luminosities.

The Compton spectral reflection in ObsID 23441 could suggest that the emission we see is only a fraction of the X-ray photons reflected off another object, and thus is only a small percentage of the true luminosity. This object could be the disk, clouds, or the face of the other eclipser. This suggests that the low state could be a state in which the direct emission is permanently occulted (similar to as during an eclipse, but by a different structure, such as a precessing disk).

\begin{table}
\caption{Best-fitting parameters of the {\it Chandra}/ACIS spectra from ObsID 21218, split between out-of-eclipse and eclipse sub-intervals. The fitting model is {\it tbabs} $\times$ {\it po}. Uncertainties are 90\% confidence limits for one interesting parameter.} 
\vspace{-0.3cm}
\begin{center}  
\begin{tabular}{lcc} 
 \hline 
\hline \\[-8pt]
  Model Parameters      &      \multicolumn{2}{c}{Values} \\
  & Out of Eclipse       &       Eclipse       \\
\hline\\[-9pt]
   $N_{\rm \rm {H}}^a$   ($10^{22}$ cm$^{-2}$)   &  \multicolumn{2}{c}{$ 1.6^{+0.2}_{\rm -0.2}$} \\[4pt]
   $\Gamma$      &  $0.92^{+0.09}_{\rm -0.09}$       &   $ 0.05^{+0.24}_{\rm -0.24}$   \\[4pt] 
   $N_{\rm \rm {po}}^b $ 
              &  $ 380^{+60}_{\rm -50}$   &   $ 1.5^{+0.6}_{\rm -0.5}$ \\[4pt]
   $\chi^2$/dof     &   $250.1/282$ (0.89)   & $21.9/19$ (1.15) \\[4pt]
   $f_{\rm 0.5-7}^c$
       & $340^{+7}_{\rm -8}$ & $4.9^{+0.6}_{\rm -0.5} $\\[4pt]
   $L_{\rm 0.3-10}^d$ 
      & $ 720^{+30}_{\rm -30}$ & $12^{+2}_{\rm -2}$\\[2pt]
\hline 
\vspace{-0.5cm}
\end{tabular}
\label{tab:21218_params}

\end{center}
\begin{flushleft} 
$^a$: locked for the out-of-eclipse and eclipse sub-intervals;\\
$^b$: units of $10^{-4}$ photons keV$^{-1}$ cm$^{-2}$ s$^{-1}$ at 1 keV;\\
$^c$: observed fluxes in the 0.5--7 keV band, in units of $10 ^{-12}$ erg cm$^{-2}$ s$^{-1}$;\\
$^d$: isotropic unabsorbed luminosities in the 0.3--10 keV band, defined as $4\pi d^2$ times the unabsorbed model fluxes; units of $10 ^{34}$ erg  s$^{-1}$.
\end{flushleft}
\end{table}

\begin{table}
\caption{Best-fitting parameters of the {\it Chandra}/ACIS spectra from ObsID 23444, split between out-of-eclipse and eclipse sub-intervals. The fitting model is {\it tbabs} $\times$ {\it po}. Uncertainties are 90\% confidence limits for one interesting parameter.} 
\vspace{-0.3cm}
\begin{center}  
\begin{tabular}{lcc} 
 \hline 
\hline \\[-8pt]
  Model Parameters      &      \multicolumn{2}{c}{Values} \\
  & Out of Eclipse       &       Eclipse       \\
\hline\\[-9pt]
   $N_{\rm \rm {H}}^a$   ($10^{22}$ cm$^{-2}$)   &  \multicolumn{2}{c}{$ 1.3^{+0.2}_{\rm -0.2}$} \\[4pt]
   $\Gamma$      &  $1.54^{+0.09}_{\rm -0.08}$       &   $ 0.71^{+0.31}_{\rm -0.31}$   \\[4pt] 
   $N_{\rm \rm {po}}^b $ 
              &  $ 166^{+21}_{\rm -19}$   &   $ 0.80^{+0.35}_{\rm -0.27}$ \\[4pt]
   $\chi^2$/dof     &   $305.1/278$ (1.10)   & $6.3/8$ (1.15) \\[4pt]
   $f_{\rm 0.5-7}^c$
       & $66^{+2}_{\rm -2}$ & $0.98^{+0.16}_{\rm -0.14} $\\[4pt]
   $L_{\rm 0.3-10}^d$ 
      & $ 143^{+6}_{\rm -6}$ & $2.1^{+0.4}_{\rm -0.4}$\\[2pt]
\hline 
\vspace{-0.5cm}
\end{tabular}
\end{center}
\label{tab:23444_params}
\begin{flushleft} 
$^a$: locked for the out-of-eclipse and eclipse sub-intervals;\\
$^b$: units of $10^{-4}$ photons keV$^{-1}$ cm$^{-2}$ s$^{-1}$ at 1 keV;\\
$^c$: observed fluxes in the 0.5--7 keV band, in units of $10 ^{-12}$ erg cm$^{-2}$ s$^{-1}$;\\
$^d$: isotropic de-absorbed luminosities in the 0.3--10 keV band, defined as $4\pi d^2$ times the de-absorbed model fluxes; units of $10 ^{34}$ erg  s$^{-1}$.
\end{flushleft}
\end{table}

\begin{table}
\caption{Best-fitting parameters of the {\it Chandra}/ACIS spectra from ObsID 23443, split between out-of-eclipse and eclipse sub-intervals. The fitting model is {\it tbabs} $\times$ {\it po}. Uncertainties are 90\% confidence limits for one interesting parameter. The Cash fitting statistics was used for this fit.} 
\vspace{-0.3cm}
\begin{center}  
\begin{tabular}{lcc} 
 \hline 
\hline \\[-8pt]
  Model Parameters      &      \multicolumn{2}{c}{Values} \\
  & Out of Eclipse       &       Eclipse       \\
\hline\\[-9pt]
   $N_{\rm \rm {H}}^a$   ($10^{22}$ cm$^{-2}$)   &  \multicolumn{2}{c}{$ 1.7^{+0.4}_{\rm -0.3}$} \\[4pt]
   $\Gamma$      &  $1.28^{+0.19}_{\rm -0.18}$       &   $ 1.55^{+0.67}_{\rm -0.66}$   \\[4pt] 
   $N_{\rm \rm {po}}^b $ 
              &  $ 1.8^{+0.4}_{\rm -0.4}$   &   $ 0.27^{+0.30}_{\rm -0.15}$ \\[4pt]
   C-stat/dof     &   $423.9/385$ (1.10)   & $18.6/29$ (0.64) \\[4pt]
   $f_{\rm 0.5-7}^c$
       & $0.95^{+0.05}_{\rm -0.05}$ & $0.10^{+0.04}_{\rm -0.03} $\\[4pt]
   $L_{\rm 0.3-10}^d$ 
      & $ 2.0^{+0.2}_{\rm -0.2}$ & $0.24^{+0.09}_{\rm -0.07}$\\[2pt]
\hline 
\vspace{-0.5cm}
\end{tabular}
\label{tab:23443_params}

\end{center}
\begin{flushleft} 
$^a$: locked for the out-of-eclipse and eclipse sub-intervals;\\
$^b$: units of $10^{-4}$ photons keV$^{-1}$ cm$^{-2}$ s$^{-1}$ at 1 keV;\\
$^c$: observed fluxes in the 0.5--7 keV band, in units of $10 ^{-12}$ erg cm$^{-2}$ s$^{-1}$;\\
$^d$: isotropic unabsorbed luminosities in the 0.3--10 keV band, defined as $4\pi d^2$ times the unabsorbed model fluxes; units of $10 ^{34}$ erg  s$^{-1}$.
\end{flushleft}
\end{table}

\begin{table}

\caption{Best-fitting parameters of the {\it Chandra}/ACIS spectra from ObsID 23441, for two alternative models. The first model is {\it tbabs} $\times$ {\it po}, the second model is {\it tbabs} $\times$ ({\it pexrav} $+$ {\it gaussian}. Uncertainties are 90\% confidence limits for one interesting parameter.} 
\vspace{-0.3cm}
\begin{center}  
\begin{tabular}{lc} 
 \hline 
\hline \\[-8pt]
  Model Parameters      &      Values \\
\hline \\[-9pt]
\multicolumn{2}{c}{{\it tbabs} $\times$ {\it powerlaw} }      \\[3pt]
\hline\\[-9pt]
   $N_{\rm \rm {H}}$   ($10^{22}$ cm$^{-2}$)   & $0.17^{+0.62}_{\rm \ast}$ \\[4pt]
   $\Gamma$      &  $-0.44^{+0.28}_{\rm -0.22}$ \\[4pt] 
   $N_{\rm \rm {po}}^a $ 
              &  $ 0.18^{+0.10}_{\rm -0.05}$  \\[4pt]
   $\chi^2$/dof     &   $28.4/31$ (0.92)\\[4pt]
   $f_{\rm 0.5-7}^b$
       & $1.32^{+0.12}_{\rm -0.11}$ \\[4pt]
   $L_{\rm 0.3-10}^c$ 
      & $ 3.5^{+0.5}_{\rm -0.5}$ \\[2pt]
      \hline \\[-9pt]
\multicolumn{2}{c}{ {\it tbabs} $\times$ ({\it pexrav} $+$ {\it gaussian})}      \\[3pt]
\hline\\[-9pt]
   $N_{\rm \rm {H}}$   ($10^{22}$ cm$^{-2}$)   & $1.2^{+0.8}_{\rm -0.7}$ \\[4pt]   $\Gamma$      &  $[1.5]$ \\[4pt] 
   $E_{\rm \rm fold}$ (keV)     &  $[100]$ \\[4pt] 
   ${\rm Rel}_{\rm \rm refl}$   &  $77^{+126}_{\rm -38}$ \\[4pt]
   $\cos i$   &  $[0.45]$ \\[4pt]
   $N_{\rm \rm {po}}^a $ 
              &  $ 0.58^{+0.40}_{\rm -0.33}$  \\[4pt]
   $E_{\rm \rm {line}}$ (keV)
              &  $[6.4]$  \\[4pt]
   $\sigma_{\rm \rm {line}}$ (keV)
              &  $[0.0]$  \\[4pt]
   $N_{\rm \rm {line}}$ ($10^{-6}$)$^d$
              &  $8.9^{+7.1}_{\rm -7.1}$  \\[4pt]
   $\chi^2$/dof     &   $25.2/30$ (0.84)\\[4pt]
   $f_{\rm 0.5-7}^b$
       & $1.40^{+0.13}_{\rm -0.11}$ \\[4pt]
   $L_{\rm 0.3-10}^c$ 
      & $ 2.9^{+0.3}_{\rm -0.3}$ \\[2pt]
\hline 
\vspace{-0.5cm}
\end{tabular}
\label{tab:23441_params}

\end{center}
\begin{flushleft} 
$^a$: units of $10^{-4}$ photons keV$^{-1}$ cm$^{-2}$ s$^{-1}$ at 1 keV;\\
$^b$: observed fluxes in the 0.5--7 keV band, in units of $10 ^{-12}$ erg cm$^{-2}$ s$^{-1}$;\\
$^c$: isotropic de-absorbed luminosities in the 0.3--10 keV band, defined as $4\pi d^2$ times the de-absorbed model fluxes; units of $10 ^{34}$ erg  s$^{-1}$;\\
$^d$: corresponding to an equivalent width of $(0.22\pm0.17)$ keV.
\end{flushleft}
\end{table}

\subsection{Source centroid at higher and lower fluxes}

We examine the X-ray images of each \chandra\ observation to determine whether only \grs\ is visible. Due to \grs's location in a globular cluster with high stellar density, there may be other nearby X-ray transients. Previous papers have suggested the possibility of a  contaminant source that is the cause of  emission during the low state and when the source is in eclipse during the high state  \citep{2016PASJ...68S..15S}. With two \chandra\ observations apiece for both of the low and the high state, we now are able to answer this question more definitively. 

We first determine whether there is a shift in the source's position during versus out of eclipse while the source is in outburst, using the methods described in Appendix \ref{chap:astrometry}. We find that the average centroid location between in eclipse and out of eclipse differed by $\sim0.75"$ and $\sim0.2"$, for observations 23444 and 21218, respectively. While these are on the high side for ACIS, they are within spec for absolute astrometry, and not rejectable. Taking into account the increased noise due to the lower number of counts during the eclipse vs outside of it, we find that the centroid location during eclipse falls within the $1\sigma$ error bounds of the centroid location outside eclipse for ObsID 21218, but represents a close to $2\sigma$ difference for ObsID 23444. We conclude that there is no strong evidence for a shift in source position during eclipse versus out of eclipse for ObsID 21218, and that there is a suggestive but not definite one for ObsID 23444.

To determine if there is a shift in position between the high-state and low-state observations, we correct for differences in astrometry using nearby point sources.  We use a total of 12, selected by eye, the locations of which are listed in Table \ref{tab:point_sources} (see Appendix \ref{chap:astrometry} for table and methods). We find that there is a $\sim2\sigma$ difference in the positions of the high-state observations and low-state observations, corresponding to a difference of $\sim0.75"$. This is again within the astrometric uncertainties of Chandra, but the clustering of the low and high states could point to a genuine physical shift in position. The position of \grs\ during the 2.6 ks eclipse in ObsID 23444 also clusters with the low state positions, while the position of \grs\ during the 2.6 ks eclipse in ObsID 21218 clusters with the high state positions.

\subsection{NICER}

\subsubsection{Data analysis}

\textit{NICER} is a non-imaging soft X-ray spectral-timing mission mounted on the International Space Station.  It consists of 56 silicon-drift detectors, each of which is paired with a co-aligned concentrator optic.  \textit{NICER}'s field of view is approximately 30 arcmin$^2$.  Since its launch in 2017, 52 detectors have been actively used, though in any given observation, one or more may be disabled.   \textit{NICER} observed Terzan 6 intermittently over a period from 2017 Nov. 1 through 2022 Oct. 27.   Each \textit{NICER} observation was organized into its continuous good-time-interval (GTI) component pointings, each of which was screened and reduced using default criteria in {\sc nicerl2}, except for an unrestricted allowance for both undershoot and overshoot event types, which correspond primarily to particle background and optical-loading events, respectively.  
GTIs were candidates for science analysis if their average per-detector undershoot rates $<$80 s$^{-1}$, the (52-detector equivalent) 0.2--12 keV source count-rate was at least 0.5 s$^{-1}$, the geomagnetic cutoff rigidity (``COR\_SAX'' $>$ 2),
and the corresponding background was at least 2.5 times below the source signal.  A total of $\sim$60 ks among 89 selected GTIs were used. Of these 89 GTIs, eclipses were predicted to be visible during 7 of them, which are summarized in Table \ref{NICER observations}.

\begin{table}
\begin{center}
\caption{Summary of \textit{NICER} observations with eclipses } 
\label{NICER observations}

\begin{tabular}{lcc}
\hline 
\midrule
ObsID & GTI Start time & GTI Exposure (s) \\

 2050440103  & 2019/06/11 03:37:02 & $902$ \\
  & 2019/06/11 16:03:28 & $1512$ \\

 2050440104 & 2019/06/12 04:10:25. & $1154$ \\
2050440105 & 2019/06/13 04:56:04 & $1453$ \\
2050440106 & 2019/06/14 18:04:07 & $1193$  \\
 2050440107 & 2019/06/15 18:49:48 & $972$  \\
  & 2019/06/15 06:27:06 & $1613$  \\

\hline
\end{tabular}
 \end{center}

\end{table}

In each case, prior to extracting light-curves from a given GTI, the detectors were screened to cull any outliers on the basis of undershoot, overshoot, or X-ray event rates (equivalent to 10-sigma distance from the detector ensemble median).  In each case, between 0-3 detectors were excluded, with most of the exclusions owing to high undershoot rates produced by optical loading at low Sun angles.  In addition to this screening, detectors 14, 34, and 54 were always excluded owing to each having presented intermittent calibration issues in the early mission, leaving 46-49 selected detectors for all observations in question.  Following the application of this screening, light curves were extracted in 0.125s intervals (later rebinned in time) in each of several energy ranges.

\textit{NICER} background light-curves were estimated using a background model analogous to the ``3C50'' model \citep{2022AJ....163..130R}, but indexing blank-sky background fields using overshoot rate and ``COR\_SAX'' as substitutes for ``IBG'' and ``hrej''.  These two alternative quantities offer the advantage of having higher event rates (producing better statistics) and continuous mapping, respectively, while correlating very closely with the 3C50 basis parameters.  This importantly enables a higher time-resolution background estimate.  Using this approach, a background spectrum was computed per 8s time interval, and used to produce a corresponding set of background light curves.

\subsubsection{Summary of timing results}

\begin{figure}[h]
    \includegraphics[width=.48\textwidth]{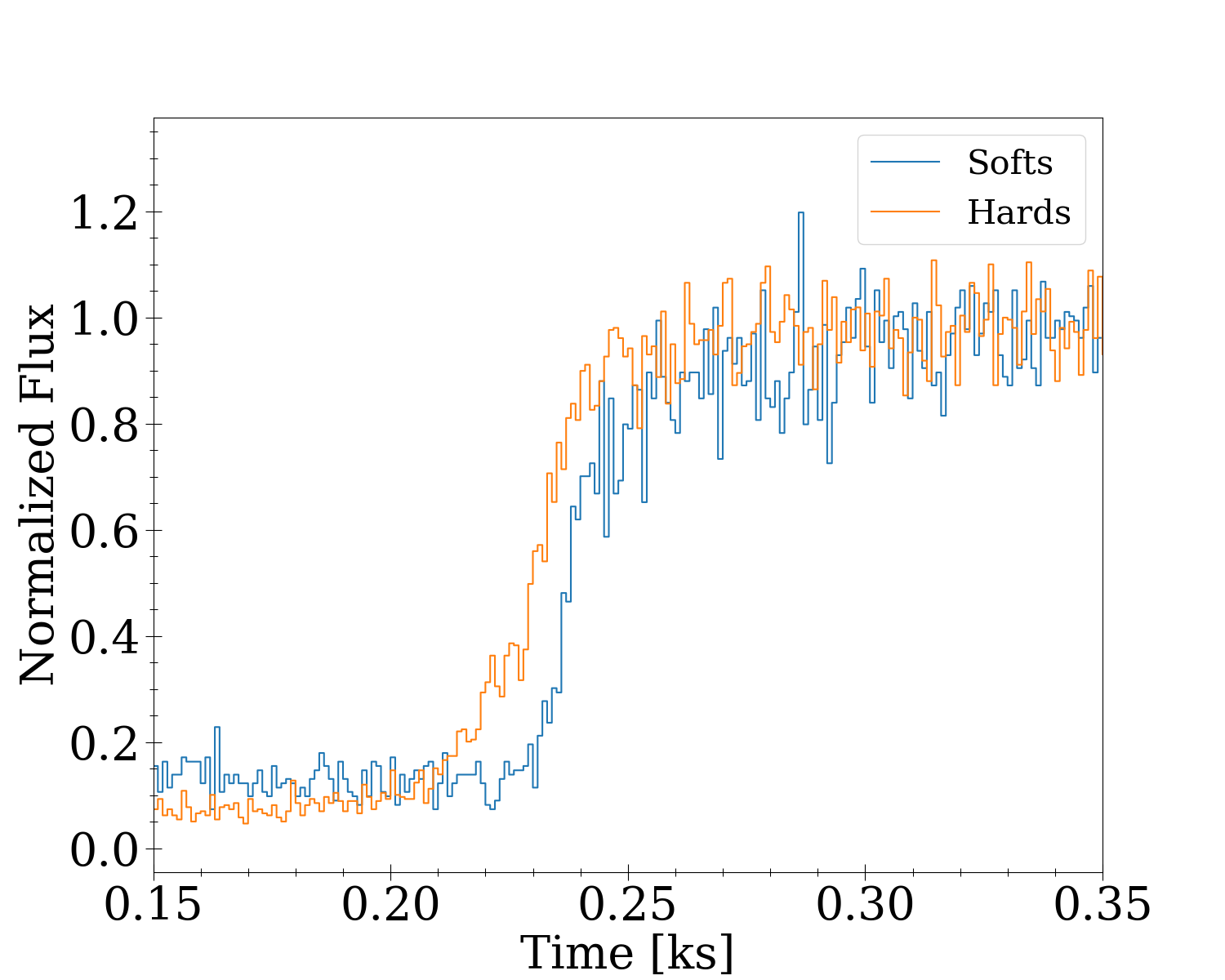}

    \caption{Stacked \textit{NICER} egress profile in the .5-2 keV and 2-6 keV bands, 1s bins. Both bands have been normalized to have a baseline flux of 1.}
    \label{fig:folded_NICER}
\end{figure}

All seven of the predicted eclipses during NICER GTIs were detected. These are summarized in Table 2, and folded lightcurves and corresponding hardness ratios are plotted in Figure \ref{fig:folded_NICER} . It is curious that only egresses were caught in the \textit{NICER} observations we examined. While we have no definitive explanation for why this is, it is likely that there is some close alignment between \textit{NICER}'s 90-minute orbit and the companion star's $.5$ day period. We did not find any evidence of any dips that were not due to the companion star.

\subsection{RXTE}

\subsubsection{Data analysis}

We analyzed all the available \grs\ X-ray individual observations ($\sim200$) performed with the Proportional Counter Array (PCA) instrument on-board the \textit{Rossi X-ray Timing Explorer} (\textit{RXTE}) satellite. Data filtering was performed using standard criteria\footnote{\url{https://heasarc.gsfc.nasa.gov/docs/xte/recipes/cook\_book.html}}, and data analysis was performed using HEASoft 6.28 tools. For this analysis, we only considered data from the Proportional Counter Unit-2 (PCU-2), as this is the only detector that is kept on throughout the duration of the \textit{RXTE} mission.

For each observation, we extracted a light-curve using Science Event data \footnote{https://heasarc.gsfc.nasa.gov/docs/xte/abc/data\_files.html\#event}, which is characterized by very high time resolution ($<500\mu$s) and moderate energy resolution (256-channel pass band). We employed the Good Xenon configuration or the Event mode configuration depending on availability of these two data modes. In the case of Good Xenon data, we used the Perl script \textsc{make\_se} to merge the matched pairs of files created by the two Event Analyzers (EA). 

Source and background light curves were then extracted with a time bin of 4s. using the \textsc{ftool} software package utilities \textsc{seextrct} and \textsc{saextrct}, respectively, where data filtering by PCU was applied through a bitmask selection.  Good Time Intervals (GTIs) were selected by choosing the times when the source elevation was $>10^{\circ}$ and the pointing offset was $<0.02^{\circ}$, following in both cases the standard thresholds. We estimated the background using \textsc{pcabackest} v3.12a and the most recent background file available on the \textsc{heasarc} website for bright sources \footnote{pca\_bkgd\_cmbrightvle\_eMv20151128.mdl}. Background-subtracted light curves were extracted for each observation using the \textsc{ftool} routine \textsc{lcmath}, where barycentric corrections were applied using the routine \textsc{fxbary}.

\subsubsection{Summary of timing results}

\begin{figure}
    \includegraphics[width=0.48\textwidth]{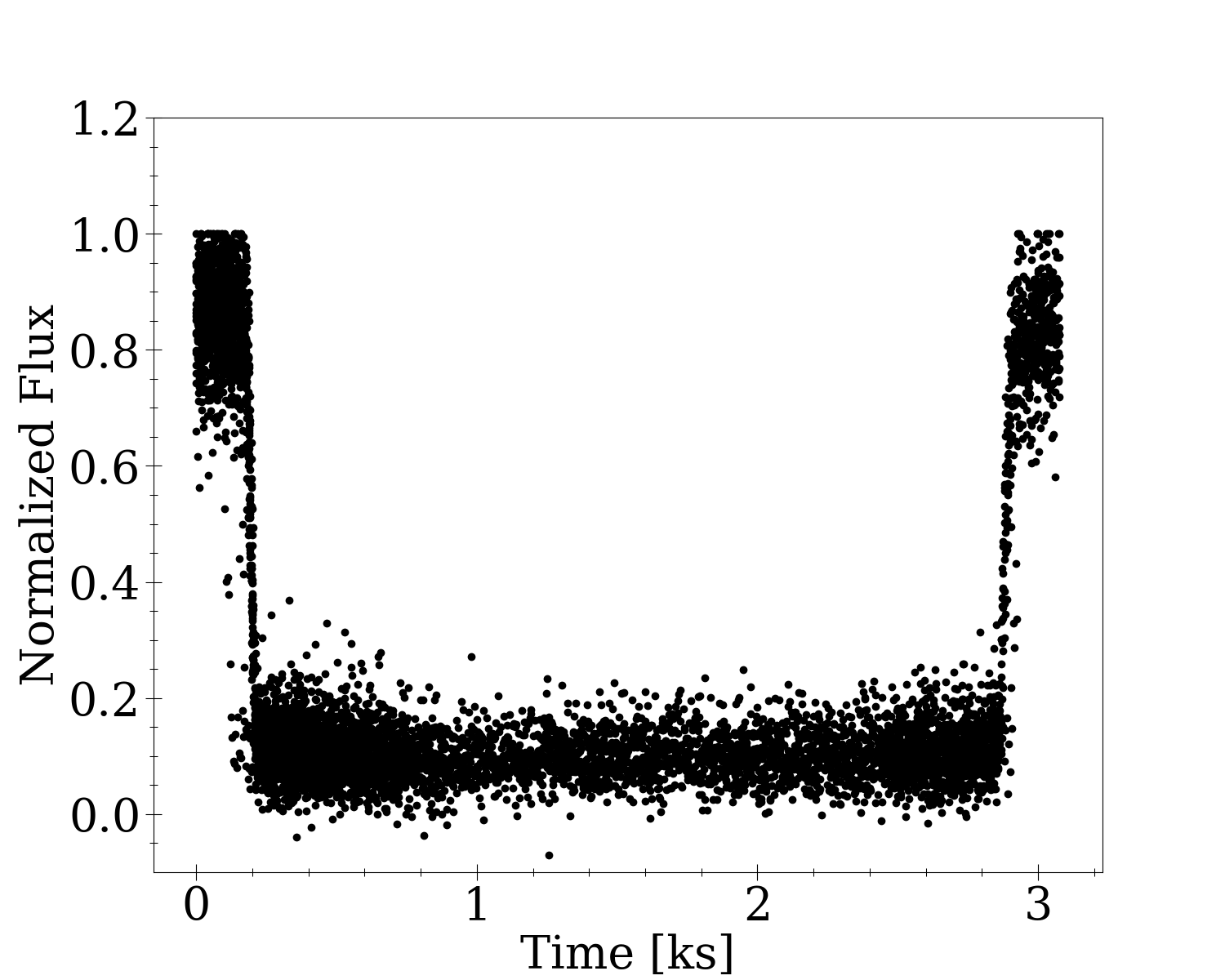}

    \caption{All \textit{RXTE} 4s binned light curves with eclipses of SNR $>5$, from 200s before ingress to 200s after egress. Baseline fluxes for each observation have been normalized to 1
    , and observations have been folded on the predicted eclipse midpoint.
    }
    \label{fig:all_RXTE_eclipses}
\end{figure}

We searched through 206 lightcurves taken with the PCA instrument, and found no examples of dips that did not occur at the predicted ingress or egress times for the companion star or which lasted longer than the expected 2.6 ks. Interestingly, there were again several observations where ingresses or egresses were predicted but were not visible. There were a total of 65 ingresses and 67 egresses that were predicted, spread across 111 observations (see Figure \ref{fig:all_RXTE_eclipses}). Of these, 19 ingresses and 25 egresses had a signal-to-noise ratio, SNR, $\leq 1$, which we define as the cutoff for whether an eclipse is visible or not.

Several shorter dips were also observed, which lasted $\lesssim 1$ minute and do not exceed 70\% depth. We found no evidence of eclipses in the \textit{RXTE} data not belonging to the companion star, and none of the accretion-related dips in the \textit{RXTE} data are nearly as long or deep as the 5.3 ks one in the \chandra\ data.






\subsection{Suzaku}
GRS 1747--312 was observed by \textit{Suzaku} on 2009 September 16 (ObsID 504092010).  The details of the observation are described in \citet{2016PASJ...68S..15S}.  \textit{Suzaku} enables us to perform imaging spectroscopy in the $0.3$--$10$ keV band, utilizing data from each of the four X-ray Imaging Spectrometer (XIS) detectors.  \textit{Suzaku} has also a non-imaging Hard X-ray Detector that covers an energy range of $10$--$600$~keV.  We focus solely on the XIS data for which we can determine a source-extraction region to minimize X-ray background contamination.

\subsubsection{Data analysis}
\begin{figure}[t]
    \includegraphics[width=0.48\textwidth]{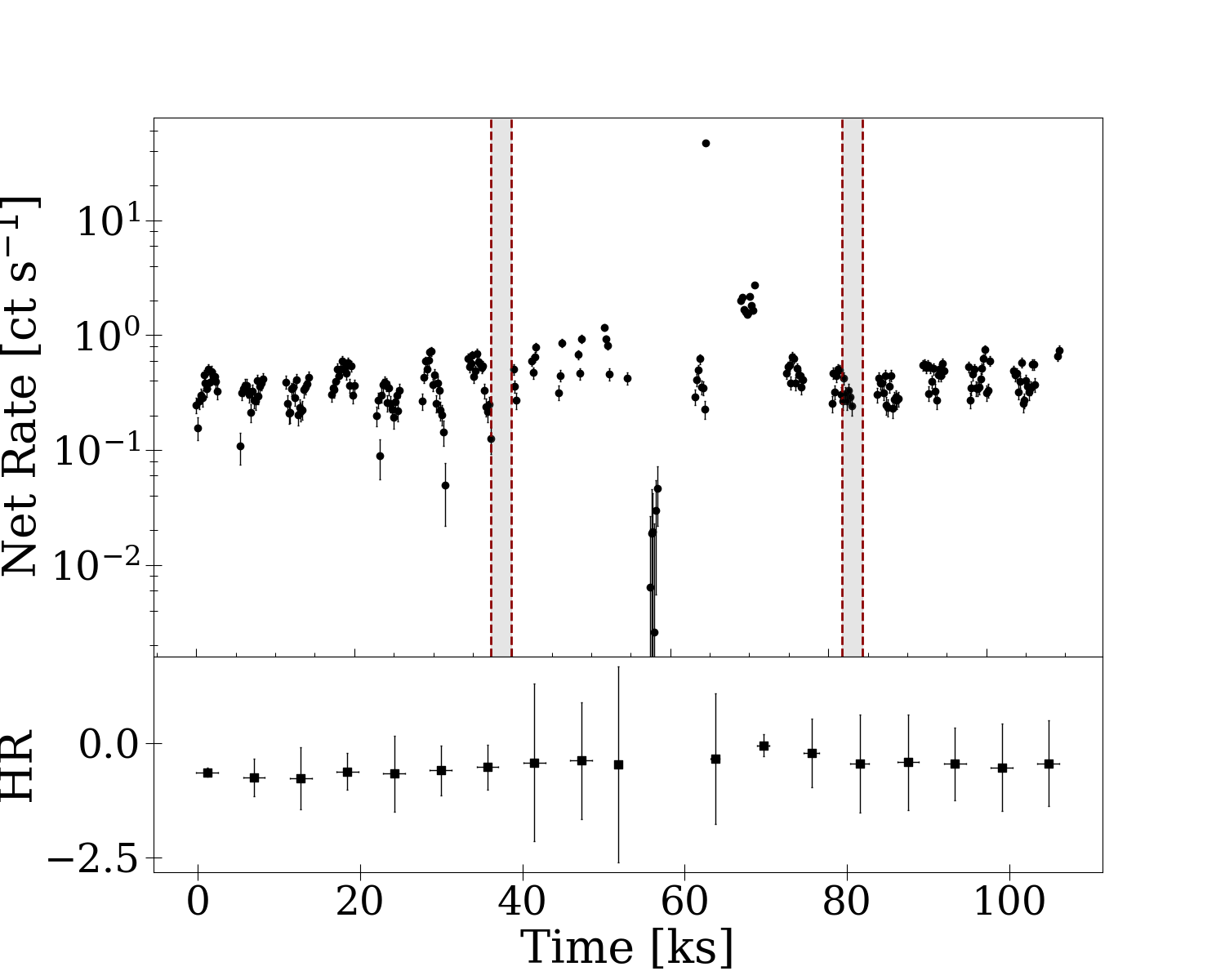}
    
    \caption{\textbf{Top:} Background subtracted \textit{Suzaku} light curve of XIS0+3 in the 0.5–10 keV band. The bin size is 192 s. The time origin is MJD 55090. Predicted eclipse times are marked with red dashed lines. \textbf{Bottom:} $log(S/H)$ hardness ratios, with $S=.5-4$ keV and $H=4-10$ keV and between $\sim 1000-3000$ photons per HR bin. The HR data point for the dip is omitted due to its very large uncertainty as a result of the small counting statistics.
    }
    \label{fig:suzaku}
\end{figure}

We analyzed the cleaned event data using the software package \texttt{Heasoft version 6.31.1}.  The \texttt{processing version of 2.4.12.27} is the same as those used in \citet{2016PASJ...68S..15S}.  The exposure time of the screened XIS data was $50.9$~ks.  We first applied the barycentric correction to the event data with \textit{aebarycen}.  We then extracted XIS0 + XIS3 light curves from the source and background regions in the $0.5$--$10$~keV band.  The source and background regions were the same as those employed in \citet{2016PASJ...68S..15S}; the source region was a circle with a $2.\!\!'3$ radius centered on (RA, Dec.) $=$ ($267.\!\!^{\circ}679$, $-31.\!\!^{\circ}276$).  The background region was an annulus with inner and outer radii of $4.\!\!'27$ and $6.\!\!'45$ where the contribution of faint X-ray sources were excluded.  The background-subtracted light curve is shown in Figure \ref{fig:suzaku}, along with the corresponding hardness ratios. The barycenter-corrected origin of the light curve is
MJD 55090.30904. There is a drop in the count rate evident at $\sim57$~ks and
a burst around $\sim64$~ks. 

\subsubsection{Summary of timing results}
The baseline count rate outside of the burst and dip is $0.43 \pm 0.01$ ct s$^{-1}$. The count rate during the dip is $0.025 \pm 0.026$ ct s$^{-1}$. The beginning and end of the dip is not visible but lasted between $\sim1.4$~ks and $\sim8.3$~ks. While two eclipses of the companion star are predicted during this \textit{Suzaku} observation, neither occur at the time of the dip. They are predicted to occur from MJD 55090.73879 to 55090.76884 and from
MJD 55091.25377 to 55091.28382. While both ingresses should be visible in the data, neither of them clearly are and their absence is statistically significant. It is possible the peculiar dip represents another example of the long $5.3$ ks eclipse visible in the \chandra\ observations. A detailed analysis of it as well as the missing short eclipses can be found in \citet{2016PASJ...68S..15S}.

\section{Detailed light curve modelling}\label{sec:detailed_lc}


\subsection{The Short 2.6~ks Eclipse}\label{sec:short_eclipse}

We first examine the short eclipse in depth, to see if its characteristics can yield information on the system as a whole.

\subsubsection{Hardness Ratios}

\begin{figure}[t]
    \centering

    \includegraphics[width=.48\textwidth]{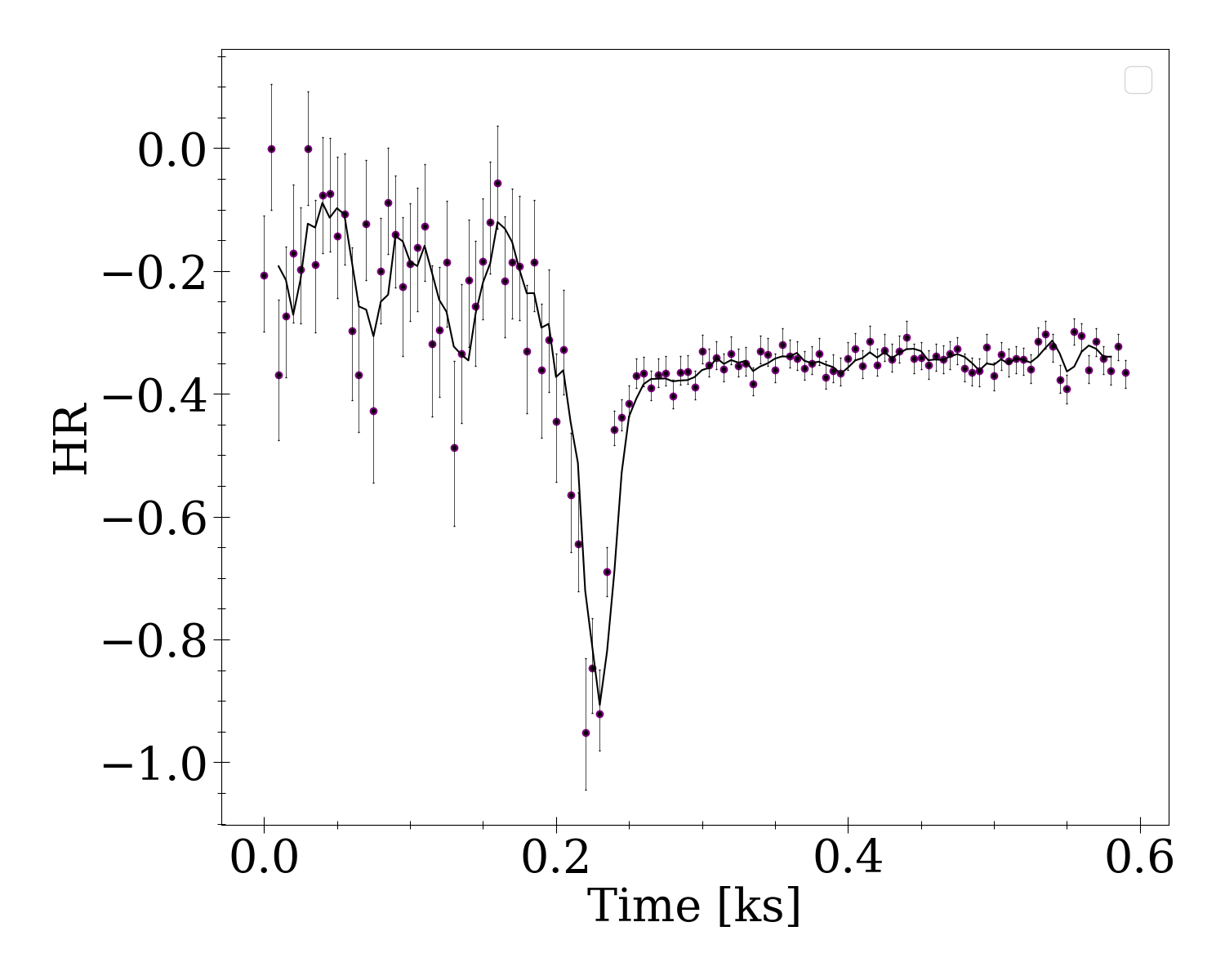}
    \caption{Changes over time in HR for 5 folded \textit{NICER} egresses. HR is calculated as $log(S/H)$ with $S=.3-2$ keV and $H=2-6$ keV.}
    \label{fig:folded_HR}%

\end{figure}

We perform a hardness ratio analysis on all of the \textit{NICER} observations which captured the egress, using a Bayesian Estimation of Hardness Ratios technique \citep[BEHR;][]{2006ApJ...652..610P}. Six out of the seven egresses showed a significant softening of the source during eclipse. The $68\%$ HPD intervals for the HR inside and outside eclipse were $[-0.274, -0.22]$ and $[-0.348, -0.342]$, respectively. This could be due to a hard, compact emission site being eclipsed. Figure \ref{fig:folded_HR} shows the folded HR curve for all observations which did not have a strong background and lasted $>100s$ before and after egress.

There is a significant drop in the hardness ratio as the source comes out of the eclipse. This is due to the eclipse lasting $\sim20$s longer in the lower energy band, as seen in Figure \ref{fig:folded_NICER}, suggesting absorption in the atmosphere of the eclipsing companion star. This is also seen in the \textit{RXTE} data and first pointed out by \cite{2003A&A...406..233I}. They model this effect using an isothermal spherical atmosphere in hydrostatic equilibrium with the density decreasing as an exponential function of height, and find that it provides a good representation of the data.

\begin{figure}[t]
    \centering

    \includegraphics[width=.48\textwidth]{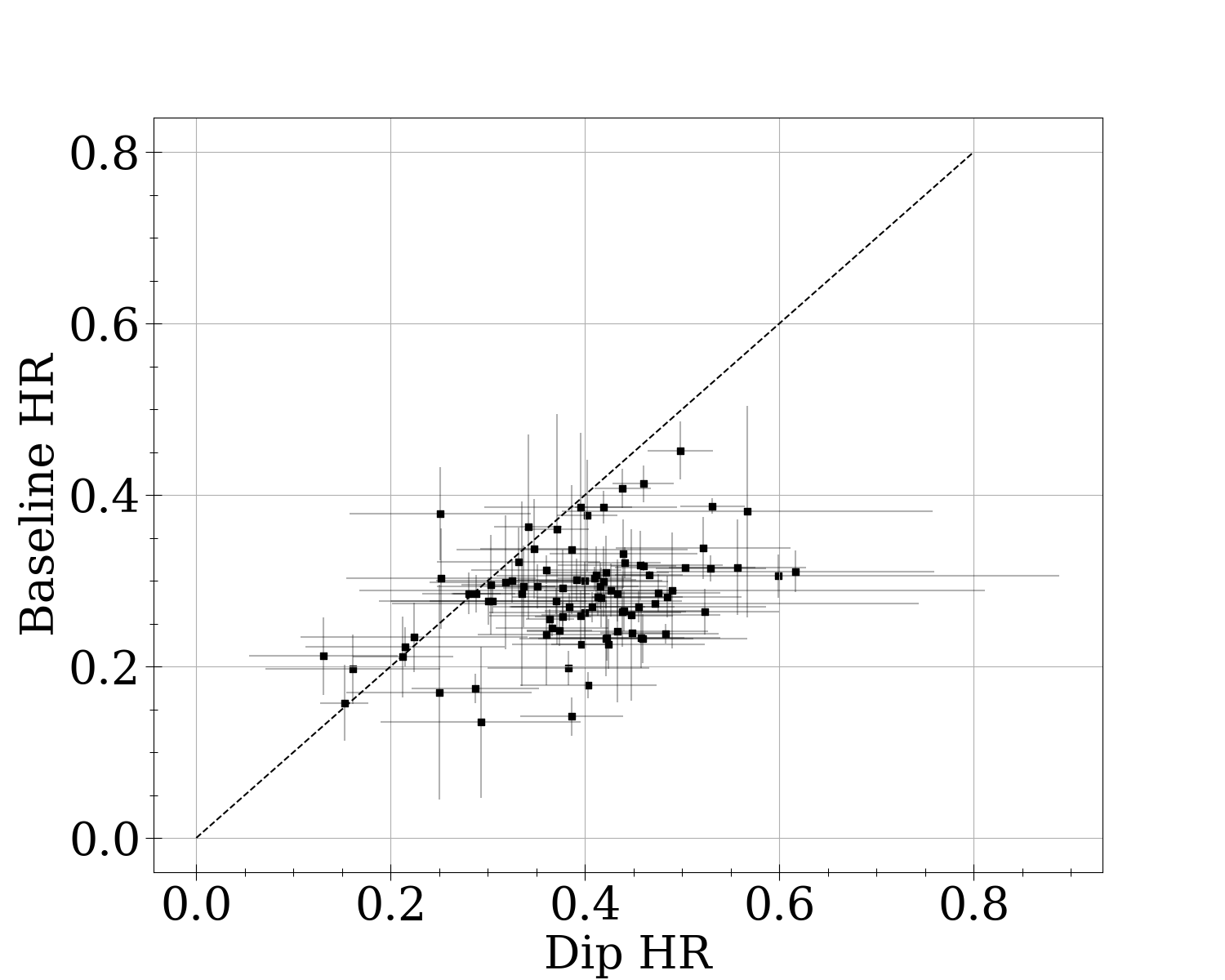}
    \includegraphics[width=.48\textwidth]{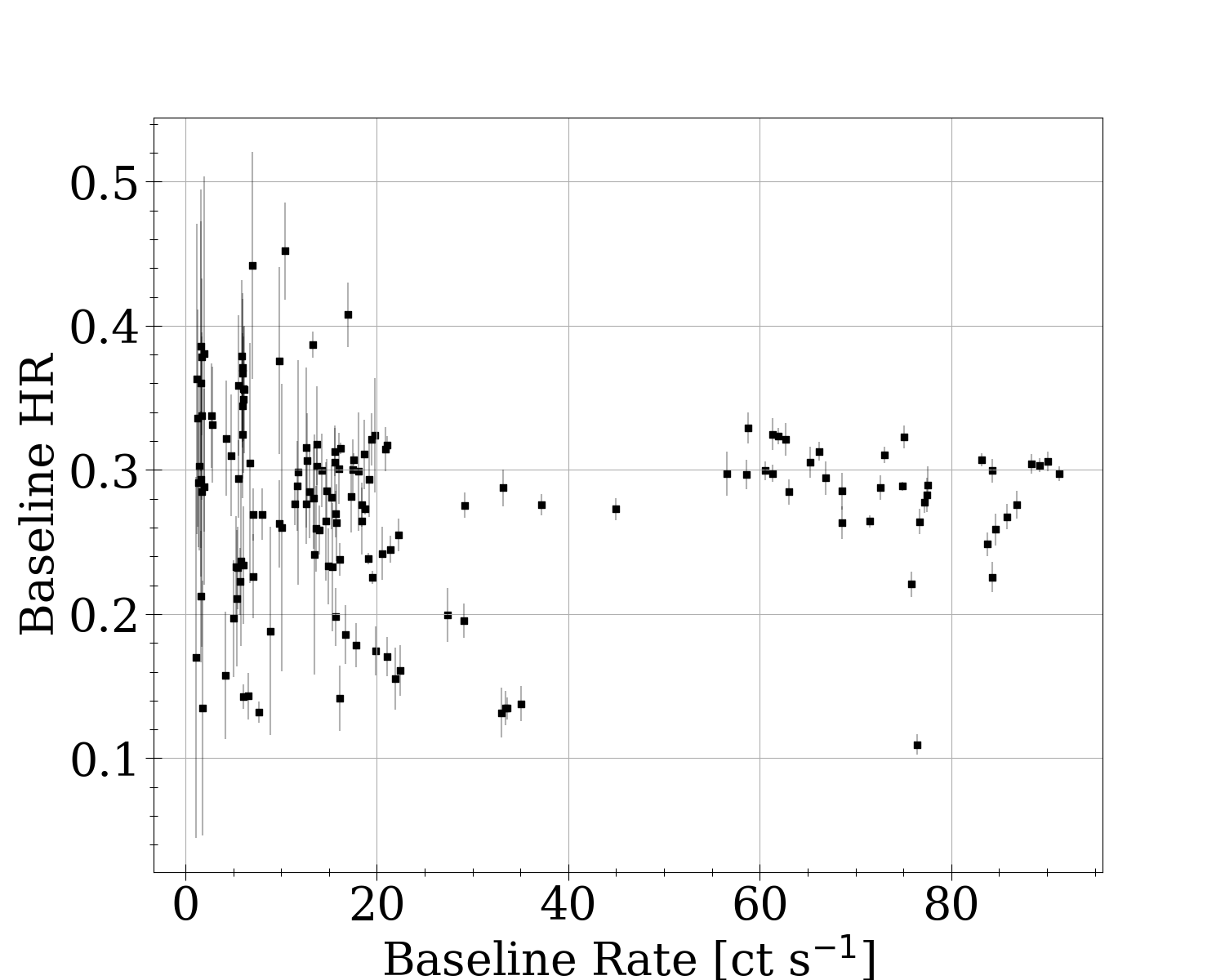}
    \caption{\textbf{Top:} Hardness Ratios for counts inside vs outside of eclipse in \textit{RXTE} data, with a $\pm40$s buffer at the eclipse transition midpoint, as before. The dashed line shows a 1:1 relationship between eclipse and baseline HR.  HR is calculated as $log(S/H)$ with $S=2-7$ keV and $H=7-20$ keV. \textbf{Bottom:} Baseline HR vs baseline rate. }
    \label{fig:rxte_HRs_fig}

\end{figure}

We also analyze the hardness ratios over the course of the ingresses and egresses visible in the \textit{RXTE} data (Figure \ref{fig:rxte_HRs_fig}). In the 100 observations capturing some part of eclipse, we find a clear trend of the source softening during its eclipse. This matches with the \textit{NICER} observations, and could be suggestive of a compact emission region with harder emission being eclipsed, with the residual count rate dominated by the softer corona. Alternatively, these changes in hardness ratio could instead suggest there is indeed a secondary, softer source, whose emission dominates the residual counts received during the short eclipse. It is curious that the \textit{RXTE} and \textit{NICER} data show a softening of the source during eclipse, while the \textit{Chandra} observations show a hardening.

We also examine how the baseline HR changes as a function of the baseline count rate. If spectral softening during the eclipse comes from a higher contribution of counts due to a secondary, softer source, then as the baseline count rate drops the baseline HR should increase (as a larger portion of the counts originate from the secondary source). There is no clear evidence of such a trend in the data, though the range of the baseline HR becomes more scattered as the count rate drops.

It is necessary to note that \textit{NICER} and \textit{RXTE} are non-imaging while sampling large fields of view (30 arcmin$^2$ for \textit{NICER} and 1 degree$^2$ for \textit{RXTE}).  \textit{RXTE} has a linear response versus offset angle, up to $1\degree$ from aimpoint, and \textit{NICER}'s response is relatively flat over an inner 1' radius, dropping off only $2-4$\%. \citep{2021ApJ...918L..26W}. It's possible that during the short eclipse in the \textit{NICER} and \textit{RXTE} observations the flux is dominated by nearby sources and diffuse emission, unlike the point-like eclipse emission observed by \textit{Chandra}.

\subsubsection{Disappearance of the 2.6 ks Eclipse}\label{sec:dissapearing_short_eclipse}

As is clear in both the \chandra\ and \textit{RXTE} observations, the short eclipse is not always visible and at times cannot be seen at all. This was noticed previously in a \textit{Suzaku} observation of the source in 2009 \citep{2016PASJ...68S..15S}. 

\begin{figure}[t]
    \includegraphics[width=.48\textwidth]{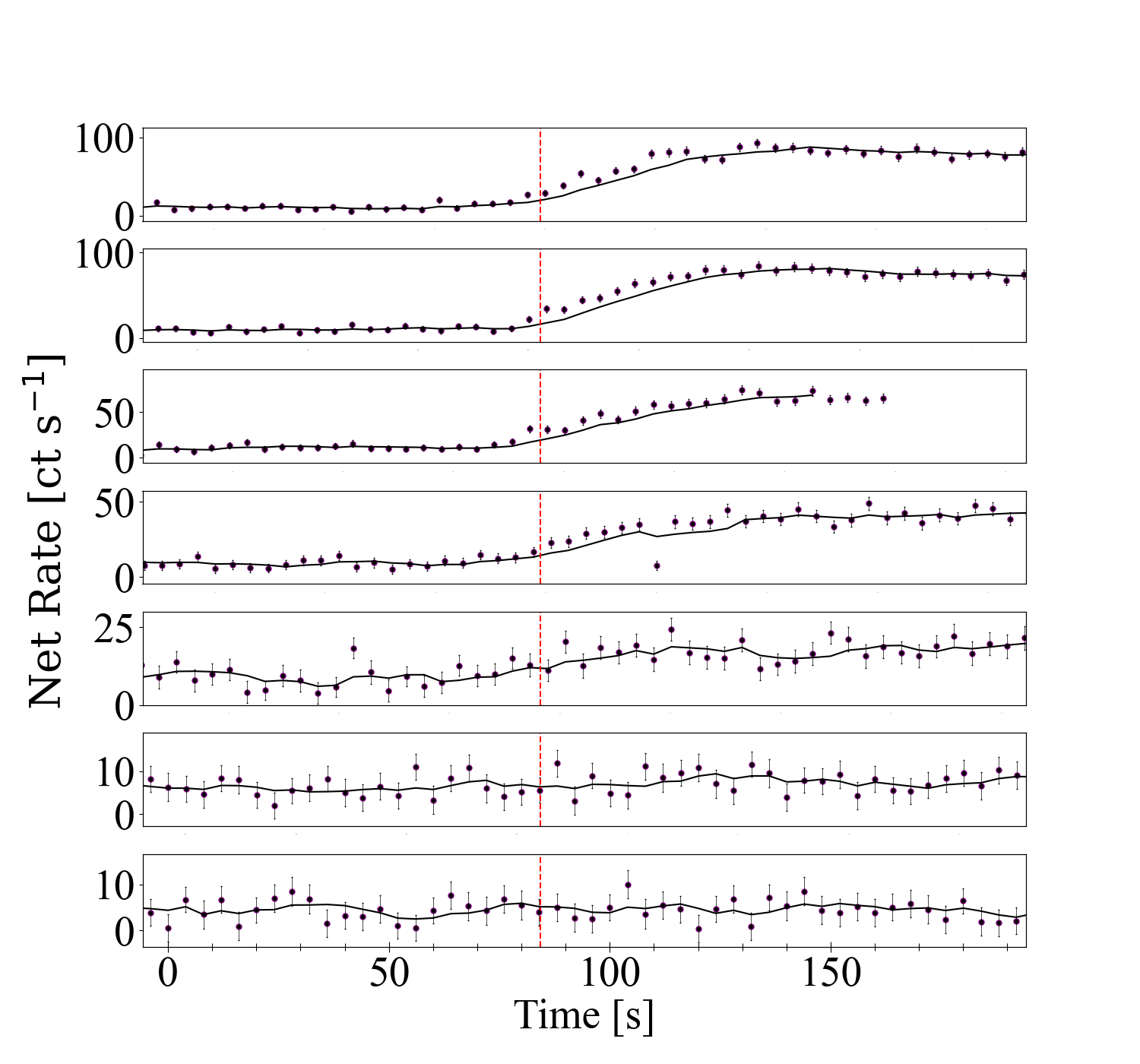}
    \caption{\textit{RXTE} Observations showing the eclipse signal slowly disappearing as the baseline count rate drops. They take place over the course of 24 days, starting at MJD 53484.02 and ending at MJD 53509.74.}
    \label{fig:egresses_rxte}
\end{figure}
\begin{figure}[t]
\includegraphics[width=0.48\textwidth]{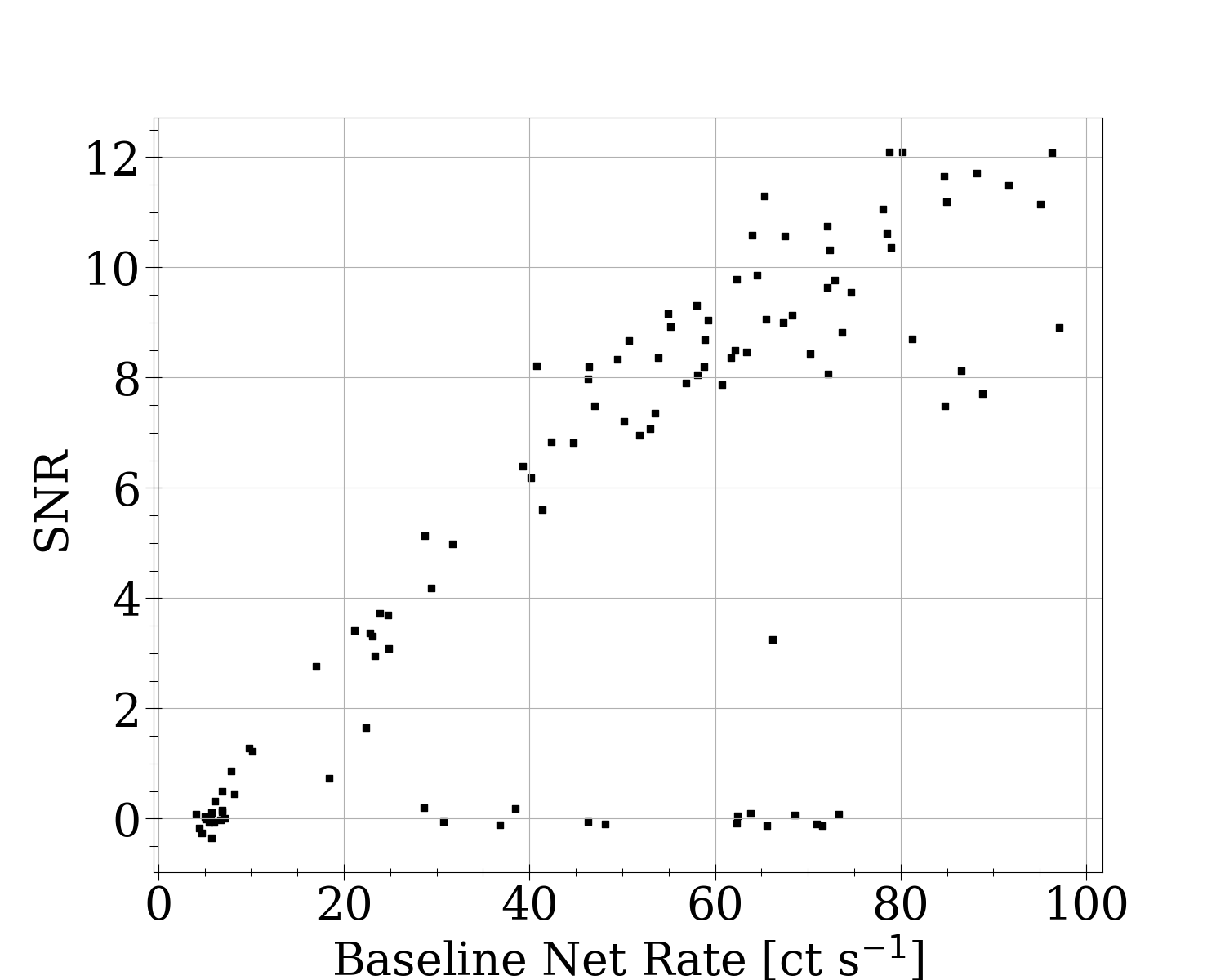}
    \caption{S/N of \textit{RXTE} eclipse depths as a function of the baseline count rate. S/N is calculated as $\frac{S}{N}=\frac{r_{\rm baseline}- r_{\rm eclipse}}{\sqrt{(\sigma_{\rm baseline}^2+\sigma_{\rm eclipse}^2)}}$, where $r_{\rm baseline}$ and $r_{\rm eclipse}$ are the baseline and eclipse count rates, respectively. The $50$s before and after ingress/egress midpoint are not included in the baseline and eclipse rate calculations, to create a buffer around the eclipse transition.
    }
    \label{fig:snr_scatter_RXTE}
\end{figure}

From the \textit{RXTE} data, it is evident that as the source's baseline decreases the depth of the eclipse decreases as well (Figure \ref{fig:egresses_rxte}). For the observations that exhibit visible eclipses, there is consistently a remaining count rate of $4-10$ ct s$^{-1}$ during eclipse. As the baseline count rate approaches this level, the eclipse begins to disappear (see Figure \ref{fig:snr_scatter_RXTE}). This in-eclipse count rate corresponds to an unabsorbed flux of $3.3-8\times 10^{11}$ ergs/cm$^{-2}$s$^{-1}$ in the $.5-7$ keV range (estimated using {\sc PIMMS}, v4.12a, with $N_H=1.5 \times10^{22}$cm$^{-2}$ and $\Gamma=1.5$). This is significantly higher than the typical in-eclipse flux of the \textit{Chandra} GRS 1747-312 point source, suggesting a contribution of contaminating emission from nearby sources in the \textit{RXTE} observations. This could also explain why the \textit{RXTE} and \textit{NICER} observations show a softening of the source during eclipse, while the \textit{Chandra} observations display a hardening. 

Curiously, there were 14 predicted eclipses during observations while the source was not in a low state, during which there was no evidence for an eclipse.
The observations that showed a lack of eclipse during a "high" state took place during a 20-day interval between MJD 55089.708878881 and MJD 55109.278130699. We searched SIMBAD for possible contaminating sources, and identify the accreting millisecond X-ray pulsar IGR J17511-3057 at $(\alpha,\delta)$ J$2000.0 = (267.7861\degree, -30.9614\degree)$, $19.4'$ away from \grs.

At a distance of $19.4'$, the PCA's sensitivity is about $68\%$. It is likely that there was appreciable contamination during these observations from the transient accreting millisecond pulsar IGR J17511-3057, which was also reported by \citet{2016PASJ...68S..15S}, in their \textit{Suzaku} observation during the same time period. The source was reported to be in outburst during the same time as these observations were taken \citep{Ibragimov_2011}, and all of the observations presented strong spikes in their power-density spectra around 245Hz, which is the frequency of IGR J17511-3057. Additionally, another nearby transient pulsar, XTE J1751-305, may also have been active during this time period \citep{2009ATel.2235....1C}. Since these observations were almost certainly contaminated by pulsar emission, we exclude them from our analysis. 

We derived the Observed-Calculated (O-C) timing variations for the 40 clear ingresses and 32 clear egresses with a SNR $\geq 2$.  We identified
the eclipses in a light curve of the observation at a resolution of 4s and selected data stretches of 240 s centered
on the initial midpoint determinations (by eye) of ingress
or egress. All of the visual determinations of the midpoint agreed with the calculated midpoints within $\sim20s$. We then fit a ramp and step function to the data, initially centered on the calculated midpoint from the ephemeris, and with an initial width parameter of the ramp of $27.3$s for ingress and $33.8s$ for egress (the average ingress and egress duration from \cite{2003A&A...406..233I}). We used the python package \texttt{lmfit} to fit the model, and interpolate the resulting fit to 0.1s resolution. One ingress (MJD 51614.1) had to be excluded from analysis due to particularly noisy data during ingress, which resulted in poor fits of the ramp-and-step model. We then calculate the average of the initial and final 80s of the 240s data snippet to determine the flux halfway through ingress or egress. This creates a $\pm40$s buffer region centered on the eclipse transition to exclude it from the calculation. We calculate the standard deviation of the midpoint flux as $\sigma_{\rm flux\_mid} = \sqrt{\sigma_{\rm initial}^2+\sigma_{\rm final}^2}/2$, and find the corresponding upper and lower bounds on the midpoint flux. The time when the flux reaches the halfway point is found by determining the data point closest to this flux from the interpolated ramp and step fit. While there may be bias from the variability of the source outside of the eclipse transition, we believe it to be negligible. The times found using this method represent the observed times of the eclipse. The full tables of $O-C$ transit timing variations can be found in Appendix B. 

\begin{figure}[t]
\centering
\includegraphics[width=.45\textwidth]{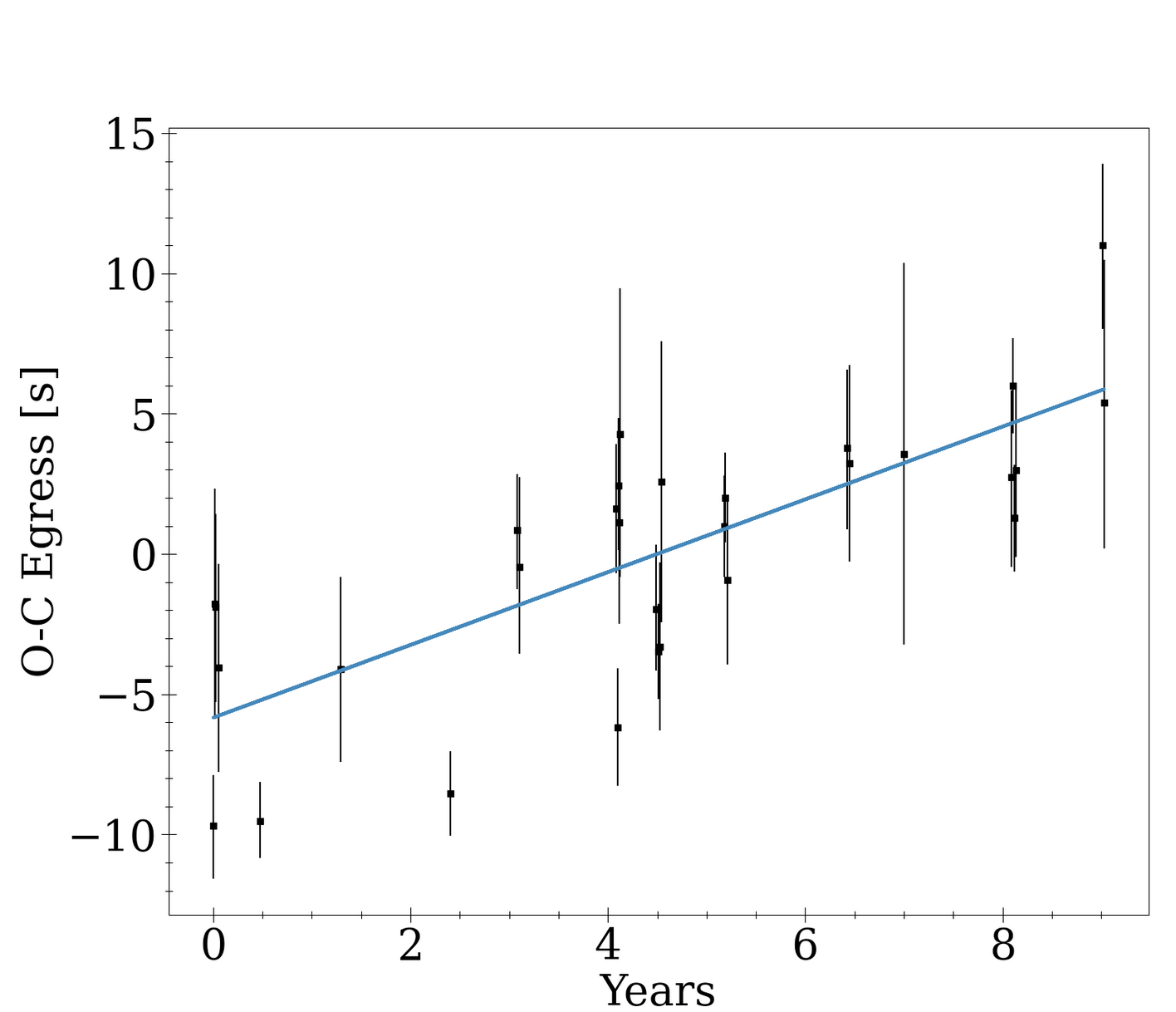}
    \caption{Observed minus calculated timing variations for egresses in the \textit{RXTE} data. Ingress variations can be found in Appendix B.}
    \label{fig:egress_OC}
\end{figure}

Investigation of the ingress and egress times show a slow nearly-linear drift of about 1 second per year that are statistically significant with the rest of the signal consistent with scatter. The ingress times are getting earlier and the egress times are getting later (see Figure \ref{fig:egress_OC}), suggesting that the main cause is a $\sim$2 second per year increase in the duration of the eclipses. The implications of these variations is discussed further in Section 
\ref{classic_triple_body}.

The constant residual count rate during eclipse while the baseline decreases may suggest that another source is located close to \grs. In such a scenario, as \grs\ decreases in brightness, the depth of the dip would decrease owing to an increasing proportion of the detected X-rays come from the contaminant source. If this contaminant source is the one being eclipsed during the long 5.3ks dip, other instances of the 5.3 ks dip might be present in the \textit{RXTE} lightcurves, albeit with a shallower depth. Such shallow dip events could be easily missed if they only correspond to a very small decrease from baseline while the source is in its high state. As \grs\ dims, the secondary source would make up a larger contribution of the X-ray events and such dips should become deeper. Despite searching, we did not find any evidence of shallow 5.3 ks dips in the \textit{RXTE} lightcurve. However, this could be due to heavy contamination from nearby sources while \grs\ is in its low state.    

It is unlikely that the disappearance of the eclipse is due to a precession in the orbit of the companion star. One would expect gradual shortening of the width of the eclipse until it would vanish and then begin to increase again, with an average duration staying relatively constant across observations. No such pattern is found.

One possible explanation is that as the binary systems cycles between its high and low states, the dominant source of emission moves from the inner disk and neutron star surface to an extended corona or jet emission. While the companion star may eclipse the source of emission in the outburst state, it no longer does so in the low state, or else the signature is too weak to detect.

\subsection{The Long 5.3 ks Dip}\label{sec:long_eclipse}

Apart from the missing short eclipses, the most striking feature in the combined archival data is the distinct dip in \chandra\ Observation 23443. The dip begins at $\sim13$ks into the observation and lasts until $\sim18.3$ks. This duration of $\sim5.3$ks is twice as long as the duration of the companion star eclipse, and is out of phase with the known eclipse period.


\subsubsection{Dips Caused by Gas/Dust}

\begin{figure}[t]
    \centering
    \includegraphics[width=.5\textwidth]{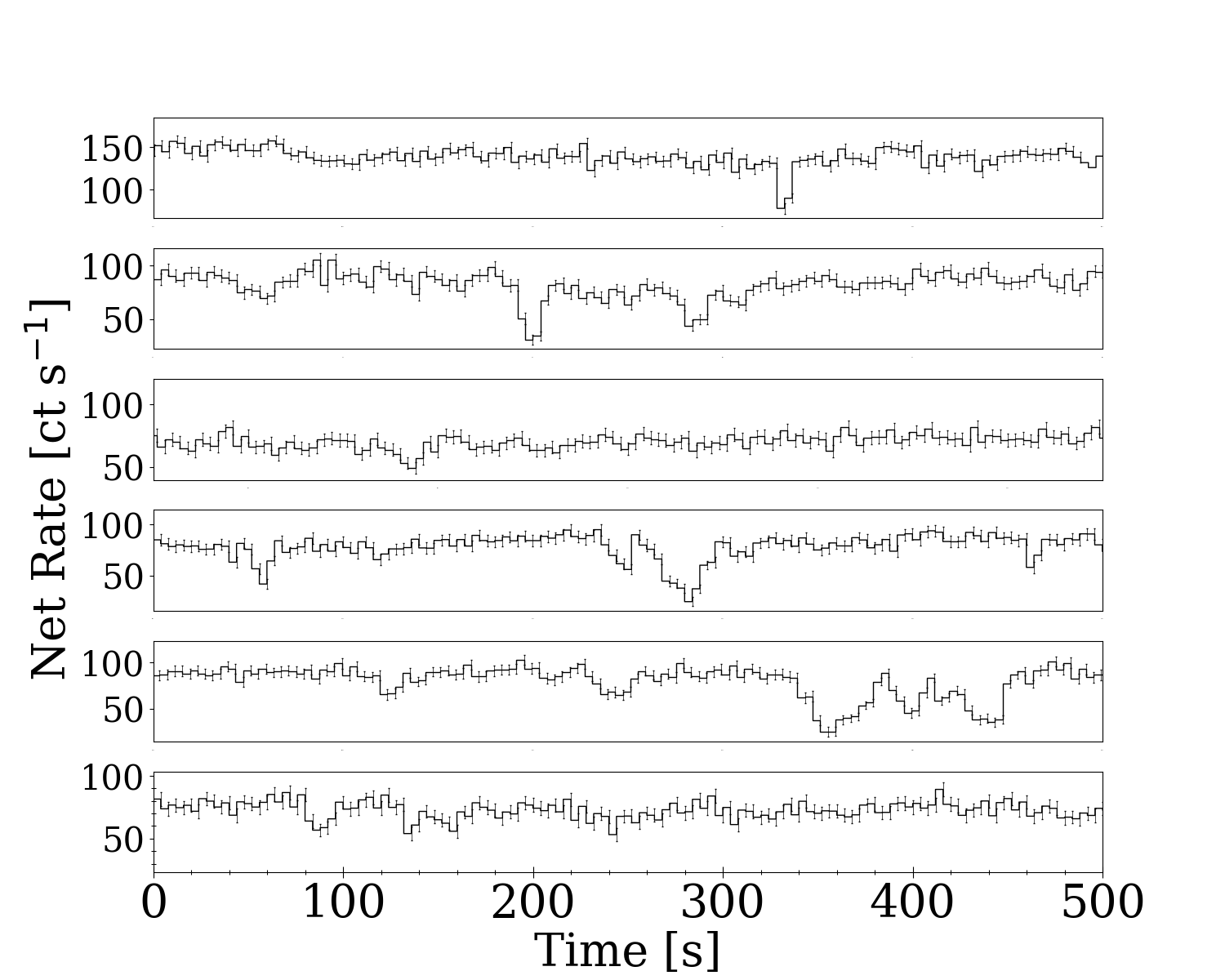}
    \caption{4s binned lightcurves of all \textit{RXTE} observations with non-eclipse related dips. Dips here are defined as having an average count rate of more than 2 standard deviations below baseline for a minimum of 12 seconds. }
    \label{fig:all_dips_RXTE}

    \includegraphics[width=.48\textwidth]{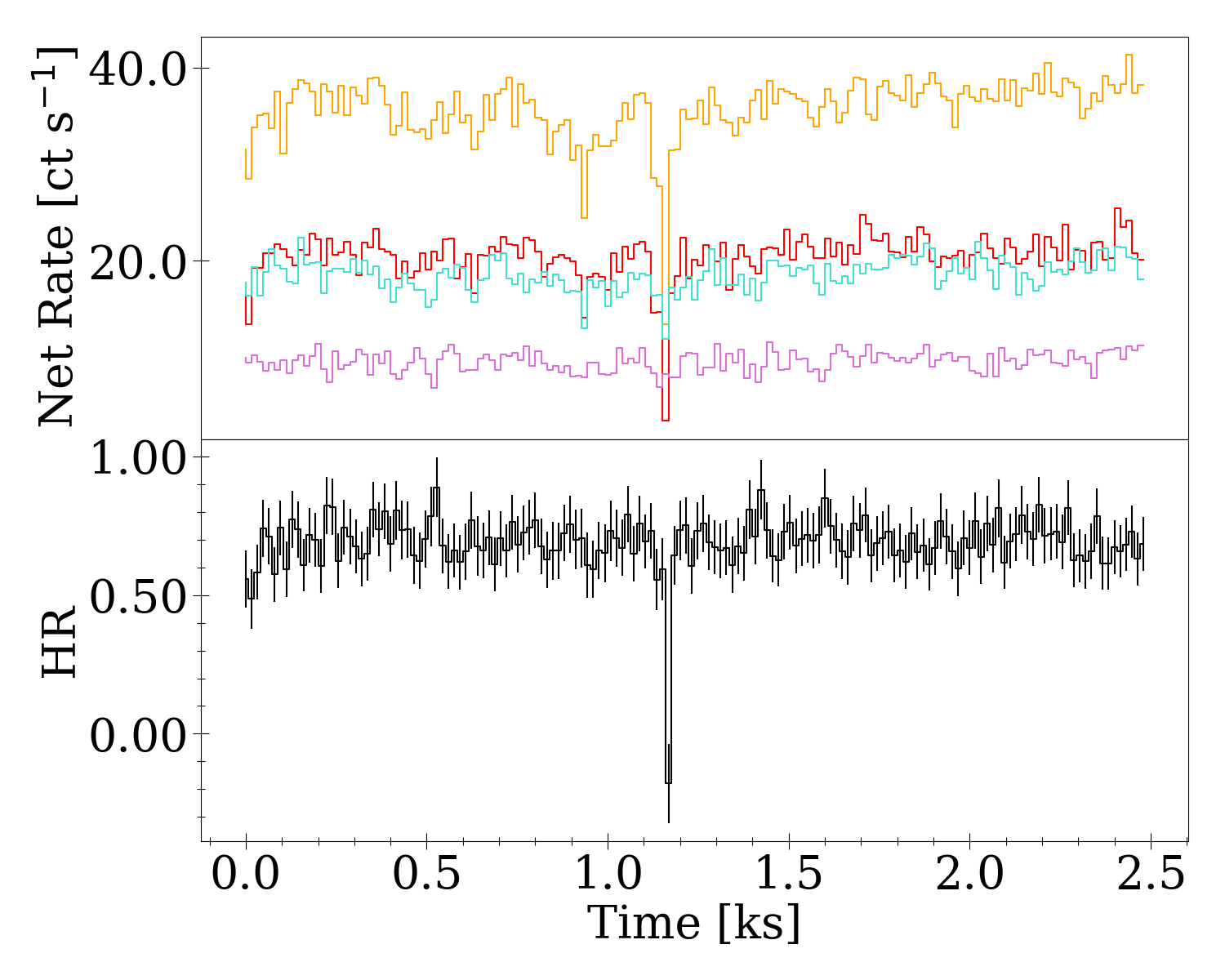}
    \caption{\textbf{Upper Panel}: 16s binned 2-20 keV light curve during a period of dipping activity (MJD 52203.096591). Color scheme is as follows: Red=2-4 keV, Orange=4-7 keV, Turquoise=7-10 keV, Purple=10-20 keV. \textbf{Lower Panel}: Log(S/H) Hardness Ratio, with $S=2-7$ keV and $H=7-20$ keV. }
    \label{fig:Zand_Chandra}

\end{figure}
As is visible in other eclipsing LMXBS (such as EXO 0748-676, MXB 1658-298, XTE J1710-281), clear dipping activity is present in many observations of \grs. These dips are thought to be caused by clumps of gas or dust created by irregular thickening of the accretion disk, and have been recognized across many X-ray sources over decades (e.g.,  \citealt{1982ApJ...253L..61W}). These density changes in the accretion disk do not have hard surfaces, and as a result create changes in the X-ray hardness. Dips caused by transitions of opaque bodies, however, will not produce this change in hardness ratio. As such, it is possible to distinguish dips caused by gas or dust from genuine transits by examining the hardness ratio during and out of the dip. \textit{RXTE} has observed accretion-related dipping activity in several observations of \grs. These dips typically last a fraction of a minute, never exceed $70\%$ depth, and exhibit significant hardening  \citep{2003A&A...406..233I}. A compilation of observations showing dipping activity and corresponding changes in hardness ratio in Figure \ref{fig:all_dips_RXTE} and \ref{fig:Zand_Chandra}.


When comparing the dip present in \chandra\ observation 23443 to such dipping activity, we argue that there are three main arguments against the \chandra\ dip being produced from accretion. Namely, (1) The 5.3 ks dip does not exhibit the abrupt hardening present in other gas/dust related dips, (2) The 5.3 ks dip is much longer than all other gas/dust related dips, (3) The 5.3 ks dip is deeper than all other gas/dust related dips. In this section we will examine this first claim in more detail.

\subsubsection{Hardness Ratio Analysis of the $5.3$~ks dip}

We examine the evolution of the source spectrum over the course of the dip to determine if it was due to accretion-related effects or instead due to the transit of an opaque body. As there are too few counts to carry out time-resolved spectroscopy, we instead analyze the changes in hardness ratio 
$HR=\log\frac{soft}{hard}$, where $soft$ and $hard$ represent source intensities in disjoint passbands across the \chandra\ response.  In the following, we adopt the passbands defined by the \chandra\ Source Catalog\footnote{\url{https://cxc.cfa.harvard.edu/csc/columns/ebands.html}}, $S=0.5-1.2$~keV, $M=1.2-2.0$~keV, $H=2.0-7.0$~keV, but combing the soft and medium passbands due to a lack of counts $<1.2$ keV (see Table \ref{table:passband_counts}).  We compute $HR$ again using BEHR for counts collected in a variety of time bins ranging from a regular grid to adaptively-optimized bins using Bayesian Blocks \citep[{\sl BB}][]{1998ApJ...504..405S,2013ApJ...764..167S}.  In all cases, we run a Markov-Chain Monte Carlo (MCMC) chain for $40,000$ iterations and discard the first $5,000$ as burn-in.
We perform our analyses on both ObsIDs 23443 and 23444, in order to have a known eclipse as a reference point. 

\begin{table}[t]
\centering
\begin{tabular}{p{3cm}|p{2cm}|p{2cm}  }
  Energy Band & ObsID 23443 & ObsID 23444 \\
 \hline
 \hline
 $.5 - 1.2$ keV & 31 & 105\\

 \hline
  $1.2 - 2$ keV & 342 & 1295\\
 \hline
   $2 - 7$ keV & 1078 & 3962\\
\end{tabular}
\caption{Counts in relevant passbands for observations 23443 and 23444.}
\label{table:passband_counts}

\end{table}

\subsubsection{Statistical Analysis of Hardness Ratios}





We compare the cumulative hardness distribution of the photons received during the 5.3 ks dip to the hardness distribution of the photons flanking either side of it, to see if there are significant changes between these.

\begin{figure}[t]
    \centering
    \includegraphics[width=.48\textwidth]{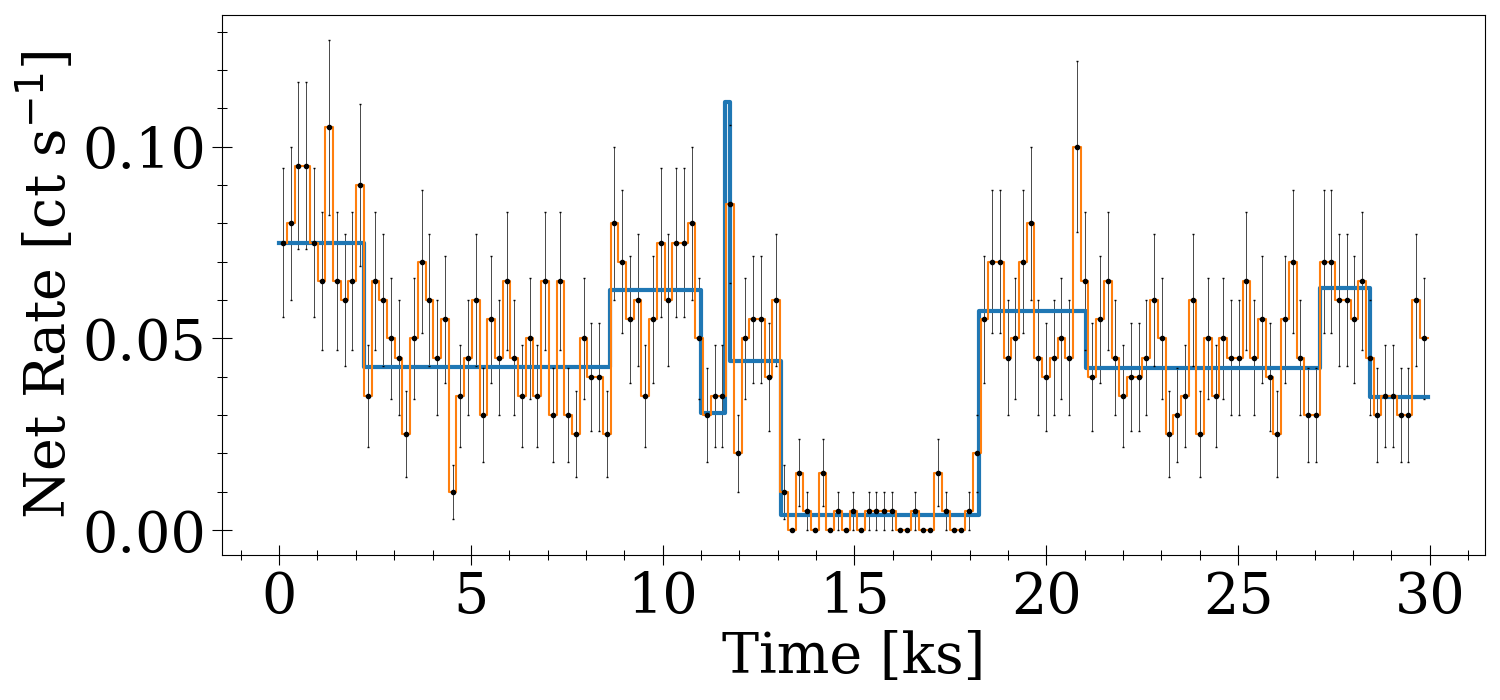}
    \caption{Source (top) and background (bottom) light curves, using Bayesian Blocks in blue and presenting the 200s-binned light curve in orange, for observation 23443.}
    \label{fig:bb_23443}
\end{figure}

To determine appropriate time intervals for the dip and flanks for comparison, we define the edges of our time blocks based on the change points found in the light curve by the Bayesian Blocks algorithm \citep{2013arXiv1304.2818S}. We use the Python implementation developed by the Astropy Collaboration et al. \citeyear{2022ApJ...935..167A}, setting $p_0=1$ for a $1\%$ false alarm probability. We then apply a filtering on the change points identified by discarding all blocks whose duration is $<10\tau$ (where $\tau\approx3.24$ is the CCD readout time for the ObsIDs). These blocks are unlikely to represent physically meaningful changes in the source. For observation 23443, Bayesian Blocks created one bin for the 5.3 ks dip from 13076s to 18229s, as well as a left flank from 11760s to 13076s and a right flank from 18229s to 29941s (Figure \ref{fig:bb_23443}). Bayesian Blocks produced only one block for the background rate, suggesting a constant background level over the course of the observation.

\begin{figure*}
    \centering
    \includegraphics[scale=.63]{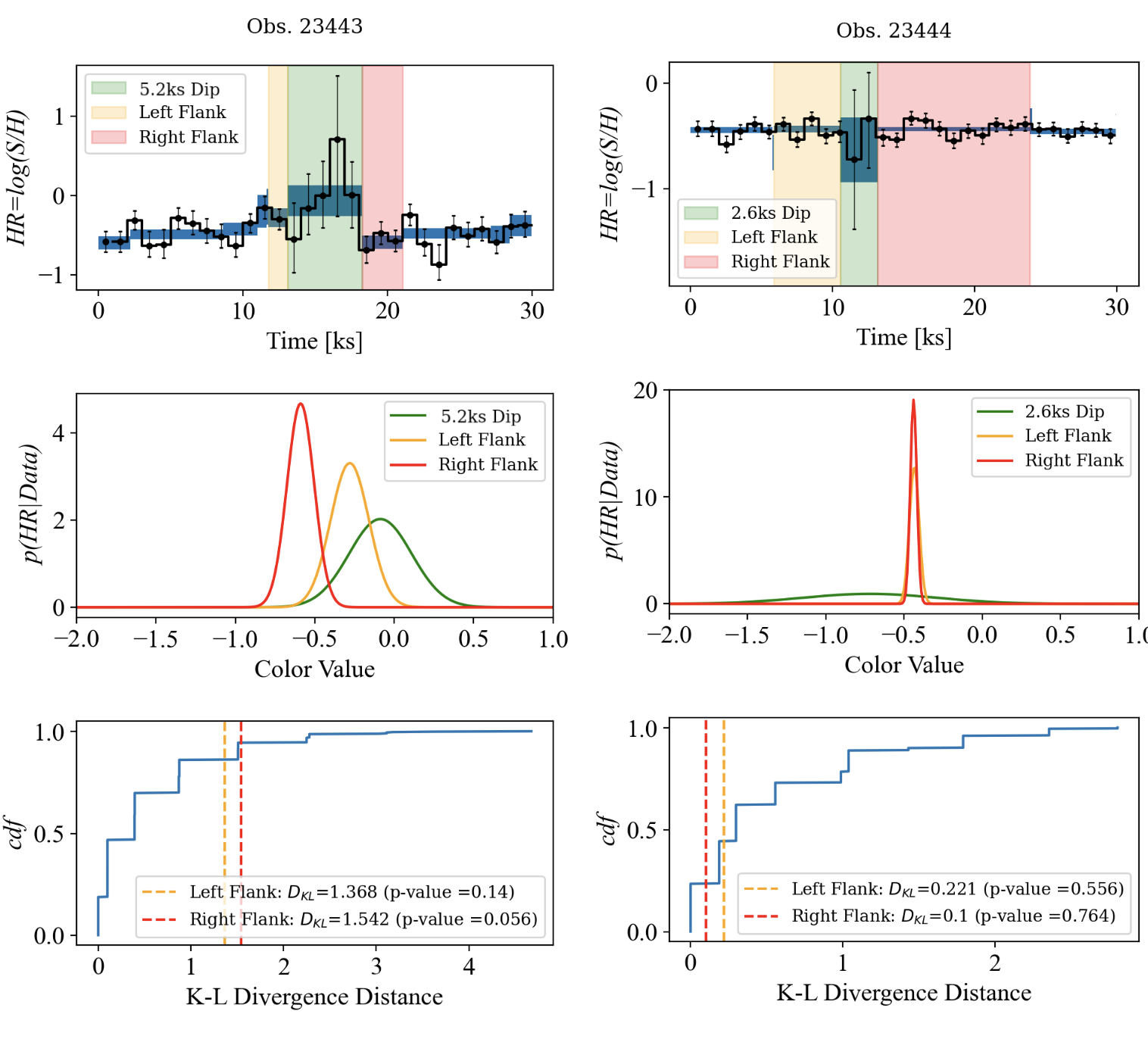}
    \caption{Searching for spectral variations across the dips.  \textbf{Left} panels show the results for the long duration 5.3~ks dip in ObsID 23443, and the \textbf{right} panels show the results for the short 2.6~ks eclipse in ObsID 23444.  \textbf{Top:} $HR$ light curves, obtained for Bayesian Blocks intervals (68\% uncertainty bands marked in dark green) and for 1~ks bins (black stepped curves with 68\% uncertainties marked with vertical bars).  The dip regions are shaded in light green, with the left flanking block shaded in yellow and the right flanking block in light red.
    \textbf{Middle}: Posterior probability distributions $p(HR|{\rm data})$ of the three shaded regions from the top row.
    \textbf{Bottom}: Testing for the existence of spectral differences between the dip and the flanking blocks.  Bootstrapped cumulative distributions of $D_{\rm KL}({\rm dip}||{\rm left,right})$ are shown as the blue stepped curves, with the observed $D_{\rm KL}$'s shown as vertical dashed lines in the colors corresponding to the flanking blocks.}
    
    \label{fig:statistics}%
\end{figure*}
\begin{figure}[t]
    \centering
    \includegraphics[width=.4\textwidth]{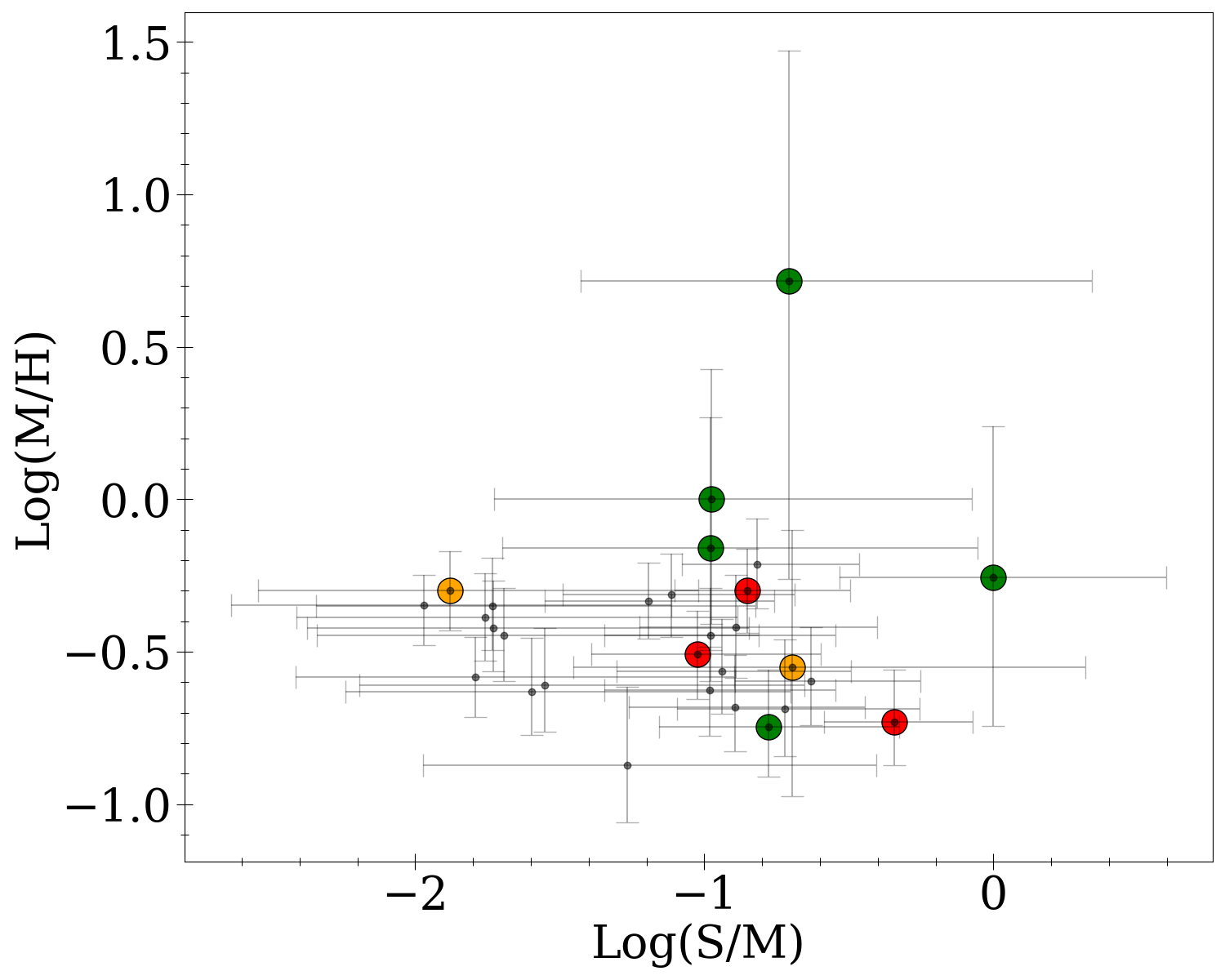}
    \caption{HR-HR scatter plot in 1~ks bins.  HRs of $\log\frac{S}{M}$ and $\log\frac{M}{H}$ and corresponding uncertainties for each bin are computed using BEHR \citep{2006ApJ...652..610P}.  The bins during the dip are marked green, and those which overlap the left and right flanking blocks (see Figure~\ref{fig:statistics} top left).  While there appears to be a tendency for the source to have excess medium band counts during the dip, the variations observed in the hardness ratios are consistent with the overall level of variations seen in the source.
    }
    \label{fig:time_series}
\end{figure}

We estimate the significance of the difference in $HR$ during the dip compared to the intervals just before and after the dip using a metric based on the Kullback-Leibler divergence ($D_{\rm KL}(p||q)$). We bootstrap the counts during the eclipse to find approximate \textit{p}-values for the  probability of seeing as extreme a difference between dip and flanking color distributions as a consequence of statistical fluctuations in the photons during the dip (see Table \ref{table:HR_p_values}). The final results are shown in Figure \ref{fig:statistics}, and a full explanation of the method can be found in Appendix \ref{chap:more_HR}. There is no evidence for any change in spectrum for the short 2.6~ks dip ($p$-values$\gg$0.05), as is expected for an eclipse by a solid body.  Similarly, there is no evidence for a change in spectrum as the source goes into the longer 5.3~ks dip ($p$-value$\approx$0.65).  As it comes out of the dip, however, there is a suggestion of increased hardness during the right flank interval, with $p$-value$\approx{0.06}$.  While this $p$-value does not meet threshold requirement for a significant change, it suggests that a change may have occurred in the source, as indicated by the apparent drift of the source seen in the HR-HR diagram in Figure~\ref{fig:time_series}.


\begin{table}[htb]
\centering
\begin{tabular}{p{3cm}|p{2cm}|p{2cm}  }
  Approx. p-value & ObsID 23443 & ObsID 23444 \\
 \hline
 \hline
 Left Flank & 0.15  & 0.6\\
 \hline
 Right Flank & 0.06 & 0.8\\
 \hline
\end{tabular}
\caption{Approximate $p$-values for the probability of seeing as extreme a difference between dip and flanking hardness ratio distributions as a consequence of statistical fluctuations in the photons during the dip.
}
\label{table:HR_p_values}

\end{table}


There are a number of physical reasons why the source might appear harder coming out of the eclipse. These include the source's spectrum naturally varying while it was hidden and then being in a different state when it was revealed, as well as the possibility of the transiter leaving a cometary wake behind it, leading to increased absorption. Figure \ref{fig:time_series} shows that the increase in HR coming out of the dip is not an outlier, arguing for the change in HR that takes place following the dip being due to fluctuations intrinsic to the source that are revealed as the source exits the eclipse. 

We conclude that there was not a significant change in hardness ratio during the dip, for an $\alpha=.05$ significance level. Furthermore, any change in hardness ratio associated with the dip is in the softer direction, in strict contrast to what would be expected for an absorption dip. This could be due to a more compact hard emission region being eclipsed, and is similar to the HR trends seen in the \textit{NICER} and \textit{RXTE} short eclipses. Since we do not see a significant hardening of the source, we conclude that the dip was not caused by gas or dust passing in front of the source, and that we are indeed seeing a genuine eclipse. This agrees with the spectral analysis from Section \ref{sec:spectral_analysis}, which shows there is no significant increase in $N_{\rm H}$ over the eclipse duration.

\section{Physical interpretation}\label{sec:models}

\subsection{Theoretical Background}

We have a limited amount of information about the system. We do have, however, several powerful
principles we can apply in order to learn more about it. 

\smallskip

{\bf Dynamics:} The first principle derives from Kepler and Newton, whose laws allow us to express  the orbital separation and orbital speed within a binary in terms of the masses of the components and the orbital period.

\begin{equation}
    a = \Big(\frac{G}{4\pi}\Big)^{1/3}P^{2/3}M^{1/3} 
    \label{eq:semi_axis}
\end{equation}

\begin{equation}
    v=\frac{2\pi a}{P}
    \label{eq:v_p}
\end{equation}
Equation~(2) assumes a circular orbit.

If the companion star is filling its Roche Lobe, we can also use the formula given by Eggleton to express its radius as a function of the mass ratio and orbital separation:

\begin{equation}
    \frac{r_{\rm \mathrm{L}}}{a}=\frac{0.49 q^{2 / 3}}{0.6 q^{2 / 3}+\ln \left(1+q^{1 / 3}\right)}
    \label{eq:eggleton}
\end{equation}

where $0<q\equiv M_2/M_x < \inf$ (where $M_x$ is the mass of the X-ray source), and $a$ is the separation of the two stars.

Because the system is in a globular cluster we know that the masses of unevolved stars are typically smaller than roughly $0.8\, M_\odot,$  although stellar interactions can yield a relatively small number of  ``blue stragglers'' of somewhat higher mass.  When the accretor is a NS, its mass is close to $1.4\, M_\odot,$ while BH-accretor masses may be significantly higher.  For the application to GRS~1747-312 it is convenient to write the above relations as follows, where 
the total binary mass is $M_t=M_1+M_2$, and the subscripts ``\textit{in}'' and ``\textit{2}'' refers to properties of the $0.515$~d period binary. In following sections we use the subscripts ``\textit{out}'' and ``\textit{3}'' to refer to the properties of any tertiary body.
\begin{equation}
a_{\rm in}= 3.51\, R_\odot \Big(\frac{{P}_{\rm in}}{0.515~d}\Big)^\frac{2}{3} \Big(\frac{M_t}{2.2\, M_\odot}\Big)^\frac{1}{3}.
\label{eq:inner_binary_semimajor_axis}
\end{equation} 

\begin{equation}
\begin{split}
    v_{\rm \mathrm{in}} 
    & = 345 \mathrm{km/s}\, \Big(\frac{0.515}{{P}_{\rm in}}\Big)\, \Bigg(\frac{a_{\rm in}}{3.51\, R_\odot}\Bigg) \\
    & = \quad 345 \mathrm{km/s}\, \Big(\frac{0.515}{{P}_{\rm in}}\Big)^\frac{1}{3}  \Big(\frac{M_t}{2.2\, M_\odot}\Big)^\frac{1}{3}.
    \label{eq:velocity}
\end{split}
   \end{equation} 


\smallskip

\begin{figure}[t]
    \centering
    \includegraphics[scale=.4]{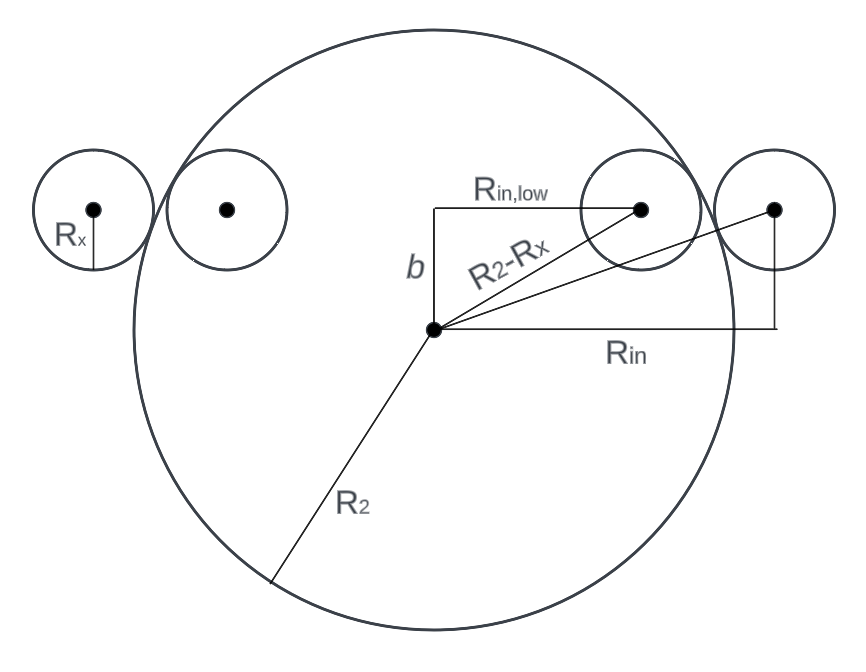}
    \caption{Diagram illustrating companion star (larger circle) transiting over the compact object (smaller circles). $R_x$ is the radius of the XRS, $R_2$ is the radius of the eclipser. $b$ is the projected distance of closest approach between the center of the XRS and the center of the eclipser. $2R_{\rm in}$ is the distance traveled by the eclipser from the start of ingress to the end of egress, with $2R_{\rm in,low}$ being the distance travelled by the eclipser while the XRS is in its lowest flux state. We define $\beta$ as the distance of closest approach divided by the radius of the eclipser ($\beta=b/R_2$ when modeling the 2.6 ks eclipse, and $\beta=b/R_3$ when modeling the 5.3 ks eclipse).
    }
    \label{fig:diagram2}
\end{figure}

{\bf Geometry:} The other principle that informs our calculations is the geometry of the eclipse itself, which can be related to the eclipse duration and to the times taken for ingress and egress.
If we assume that the shapes of the eclipser and the object it traverses are spherical, then 
both can be mathematically modeled as circular disks. 
We can derive a ratio between the disk radii, $\rho$, using the ingress and egress durations for the eclipse (see Figure \ref{fig:diagram2}):

\begin{equation}
  \rho = \frac{R_x}{R_2} =  \Big(\frac{v_{\rm in}}{4\, R_2}\Big)^2 \Big(\tau^2-\tau_{\rm \mathrm{low}}^2\Big)
   \label{eq:rho_v}
\end{equation}

where $\tau$ is the time from the start of ingress to the end of egress, and $\tau_\mathrm{low}$ is the time 
during which the flux is lowest, generally during the interval stretching 
from the end of ingress to the start of egress. Here $v_{\rm rel}$ is the relative velocity between the eclipser and the eclipsed object, perpendicular to our line of sight.  This equation shows that for a given dynamical system, measurements of the eclipse timings provide a relationship between the radius of the X-ray Source (XRS) and the radius of the eclipser. This relationship is helpful in determining the nature of the system. The full derivations can be found in Appendix \ref{ref:math_section}.

We can also express $\rho$ purely geometrically in terms of $\beta=b/R_2$, where $b$ is the projected distance of closest approach between the center of the X-ray source and the center of the eclipser:

\begin{equation}
    \rho = \kappa \pm \sqrt{\kappa^2 - (1-\beta^2)}
    \label{eq:kappa}
\end{equation}
where
\begin{equation}
    \kappa = \frac{\Big[\Big(\frac{\tau}{\tau_{\rm low}}\Big)^2+1\Big]}{\Big[\Big(\frac{\tau}{\tau_{\rm low}}\Big)^2-1\Big]}
\end{equation}
When $\rho$ has been computed from Eq~(6), the equations above provide an estimate of $\beta$.

Equation  \ref{eq:kappa} also provides a value for $\rho_{\rm max},$ the maximum value of $\rho,$ in terms of the eclipse timings. This is because the maximum value of $\rho$ corresponds to the approach where $\beta=0.$  We cannot compute $\rho_{\rm min}$ in a similar way, because $\rho$ approaches zero as $\beta$ approaches unity.  

There is, however, another way to estimate the value of $\rho_{\rm min}$,  because the value of $\rho_{\rm min}$ is achieved when the radius of $R_2$ is maximized. This happens when $R_2$ fills its Roche lobe. We have $R_L= f(q)\, a_\mathrm{in}$, where approximate values of $f(q)$ are given by the Eggleton formula. We can then write:
\begin{equation}
   \rho_{\rm min}= \Bigg[\Big(\frac{\pi}{2}\Big)\Big(\frac{1}{P}\Big)\Big(\frac{1}{f(q)}\Big)\Bigg]^2\Big(\tau^2-\tau_{\rm \mathrm{low}}^2\Big),
   \label{eq:rho_egg}
\end{equation} 
where $P$, $\tau$, and $\tau_{low}$ are in seconds.

\smallskip

{\bf Three-body Dynamics and Geometry}: If the inner binary has another object in a wider co-planar orbit, that object generates separate eclipses. When calculating the value of $v_{\rm rel}$ to include in Equation \ref{eq:rho_v}, we also must remember that this is the perpendicular component of the velocity of the third body relative to the velocity of the XRS, which is also moving. The XRS's velocity must be added or subtracted to that of the third body.  We must compute the component of each velocity perpendicular to our line of sight, and add the results.  
\begin{equation}
    v_{\rm \mathrm{rel}}=v_3 \pm v_{\rm x},
\end{equation}
where $v_x$ is the velocity of the XRS perpendicular to the line of sight, and $v_3$ is the orbital velocity of the third body. The choice of ``$+$'' or ''$-$'' corresponds to cases in which the velocities are anti-aligned or aligned, respectively.\footnote{Note that the sign of the alignment depends both on whether the orbits are prograde or retrograde and on the orbital phases.}      

We know when the eclipse generated by the outer object occurs relative to the periodic eclipses generated by the inner orbit. This potentially gives us the phase of the inner binary at the time of the outer eclipse.  Because we cannot know on the basis of a single eclipse whether the outer eclipse occurred on the near  or far side of the  orbit, the phase is uncertain by $\pm \pi$. We can nevertheless derive the instantaneous value of $v_x$ at the time of the outer-orbit eclipse.
We have $v_x=-\omega a_x sin(\theta)$, where $a_x$ is the semi-major axis of the X-ray source's orbit, $w=2\pi/P_{\rm in}$, and $\theta$ is the phase of the companion star (defining $\theta=0$ as when the distance between the companion star and barycenter of the system transverse to our line of sight is maximized). We find $a_x=a\frac{.8M_\odot}{2.2M_\odot}=1.28R_{\rm \odot}$ and that $\theta\approx.1\pi$, as the long eclipse occurs $\sim9.5$ks after the midpoint of the predicted short eclipse. This gives $v_x\approx 40$km/s. We can then estimate the velocity $v_3$ as a function of the mass and period of the third object. We scale our calculations to a total mass of $3M_\odot$, assuming an upper bound on the third body of $0.8M_\odot$.

\begin{equation}
a_{\rm 3} = 1.44 \mathrm{AU} \Big(\frac{P_3}{1 \mathrm{yr}}\Big)^{2/3}\Big(\frac{M_{\rm tot}}{3M_\odot}\Big)^{1/3}.
\end{equation} 

\begin{equation}
v_3 = \frac{2\pi a_{\rm 3}}{P_{\rm 3}} = 43 \mathrm{km/s} \Big(\frac{P_3}{1 \mathrm{yr}}\Big)^{-1/3}\Big(\frac{M_{\rm tot}}{3M_\odot}\Big)^{1/3}
\label{eq:Period_mass_velocity}
\end{equation}


Rearranging Equation \ref{eq:rho_v}, we obtain  an equation for $R_x$ in terms of the mass and period of the third body. For low-mass main-sequence stars, we can also use the relationship $R\sim (M/M_\odot)^{.85}R_\odot$ \citep{1981gask.book.....M}.
\begin{equation}
\begin{aligned}
   & R_{\rm x} = \frac{1}{16R_3}\Bigg[v_x  \pm 2\pi(\frac{GM_{\rm tot}}{4\pi P_3})^{1/3}\Bigg]^2 \big(\tau_{\rm \mathrm{3}}^2-\tau_{\rm \mathrm{3,low}}^2\big)
   \label{eq:R_x_from_pm}
\end{aligned}
\end{equation}

This can be written as
\begin{equation}
   R_{\rm x} = 
   \Big(\tau_{\rm \mathrm{3}}^2-\tau_{\rm \mathrm{3,low}}^2\Big)\, 
   \Big(\frac{40\, {\rm km/s}}{16}\Big)^2
   \frac{1}{R_3} \, \Big[1\pm 1.1\Big(\frac{1\, {\rm yr}}{P_3} \, \frac{M_{tot}}{3\, M_\odot})^\frac{1}{3}\Big]^2
\end{equation}

This  equation relates the values of the orbital period of the third (outer) star to the dynamics of the inner binary and to the ratio of the XRS and eclipser radii.   When we have information about the inner binary and about $R_x$, we can derive the relationship between the radius and orbital period of the object in the outer orbit.
\smallskip
\\

\textbf{Light Travel Time Effects:}
Limits on the mass and orbit of a third body can also be derived from the transit timing variations of the short 2.6 ks eclipse. As discussed in Section \ref{sec:dissapearing_short_eclipse}, the ingress and egress times of the short eclipse show a slow, statistically-signficant nearly-linear drift of about 1 second per year. Such a subtle effect could potentially be due to minor two-body effects like apsidal precession caused by the distorted shape of the donor star.  Or perhaps the size or orientation of the emitting region is slowly changing. The fact that the trend is almost constant with a variation of less than 10 seconds over 10 years indicates that the position of the eclipse has been almost uniform for that entire time.
Thus, the relative constancy of the ingress and egress times allow us to  rule out models in which there are significant dynamical interactions between the inner orbit and the outer companion. 
Furthermore, a Lomb-Scargle periodogram analysis \citep{1982ApJ...263..835S} does not show any significant signals. 

Even in the absence of three-body dynamical effects, light travel time effects are expected. Considering only light-travel time variations from a circular companion, we check for sinusoidal signals in the observed times. As demonstrated in \cite{2022MNRAS.510.1352B}, the light travel time effects (LTTEs) can be expressed as a function of the semi-major axis of the inner binary around the center of mass of the triple system ($a_{\rm AB}$):

\begin{equation}
    A_{\rm \mathrm{LTTE}}=\frac{a_{\rm \mathrm{AB}} \sin i_{\rm \mathrm{out}}}{c} \sqrt{1-e_{\rm \mathrm{out}}^2 \cos ^2 \omega_{\rm \text {out }}},
\end{equation}

where $i$ is the inclination angle of the system, $c$ is the speed of light, $e$ is the eccentricity of the orbit, and $\omega$ is the argument of periastron of the outer body. We assume a circular orbit for simplicity, as well as an inclination $>75\degree$. Under these assumptions, a $10$s semi-amplitude corresponds to a value of $a_{\rm AB}$ of $\approx$0.02 AU. We can thus impose constraints on the mass and semi-major axis of any third body as

\begin{equation}
    a_{\rm AB}=\left(\frac{M_3}{M_1+M_2}\right)a_3 < 0.02\ \mathrm{ au}
\end{equation}

which can be rewritten in terms of the mass and period:

\begin{equation}
    a_{\rm AB}=\left(\frac{M_3}{M_1+M_2}\right)\,\big(\frac{P_3}{1 \rm{yr}}\big)^{2/3}\big(\frac{M_{\rm \rm tot}}{M_\odot}\big)^{1/3} \rm{au}< 0.02\ \rm{au}
    \label{eq:a_ab}
\end{equation}

where $M_{\rm \rm tot}=M_1+M_2+M_3$. 

\vspace{.5cm}
\subsection{Models}

The system we are studying is potentially complex and may be comprised of distinct subsystems.  We therefore start simply, with the things we know. 
The system contains a periodically eclipsing XRB with a second, longer eclipse that occurs at a different phase from the periodic eclipses. The recurrence time of the longer $5.3$~ks eclipse is not measured.  While it is entirely possible that we have missed recurrences, the one detected for certain occurred during a time when \grs\ was in a low state. In contrast, the shorter 
2.6 ks eclipses are sometimes missing during X-ray low states. 

There is at least one compact object accreting matter. We call this accretor $A_1$. The occurrence of Type I X-ray bursts indicates that the accretor is a NS. 
The system also contains an eclipser, $E_1$. The eclipser $E_1$ is the object creating $2.6$~ks eclipses of $A_1$ at a period of $0.515$~d. In the literature it has been assumed that $E_1$ is the donor star. While this is the most likely explanation, for completeness we also consider models where $E_1$ is not the donor star. The nature of $E_1$ is modeled but has not been verified, as there is no optical counterpart to \grs. 
There exists also a secondary eclipser, $E_2$. This eclipser produces $5.3$ ks eclipses, but little else is known of its nature.

We explicitly consider the following three overarching system architectures, in which there exist:
\begin{enumerate}
    \item One compact object in orbit with two non-compact objects.
    \item Two compact objects orbiting each other in the same system which also contains a donor star.
    \item Two separate compact objects, each in its own LMXB.
\end{enumerate}

\smallskip

\subsubsection{(1) The Single Compact Object System}\label{classic_triple_body}

We first consider models in which the system is comprised of one compact object and two non-compact objects.  At least one of the latter  must be stellar in nature,  providing matter to the compact object through Roche lobe overflow. The other non-compact object may be either stellar or substellar.  Because of the detection of Type I X-ray bursts, we start with an inner binary composed of the NS and the 
2.6
 ks eclipser,  $E_1$.  The 
5.3
 ks eclipser, 
$E_2$,
 is in a wider circumbinary orbit.  Dynamical stability requires that the ratio 
$a_{out}/a_{in}$
 is higher than a critical ratio, which we take here to have a value of 
3 \citep{mardling1999dynamics}. 

\smallskip

{\bf (1.a.) $E_1$ fills its RL and $E_2$ doesn't fill its RL.} 
Models in the literature already consider the XRB to be an LMXB in which the donor is a subgiant \citep{2003A&A...406..233I}.  We find the same result for the inner binary. The new element here is that there is a second object orbiting the compact accretor.  We will consider case {\bf 1.a.} in detail, as it provides the tools needed for the other cases.

 
The orbital radius and velocity can be computed from Equations \ref{eq:inner_binary_semimajor_axis} and \ref{eq:velocity}. Scaling our calculations to a mass of $0.8M_\odot$ for the donor, $1.4M_\odot$ for the accretor, we find that $a_{\rm in}=3.51R_\odot$ and $v_{\rm in}=345$km/s. We can compute the Roche-lobe radius of the donor star using Eggleton's formula, and find that $R_{\rm RL} = R_2 = 1.15R_\odot [M_{\rm bin}/(2.2\, M_\odot)]^\frac{1}{3} [f(q)/0.332]$ (assuming a NS accretor). This radius indicates the the donor is a subgiant, as a $<0.8M_\odot$ main-sequence star would be smaller.  A subgiant donor can provide mass at the rate  needed to achieve the highest luminosities attained by the accretor.  Note in addition that if the accretor is a BH, $M_{\rm bin}$ is larger and $f(q)$ may be smaller by a modest factor.  Thus,
 even for the case of a BH accretor, the donor is a slightly evolved subgiant. 
 
 To find a value of $\rho,$ we use the ingress and egress durations derived from \cite{2003A&A...406..233I} to estimate $\tau_{\rm \mathrm{in}}=2626.6\pm 2.1{\mathrm{s}}$, and
$\tau_{\rm \mathrm{in,low}}=2565.5\pm 2.1{\mathrm{s}}.$ Using Equation \ref{eq:rho_v} and \ref{eq:rho_egg}, we find
    $\rho=0.0036$.
  With $R_x=\rho \, R_2,$ we have  $R_x = 2.9 \times 10^8$~cm for the RL-filling NS case with $M_2 = 0.8\, M_\odot$.  Note that the set of values $\{v_2, R_2, R_x,\beta\}$  determine the solution and should produce a light curve consistent with the observed $2.6$~ks eclipse.


In principle, the X-ray observations can provide a check on the value of $R_x.$  Unfortunately, however, the observed X-ray spectrum does not constrain the size of the emitting region. The loss of most photons below 1 keV, and the dominant power-law shape above 1 keV, preclude the identification for example of blackbody components from the inner disk or the surface of the compact object, which would have constrained the origin and size of at least one component. Our inferred size $R_{\rm \rm X} \approx 3 \times 10^{8}$ cm, is larger than what we may expect for emission solely from the inner disk,  but it is consistent with the coronal sizes inferred by \citep{church04} from the study of dipping NS LMXBs, for comparable luminosities of $\sim$ a few $10^{36}$ erg s$^{-1}$. We caution that there are several uncertainties in our numerical calculation of $R_{\rm \rm X}$, due to uncertain system masses and accretion geometry.

\subsubsection{Characteristics and Orbit of $E_2$: Non-RL-filling}
 Because the $5.3$~ks eclipse doesn't repeat over short time intervals,  the orbital period is not  as short as the longest exposure. Furthermore, only one example of it is detected in the full set of exposures, so that the period is likely comparable to or longer than the total exposure, $T_{\rm exp} \approx 10^6$~s,  of the X-ray observations we have analyzed.  We therefore  expect that $P_3$, the period of the outer orbit, is not likely to be much shorter than $11.6$~d, and that it may be significantly longer. If we take the peculiar dip in the \textit{Suzaku} observation as another example of the long 5.3 ks eclipse, we could put a lower bound on the period of $\sim6$ d. Any lower bound derived through exposure time arguments is very soft, and the condition of dynamical stability alone would permit a much shorter period of about $2.5$~d.  \footnote{Note that this system is undergoing mass transfer, so that dynamical stability arguments derived for purely gravitational interactions may need to be revised.} 

An upper limit on $P_3$ can be derived if we consider that the \textit{Suzaku}-observed eclipse may be a second occurrence of the \chandra-observed $5.3$~ks eclipse. This would suggest that the orbital period is at most about $11$~yrs, and could be significantly shorter if our observations missed other eclipses.  

In addition, the high stellar density in globular clusters leads to frequent interactions that have the effect of destroying wide binaries. 
While a soft orbit (one in which the orbital velocity is smaller than the ambient velocity of nearby cluster stars) can survive for a limited time, harder orbits are more likely to persist. This means it is likely that $P_3 < 7$~yr\footnote{Seven years corresponds to the orbital period of a $2.2\, M_\odot$ system with an orbital speed of $20\, $~km/s. This should be viewed as a very rough estimate of the maximum orbital period.}.   Note, however, that this dynamical argument cannot yield exact results:  the time taken for dynamical interactions to occur is governed by statistical processes. Globular clusters therefore include some soft systems that survive over longer-than-average times.  An example illustrating this point is 
PSR B1620-26 b in the globular cluster M4. PSR B1620-26 b is a $2.6\, M_J$ planet in a wide orbit with a close-orbit white-dwarf/millisecond-pulsar. Its semimajor axis is $23$~AU, and its existence shows that it is possible for wide-orbit planets to live for considerable times in ``soft'' orbits.

\begin{figure}[t]
    \centering
    \includegraphics[width=.48\textwidth]{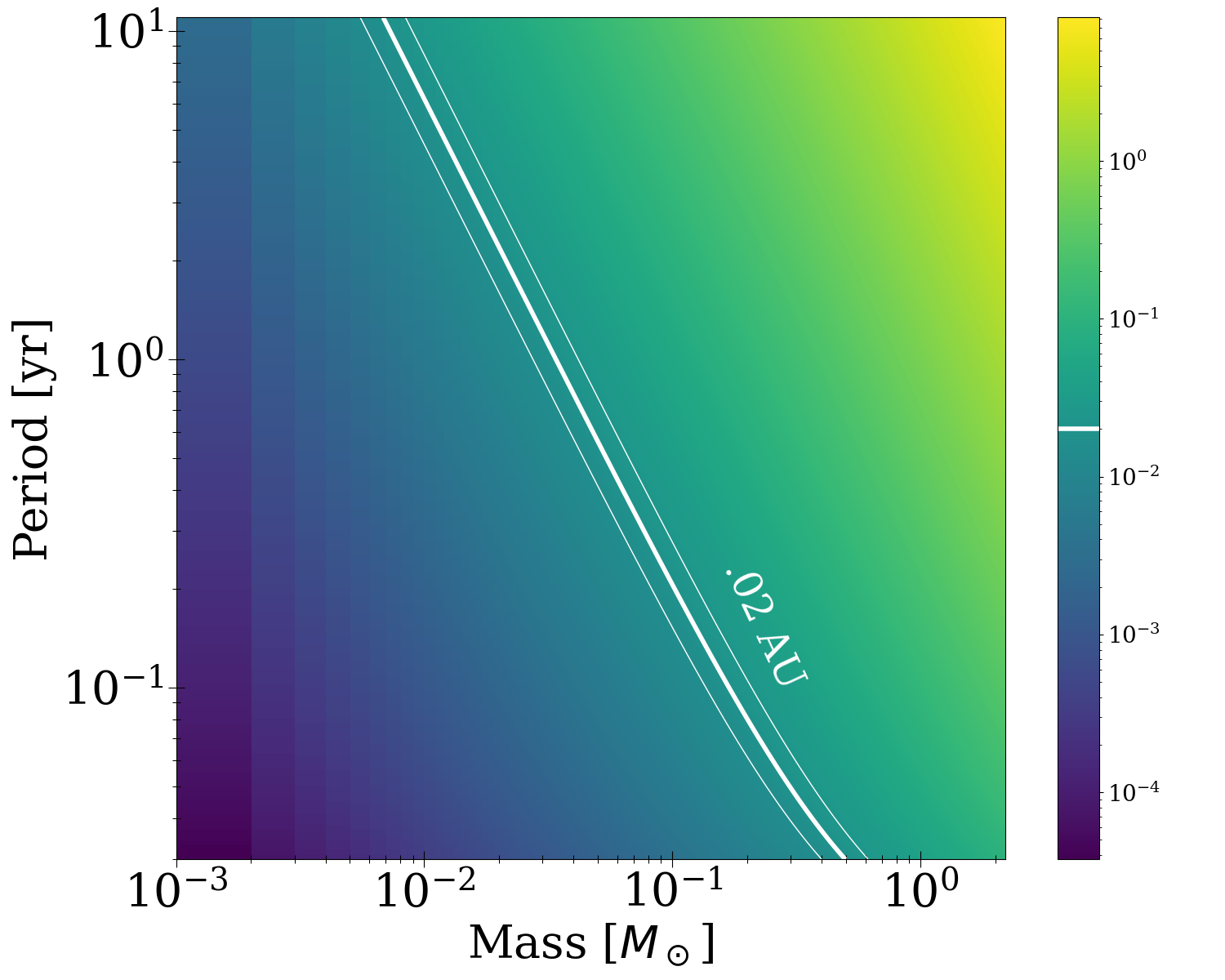}
    \caption{Log of the period versus the log of the mass. In this representation, each point corresponds to a value of $a_{\rm AB}$ (using Equation \ref{eq:a_ab}), which must be $<0.02$ au. Periods range from 6 d to 11 yrs, and masses range from $10^{-3}-1M_\odot$. The thick white contour line is at $a_{\rm AB}=0.02$ au, assuming a NS-accretor inner binary with mass $M_{\rm bin,in}=2.2M_\odot$. Points to the left of the white line correspond to masses and periods consistent with the observed LTTEs. Thin white contour lines show $a_{\rm AB}=0.02$ au for inner binary masses of $M_{\rm bin,in}=1.6M_\odot$ and $M_{\rm bin,in}=3M_\odot$. Color bar to right shows increasing values of $a_{\rm AB}$ on a log-scale, from $10^{-4}$ au to $7$ au.}
    \label{fig:lttes}
\end{figure}

The lack of observed LTTEs during the many repetitions of the 2.6~ks eclipse places clear limits on the allowed mass of the third body.  Assuming a mass of the inner binary, we can identify the period-mass combinations of a third body producing LTTEs consistent with  the \textit{RXTE} data. Figure \ref{fig:lttes} shows the results for a NS-accretor inner binary.
Assuming a likely lower bound on the period of $\sim6$d, only donor masses under $\sim0.5M_\odot$ down to planets are allowed. As the period ($P_3$) increases, higher-mass objects are eliminated, so that only planets are allowed for the largest periods shown here.  If the inner binary is more massive, then slightly larger tertiary masses and outer orbits are plausible.

Next, we  employ analogs of the geometrical arguments we used for the inner orbit.  Using estimates on the values of $\tau_3$ and $\tau_{\rm 3,low}$ from the 5.3 ks \textit{Chandra} eclipse, we use Equation \ref{eq:R_x_from_pm} to test what period and mass combinations for the third body would produce realistic sizes of the X-ray source emission region.  The value of $R_x$ computed by considering the inner orbit should be roughly equal to the value computed by  considering the outer orbit.

From the fits to the long eclipse, we have that the most likely values of these parameters are $\tau_{\rm out}=5.30\mathrm{ks}$ and $\tau_{\rm out,low}=5.08$.  This corresponds to a total of the ingress and egress time of $220$~s. 
These are the values we employ in our fiducial calculations. Note however, that, since this eclipse was observed only once by {\sl Chandra}, and since we have no \textit{RXTE} or \textit{NICER} data on the long eclipse, there
are significant uncertainties in the ingress and egress times and, consequently in the values $\tau_3$ and $\tau_{\rm 3,out}$. We therefore also include a calculation in which the value of $\tau_{\rm 3,low}$ is set to $5.18$~ks, which can be found in Appendix \ref{sec:more_3_body_plots}.

\begin{figure}
    \centering
    \includegraphics[width=.48\textwidth]{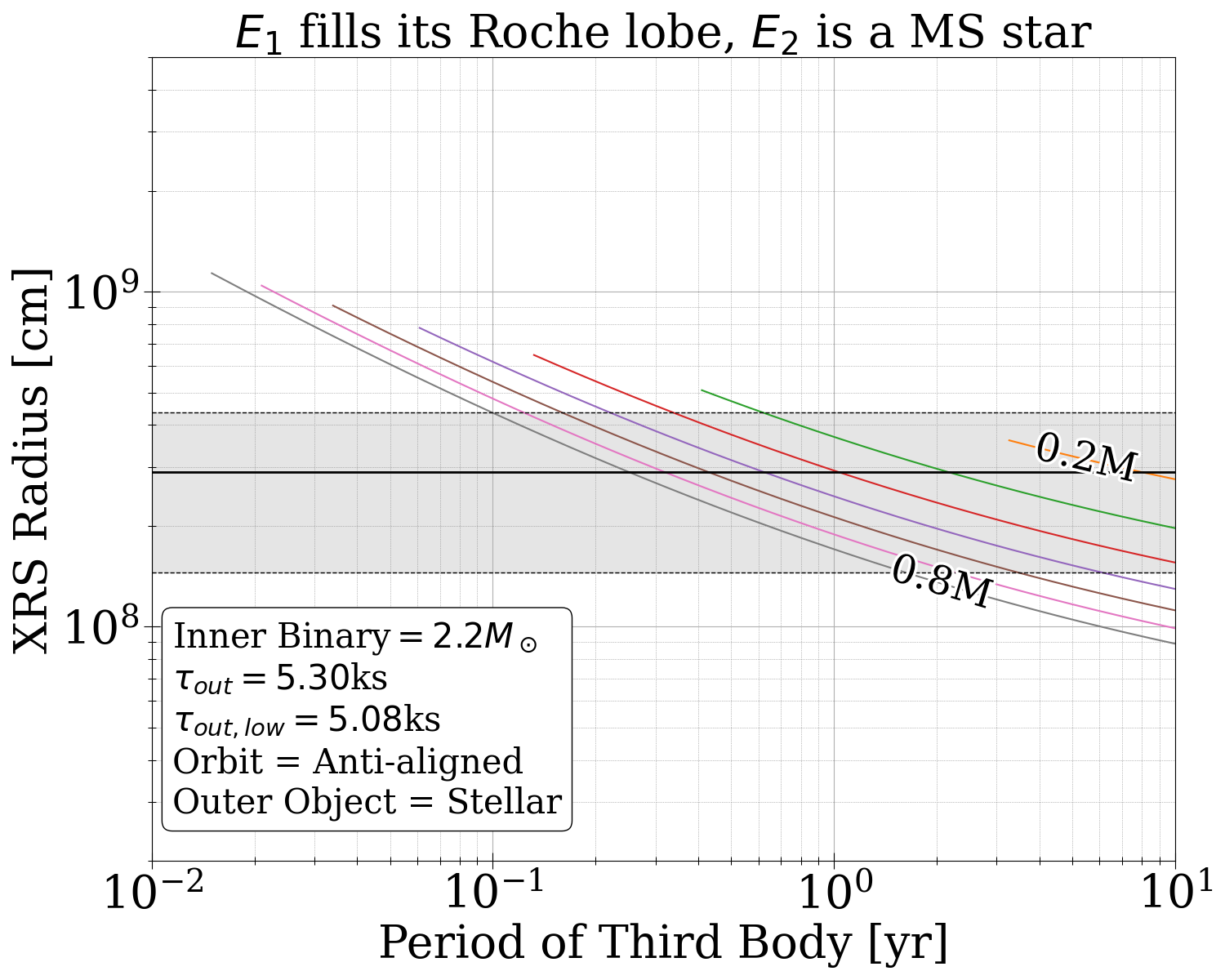}

    \includegraphics[width=.48\textwidth]{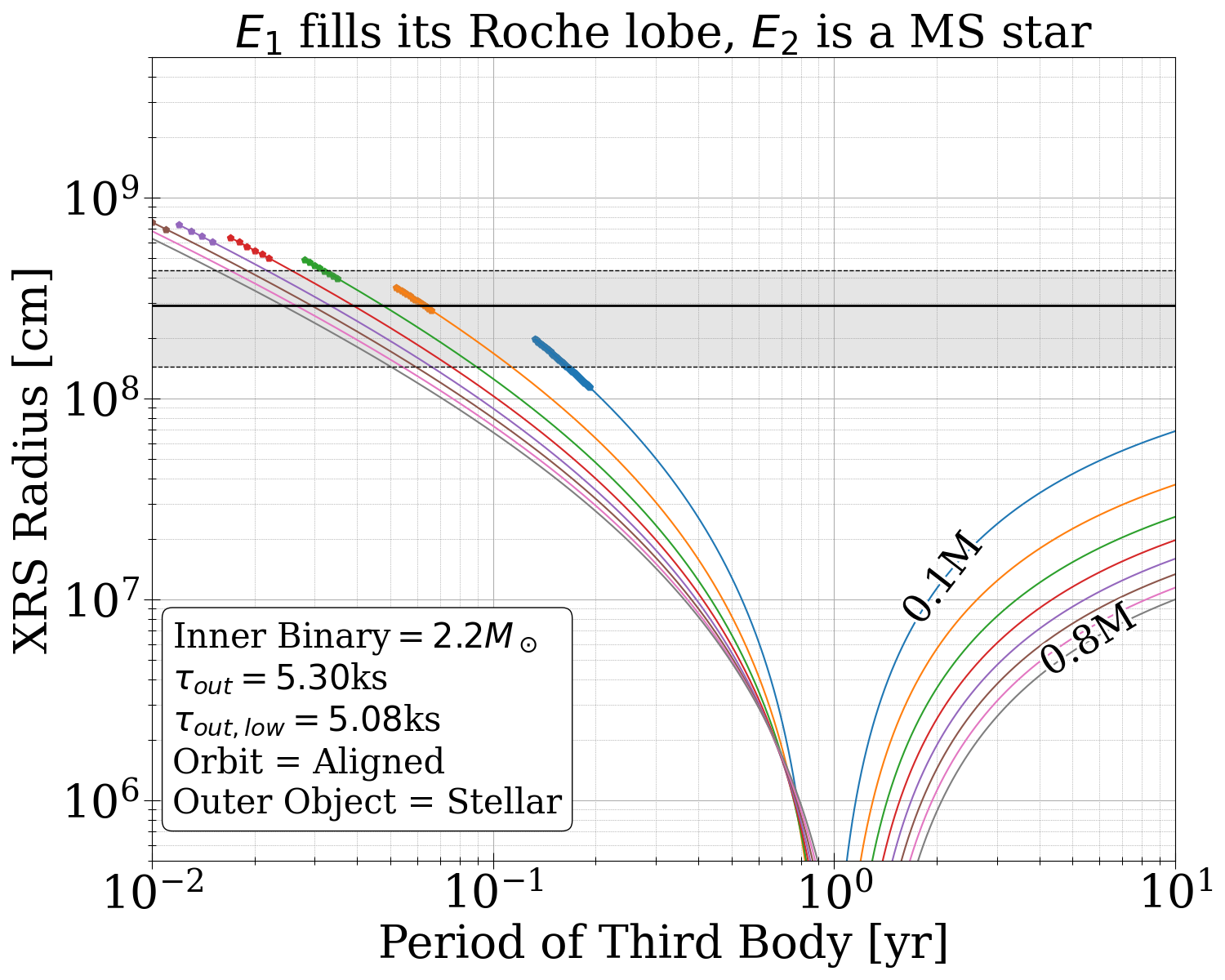}


    \caption{Plotted are the logarithm of the XRS size, in cm, versus the logarithm of the outer orbital period, in years, for main-sequence stars of masses between $0.1-.8M_\odot$ on circumbinary orbits. Masses correspond to different color curves (in increments of $.1M_\odot$). Dots/thicker lines on curves  show regions where the period and mass are consistent with the lack of strong LTTEs in the data. The horizontal black line is the XRS radius derived in section 4.3, with an error bound in grey of $\pm 50\%$. Because $M_3$ orbits the same XRS as the 2.6 ks eclipser, we expect this quantity to be the same. {\bf Top panel:} Anti-aligned orbit. {\bf Bottom panel}  Aligned-orbit.}

    \label{fig:3_body_models}

\end{figure}

\begin{figure}
\centering
    \includegraphics[width=.48\textwidth]{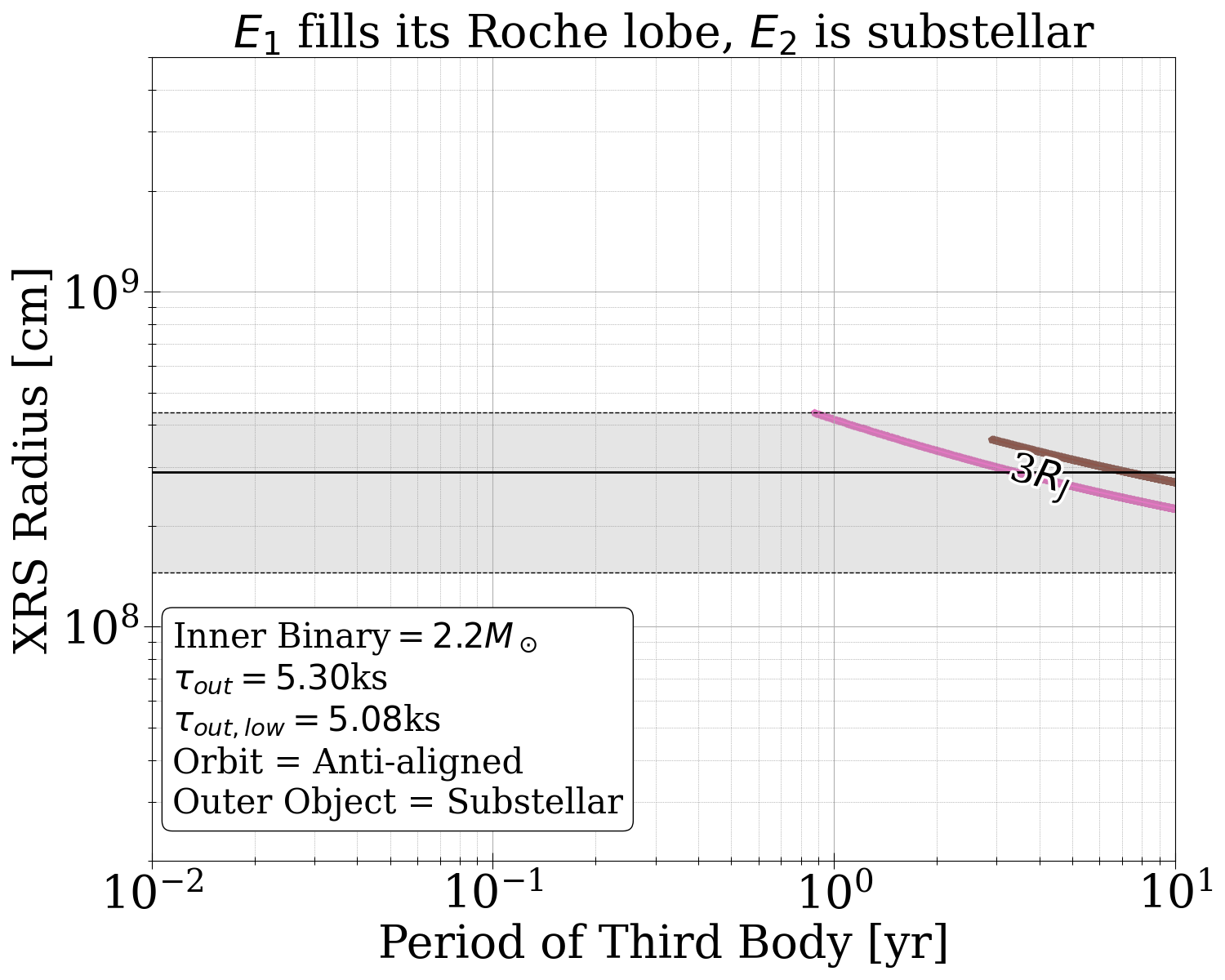}
    \includegraphics[width=.48\textwidth]{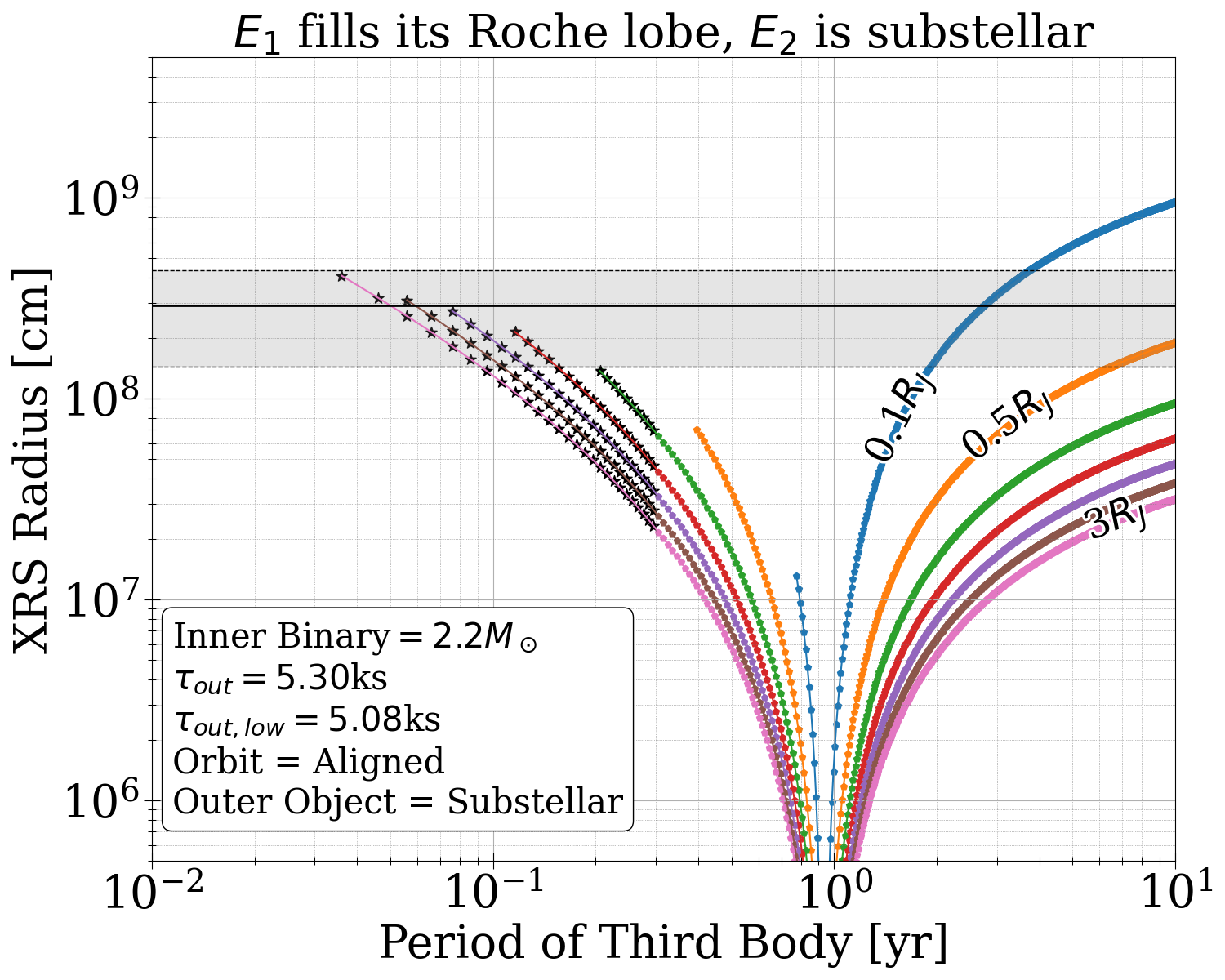}


    \caption{The quantities plotted on each axis are the same as in the previous plot, as are the positions of the horizontal black lines.  The outer orbiter is a planet. Its radius is chosen as shown: $0.1 R_J$ (blue) in the upper curve,
$0.5~R_J$ in the next lower (orange) curve, increasing by a further $0.5\, R_J$ for each subsequent curve proceeding downward.  The mass is chosen by computing the density, which is taken to be $5.5$~ g/cm$^3$ for $.1R_{\rm J}$ planets and $1.4$~g/cm$^3$ for larger ones. As before, dots/thicker lines mark regions consistent with the lack of LTTEs, and we mark with black stars where $\geq1R_{\rm J}$ sized objects with brown dwarf masses are consistent as well. {\bf Top panel:} Anti-aligned orbit. {\bf Bottom panel}  Aligned-orbit.}

    \label{fig:3_body_models_planets}

\end{figure}

\textbf{Results:} The computed X-ray source radii are plotted as functions of period, $P_3$,  in Figures \ref{fig:3_body_models} and \ref{fig:3_body_models_planets}.    A solution consistent with the orbital dynamics and eclipse characteristics must pass through the gray shaded region.  In Figure \ref{fig:3_body_models}, each curve corresponds to a main-sequence star of a given mass.  The curves have points superposed, making the curves appear thicker and brighter in color for cases in which the LTTEs are consistent with RXTE, and for which $\beta^2 > 0.$  Thus, it is only the dotted (thicker-looking) portions of each curve that are allowed.  In the anti-aligned panels, the velocities of the eclipser and XRS point in opposite directions at the time of eclipse; that is they are anti-aligned. In the  bottom panels, the velocities are aligned.  In the case in which the velocities are aligned, the difference in the velocities is smaller; an eclipser that produces a eclipse of duration $\tau$ can have a smaller radius than one producing an eclipse of equal duration in the anti-aligned case. 

Figure \ref{fig:3_body_models} shows the results for the case in which $E_2$ is a main-sequence star. We find that {\bf the orbiter can be a main-sequence star only when the velocities are aligned. For the assumed ingress and egress durations, values of  $M_3$ must  be in the range  $(0.1-0.3)\, M_\odot$   When $M_3$ is $0.1\, M_\odot$,  $P_3$ is roughly $50-60$~d. For larger masses, $P_3$ is smaller: roughly $6$ days for an M-dwarf of $0.3\, M_\odot$.} While direct detection of an M-dwarf is difficult, the shorter periods associated with low-mass M-dwarfs may allow future observations to detect more long eclipses. 


In Figure \ref{fig:3_body_models_planets} we consider models in which the radius of the object in the outer orbit is in the substellar regime.  {\bf For aligned orbits, planets with radii near or above a Jupiter radius are allowed at periods of $\sim 40$ days. The values allowed for $R_3$ increase as $P_3$ declines. For these short periods, both planets and brown dwarfs are allowed. Also allowed are small planets with $P_3 \sim2$ yr. For anti-aligned orbits, larger values of $P_3$ allow larger $R_3$ for planetary mass objects. Overall, brown dwarfs are allowed in short orbits, while planets are allowed in both short and long orbits.} Orbital parameters of specific cases for both the substellar and MS regimes can be found in Table \ref{tab:RL_MS} in Appendix \ref{sec:params}.

These calculations rely on several key assumptions and estimates. Specifically, these are the estimates of the time duration of the ingress and egress, as well as the assumption that the geometry of the X-ray emission region is circular. However, even if the total eclipse transition time is significantly shorter ($\tau_{\rm out,low}=5.18$ks), the conclusions drawn do not differ significantly (see Appendix \ref{sec:more_3_body_plots}). Most importantly, slightly more massive stars on aligned orbits ($\sim0.6M_\odot$) become consistent with both the geometric and LTTE arguments. It is also possible that the size of the XRS changes between the low state and the high sate, in which case the inferred size from the long 5.3 ks eclipse does not necessarily need to be consistent with that derived from the 2.6 ks one. Even if this is the case though, the LTTE arguments still place strong limits on the nature of any third body.

\smallskip


{\bf (1.b.) $E_1$ doesn't fill its RL and $E_2$ fills its RL.}  

We consider the case in which the $2.6$ ks eclipser does not fill its Roche lobe. Instead, $E_2$, the object in the outer orbit producing the longer eclipse, is the donor star. 
We first consider the possible sizes of $E_1$.  Using $\beta=0$ and assuming $R_2>>R_x$, we compute the smallest value of $R_2$ as
\begin{equation}
    R_{\rm 2} \approx \frac{2(2.6\rm{ks})}{v_{in}} =
    0.64\, R_\odot \, \Bigg[\Big(\frac{P_{in}}{0.525~{\rm d}}\Big)\, \Big(\frac{2.2\, M_\odot}{M_{bin}}\Big)\Bigg]^\frac{1}{3}
\end{equation}

We see that $R_2$ must be in the stellar range. A radius of $0.64\, R_\odot$ corresponds to a stellar mass of $\sim 0.6\, M_\odot.$  Because the star is in a globular cluster, we consider vales of $M_2$ in the range $(0.6-0.8)\, M_\odot$\footnote{If $E_1$ is a blue straggler the mass could be larger.}. For these masses, if we take $E_1$ to be a MS star than the value of $R_x$ is $4-4.8\times10^8$cm (From Equation \ref{eq:rho_v}).





 In each case, $a_{in}$  has a value roughly equal to $3.5 R_\odot$. The minimum value of $a_{out}$ consistent with dynamical stability is therefore $\sim 10.5\, R_\odot$. Assuming an inner binary mass of $2.2M_\odot$, the LTTEs place upper limits on $P_{out}\lesssim4.5$d and $a_3\lesssim16R_\odot$ for $M_3=0.6\, M_\odot$  and $P_{out}\lesssim2.5$d and $a_3\lesssim11.2R_\odot$ for $M_3=0.8\, M_\odot$. 
 

 
\begin{figure}[t]
   \centering
   \includegraphics[width=.48\textwidth]{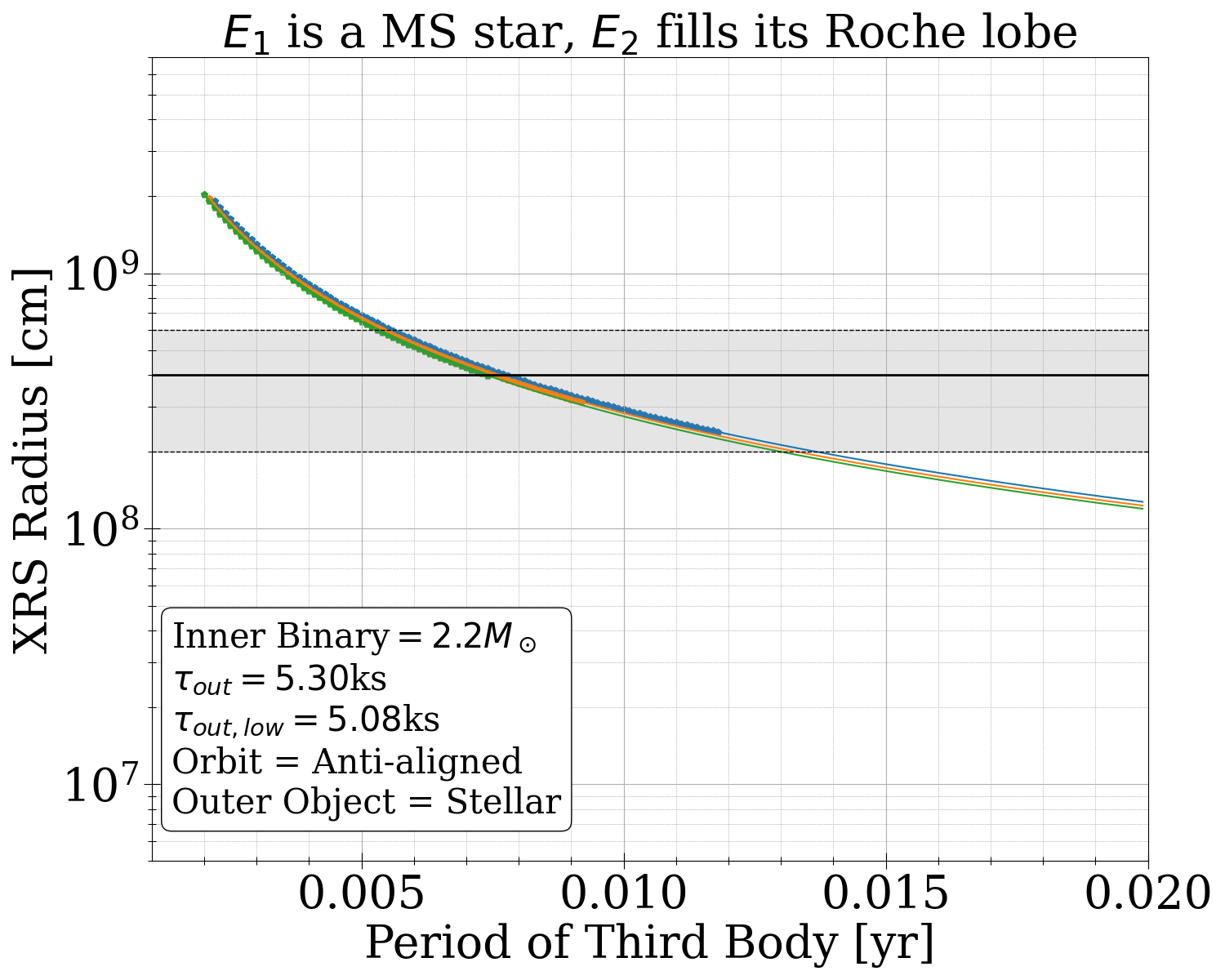}
   \centering

   \includegraphics[width=.48\textwidth]{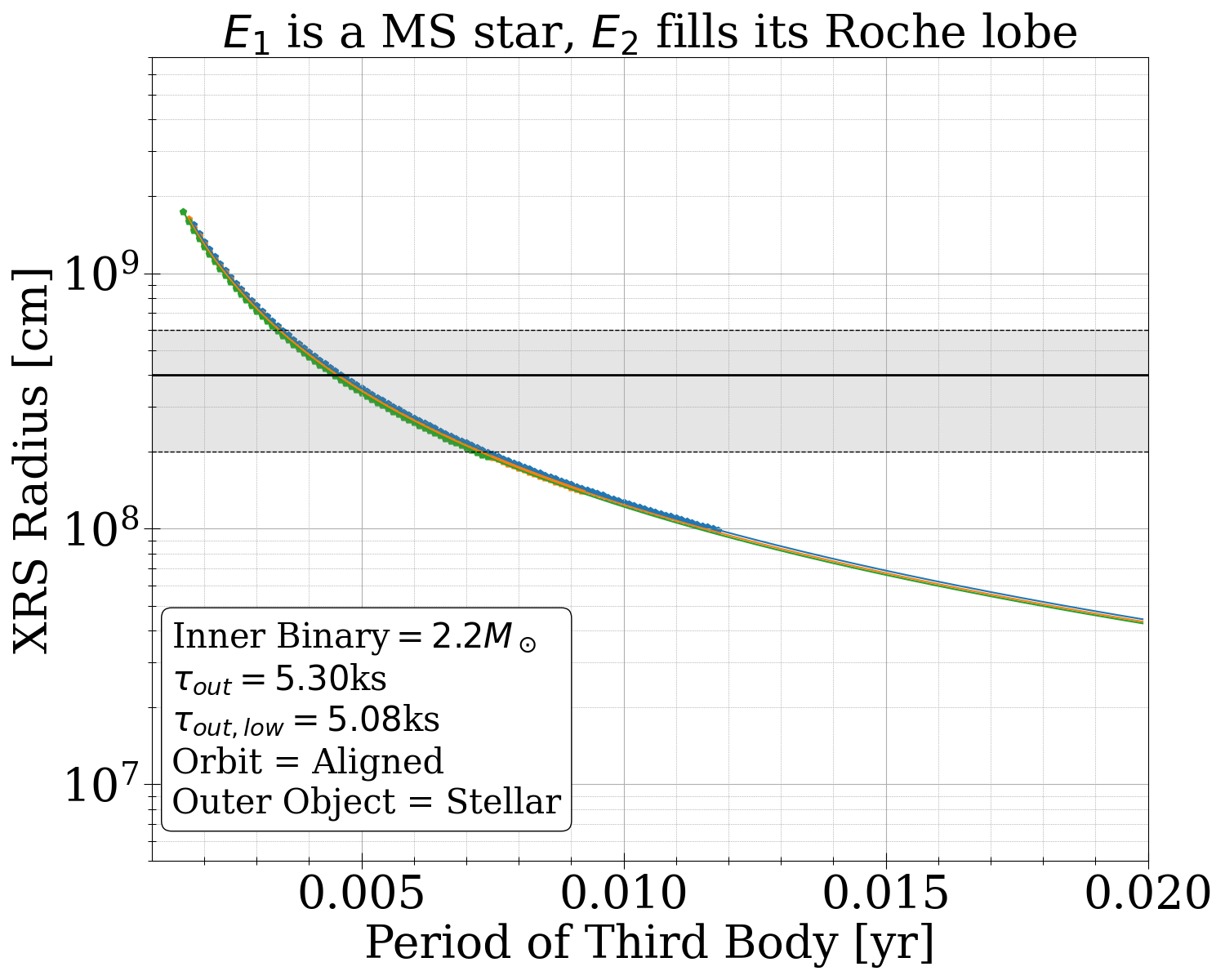}

   \caption{Logarithm of the XRS radius, in cm, versus the logarithm of the orbital period of $E_2$, in years, assuming a Roche lobe radius. Plotted are masses starting from $.6M_\odot$ (blue on the top curve) to $.8M_\odot$ (green on the bottom curve) in increments of $.1M_\odot$. The black line shows an XRS of $4\times10^8$cm, as derived from the inner orbit assuming a MS radius for $E_1$, with an error bound in grey of $\pm 50\%$. Dots on curves show regions where the period and mass are consistent with non-detection of LTTEs in the data. \textbf{Top panel:} Anti-aligned orbits. \textbf{Bottom panel}: Aligned orbits.}
   \label{fig:roche_3}
\end{figure}

 As before, we compute $R_x$ for both the anti-aligned and aligned outer orbits, now assuming a Roche lobe radius for $E_2$ (calculated using Equation \ref{eq:eggleton}). Results are shown in Figure \ref{fig:roche_3} and some specific cases are described in Table \ref{tab:MS_RL} (Appendix \ref{sec:params}). In agreement with the LTTE constraints, the period of an outer star filling its Roche lobe would need to be very short $<4.5$d, to produce consistent XRS sizes. 
 

 This suggests that the triple system could consist of a NS in orbit with a subgiant donor with mass $\sim 0.8\, M_\odot$ and radius $\sim 3.4\, R_\odot$, responsible for the $5.3$~ks eclipse. The NS would also have a closer companion, a dwarf star that could be as massive as the cluster turn off.   An interesting feature of this scenario is that the dwarf star would be orbiting within the Roche lobe of the NS, and could experience mass infall of some portion of the matter donated by the outer subgiant. If so, the accretion luminosity would likely be smaller than $10^{-4}$ times than of the NS, and accretion emerge associated with the dwarf star would likely be emitted at wavelengths longer than X-ray wavelengths.  

 The pattern of mass flow could be unusual, because the standard Roche-lobe picture doesn't hold for stellar triples. In this case, we would expect that the L1 point, where the total force (gravitational and rotational) balances, would be moving, so that mass would be sprayed toward the NS \citep{di2020dynamical}. A circumbinary disk could form. If so, mass could be channeled to the dwarf star,the least massive binary component, and thence to the NS. 
 
Note that this model has $E_1$ orbiting interior to the donor, potentially within  the accretion stream. This may imply that the 2.6 ks eclipses or other aspects of the system could have physical characteristics that make this model potentially falsifiable. The periods required ($\lesssim4.5$d) are also smaller than we would expect based on the exposure time arguments presented above. Additionally, the large radius of the RL filling $E_2$ would necessitate a very large $\beta$ value (see Table \ref{tab:MS_RL}). Overall, this scenario seems unlikely, and the generally accepted model where $E_1$ is the donor is preferred.

\smallskip

{\bf (1.c.)  $E_1$ fills its RL and $E_2$ also  fills its RL.} We also consider the case in which $E_2$ also fills its Roche Lobe. This would seem to be an unlikely situation.  As we have seen, $E_1$ must be slightly evolved in order to fill its RL. $E_2$ is in a wider orbit and therefore needs to be even larger and more evolved to fill its RL. This suggests that the two donor stars must both have masses roughly equal to $0.8\, M_\odot.$  Given the fact, however, that exchange processes in which binaries participate tend to expel the least massive stars in the system, it may  be reasonable to expect that, if a compact object has two stellar companions, those companions would be among the most massive stars in the cluster. It is further unlikely that this scenario is the case due to the lack of strong LTTEs, and the fact that for such a system a third object would produce consistent XRS sizes only for periods shorter than we would expect ($<6$d). Additionally, this model faces the same problems as \textbf{1.b.} in that the large radius of $E_2$ would necessitate very large $\beta$ values  ($>.99$) to produce the right duration eclipse.
\smallskip




\smallskip
\smallskip


\subsubsection{The Double Compact Object System}

In addition to models in which there are two non compact object eclipsers, we also consider models in which there are two accretors within the same system. 
There are two reasons to consider this model. First, we expect such systems to form within those dense stellar environments which have many compact objects and within which the stellar interaction rate is high. Terzan~6 is such an environment.  Second, this model presents several possible advantages in explaining aspects of the observations of \grs .

One advantage is that it would allow for both a NS and BH, reconciling the high radio emission and Type I X-ray bursts of GRS 1747-312.   Another is that it has the potential to explain the paucity of $2.6$~ks eclipses in the low state, and the occurrence of the $5.3$~ks eclipse in the low state.  The model is not, however, a panacea, and we describe some drawbacks at this section's end. 

In this scenario, the two accretors, $A_1$ and $A_2$  are in a close orbit.  We take $A_1$ to be a BH, and $A_2$ to be a NS.
The donor star, which is also the second eclipser, $E_2$, fills or nearly fills its Roche lobe.  It donates matter to both inner compact objects.  The mass flow from the donor is complex, as the L1 point executes a nearly periodic dance.  The less massive object, the NS, may pull in more matter directly because it passes closest to the inner edge of any circumbinary flow. Some matter may nevertheless be donated directly to the BH,  and additional matter may interact with the NS and its accretion flow before traveling to the BH.  The BH  undergoes a regular (but not periodic) sequence of high and low states, which we assume to be related to modest changes in its accretion rate.  The inner orbit has a period of $0.515$~d, and X-rays from the BH are occulted for $2.6$~ks by the disk or corona of the NS.  Low states correspond to low states of the BH, and the paucity of low-state 2.6 ks eclipses may be due to the fact that the  BH emits too few X-rays during its low state.  

 For NS masses between $1.2-2.2M_\odot$ and BH masses between $5-15M_\odot$, the orbital separation would be between $5-7R_\odot$. Considering the inner binary, the Roche lobe of the NS would be $1.3-1.6R_\odot$. A donor star in an outer orbit would need to be at least three times as far out for orbital stability, so $a_{3}\gtrsim15R_\odot$. Such a star would need to be a subgiant to fill its Roche lobe. If we take a $0.8M_\odot$ star, its Roche lobe would be $\sim.21$ the outer orbital radius, or $\sim3.15R_\odot$. Due to the range of possible masses for the inner binary, the lack of LTTEs does not restrict the mass of the third body as much as in the single NS accretor model (see Equation \ref{eq:a_ab}). For a high inner binary mass ($\sim15M_\odot$), the lack of LTTEs places an upper limit on the period of an outer donor at $<20$d, for which a $.8M_\odot$ subgiant would produce stronger than observed LTTEs ($>10$s). There is a noteworthy challenge to this model, which is that the constant duration and timing of the eclipses require that the NS's accretion structure doesn't change over the course of the 25 years that GRS 1747-312 has been observed. Specific orbital parameters for solutions to this model can be found in Table \ref{tab:2_CO} in Appendix \ref{sec:params}
 


A second two-accretor triple model places $A_1$, a BH, in a close binary with $E_1$, which is also the donor star. This mimics the first triple body scenario. The difference is that $A_2$ is a NS in a wide outer orbit. During low states of the BH, the NS produces low-flux emission, and it can be eclipsed by $E_1$. For black hole masses of $5-15M_\odot$, the semi-major axis of the donor in such a system would be in range from $4.8-6.8R_\odot$ (for a $.8M_\odot$ donor). Such a donor would need a radius of $0.9-1.16R_\odot$ to fill its Roche lobe, and its orbital speed would be between $320-670$km/s. This model is  constrained by the the lack of strong ETVs. Even with an inner binary mass as high as $\sim15M_\odot$, the lack of strong LTTEs rule out any third body with $M_3\geq1.2M_\odot$ and $P_3\geq12$d. As before, specific orbital parameters for solutions to this model can be found in Table \ref{tab:2_CO} in Appendix \ref{sec:params}. For both of the double compact object scenarios, the large radius of $E_2$ would again require very high $\beta$ values to produce the right duration 5.3 ks eclipse.

\subsection{The Double LMXB Scenario}

We consider the possibility that the radiation we detect from the direction of  \grs\ emanates from two different LMXBs.     We assume that only one of these LMXBs is responsible for the regular bright states. This source is blocked during the 2.6~ks eclipses, but it is not active during the  low state, or at
least does not provide the bulk of the X-rays detected during the low state.  This would explain the lack of
2.6~ks eclipses during the low state. 
In addition, the relatively high ratio of radio to X-ray flux could be explained if  one accretor is a BH and the other is a NS.  

In this  model, the 
orbiting mass responsible for the $5.3$~ks eclipse 
is in a different binary from the mass generating the shorter eclipses.  
The luminosity of this separate binary would reach up to  $\sim3\times10^{34}$ergs/s, as seen in the low-state \textit{Chandra} observations. Assuming that $E_2$ indeed orbits a separate compact object, the parameters of specific cases for both a NS and BH LMXB can be found in Table \ref{tab:2_seperate} in Appendix \ref{sec:params}.

The question that must be considered is whether the two LMXBs are likely to be unrelated to each other.
If there are two individual and separate LMXBs, they are separated by  $\sim 0.7$ arcseconds, about 7000 AU in projection. Table 2 in \citet{2023hxga.book..120B} lists all known total eclipsing LMXBs. With only $\sim12$ out of $>200$ LMXBs known to be eclipsing, the probability that there exist two independent ones so close to each other is small. 

If, therefore, there are two LMXBs, they are most likely to be gravitationally bound.  While we do not expect to find double LMXBs in the field, \grs\ is on the edge of the core of Terzan~6 ($1.4-2"$ from the center) and it may be possible for interactions to produce such systems.  The probability of forming double LMXBs appears not to have  computed.   Whether or not \grs\ is a double LMXB, its properties suggest that such calculations are needed, especially because processes that occur in globular clusters seem capable of producing bound binaries of binaries. 

The rate of interactions within a cluster is dominated by the cross section for gravitational focusing.  Consider the cross section for the interaction of a mass $m$ interacting with another mass, $M$.  Their relative  speed is $v$. 
In order for the interaction to be significant\footnote{By ``significant'' we mean that the interaction leads to a change of state, such as a capture of $m$ or the ejection of a star from $M$, if $M$ represents the mass of a binary.}, the two masses must pass within a distance $s$.  The interaction cross section is
\begin{equation}
    \sigma(m,v)= \pi\, \Big[ s^2 + \frac{2\, s\, G\, (m+M)}{v^2}\Big]
\end{equation}
The second term is the associated with gravitational focusing and is the dominant term for most cluster interactions. The interactions can be of many different types, and may lead to the capture of a passing star, the replacement of a star within a binary by another,  typically more massive, star,  or the creation or break-up of a triple. 

We note that the formation of a double binary need not occur within a single interaction, but may instead require a sequence of interactions that occur over time.  Furthermore, the start of an X-ray active phase may start long after the stars that form the XRB have come together.  

The overarching principle is that, during each interaction that leads to the binding of a star to others, some of the inital kinetic energy must be lost.  This can happen through the ejection of a system component, but can also happen if the components of a binary participating in the interaction are drawn into closer orbits, or if tidal interactions allow some of the initial kinetic energy to be absorbed.   Thus, close binaries tend to become closer, and if a star is ejected it is most likely to be replaced by a more massive star.  

Factors that favor interactions of binaries include that  the value of $s$ is larger,  as are the values of $M$ and/or $m$, while the values of $v$  tends to be smaller.  Sequences of interactions that can lead to bound binaries, each of which contains a compact object, should be studied. It is not unexpected that the two members of such a binary are X-ray emitters during a common epoch. This is because a sequence of interactions is likely to lead to close binaries, with each binary containing a main sequence star that is one of the most massive in the cluster, thus likely near the turn off. Calculations can determine whether the two orbits are likely to be co-planar.

\subsection{Chance Fly-by of an Unrelated Object}

For completeness, we also explictly consider the possibility that the 5.3 ks eclipse is due to the chance fly-by of an unrelated object either in our own solar system or Terzan 6. The proper motion of Terzan 6 is $\sim140$ km/s. Objects in the Oort cloud on average move at speeds $\ll1$ km/s and are $<100$ km in diameter\footnote{It is possible larger objects may exist in the Oort cloud, but these are likely to be very rare.} (\citet{2021arXiv210303289E}, \citet{2022A&A...659L...1L}). An object of this size could not produce an eclipse as long as 5.3ks. Another possibility is that there is a chance fly-by due to an unbound KM dwarf within the globular cluster itself. A point against this argument is that GRS 1747-312 is not located at the core of the cluster, where the stellar density is highest. It is possible that stars could be ejected from the core with high velocities due to a variety of possible interactions, including between a single star and a binary, between two binary stars, between a binary and an intermediate-mass black hole (IMBH) at the center \citep{lutzgendorf2012high}. The velocities of such stars have been observed to reach speeds up to 40 km/s \citep{meylan1991two}. Such stars are rare though, and even at these speeds would likely be moving too slowly relative to GRS 1747-312 to produce an eclipse as short as 5.3 ks. While a high $\beta$ value could produce the right eclipse duration, this possibility starts to become increasingly unlikely. Furthermore, there is the additional argument that there may even be two observations of the long 5.3 eclipses, if one considers the \textit{Suzaku} data as well.                                                                                                                                                            

\section{Conclusions}

\subsection{Summary of Model Results}

In Section \ref{sec:models} we use dynamical, geometrical, and timing considerations to assess a variety of physically distinct models to explain the 5.3 ks eclipse. The three overarching models we consider are as follows.


\begin{enumerate}
    \item \textbf{One compact object, with two non-compact objects orbiting it}: The most likely arrangement of this scenario keeps the $2.6$ ks eclipser as the donor, and puts the $5.3$ ks ecliper in a wider circumbinary orbit. The lack of LTTEs$>10$s in the timings of the short eclipse place strong constraints on the possible masses and periods of the third body. Third bodies $>0.6M_\odot$ are likely ruled out, as are any stellar-sized object with period $>1$yr. Kepler's laws combined with assumptions about the XRS geometry suggest that main-sequence stars as third bodies would need to be on periods $<0.2$ yr and have masses $<0.6M_\odot$ in order to be consistent with the LTTE results. Jupiter-scale third bodies on a wide range of periods would be consistent with both the dynamical/geometrical and timing arguments. Altogether, a third body around a NS-donor inner binary would likely need to be an M-dwarf or brown dwarf on a short orbit, or a planet. 
    
    
    \item \textbf{Two compact objects orbiting each other in the same system, sharing a donor star:} The consistency in the duration and timing of the 2.6 ks eclipses make the possibility of a double BH/NS accretor inner binary potentially rejectable. The lack of strong LTTEs also place strong constraints on a BH/main-sequence star inner binary with a NS in a wide outer orbit, requiring it to have a period $<12$ d. 
    
    
    \item \textbf{Two separate compact objects, each with its own star orbiting it:} It is unlikely for there to be two, entirely independent eclipsing LMXBs on the edge of the cluster core separated by just $0.75"$, and they may be gravitationaly bound. From a modeling standpoint though this scenario explains the disappearing 2.6 ks eclipse. 
    
    
\end{enumerate}


\subsection{Discussion}

In this paper we present the discovery of a long, 5.3ks eclipse in a 2021 \textit{Chandra} observation of Terzan 6. This eclipse does not match the ephemeris or duration of the known eclipsing LMXB in Terzan 6, \grs. After a thorough analysis, we verify the long 5.3 ks dip in the \textit{Chandra} data as a genuine eclipse and present the possibility that it is due to a third body in the \grs\ system, which could be either an M-dwarf, brown dwarf, or planet. The analysis included 
a comprehensive study of archival data of \grs\ from \chandra, \textit{RXTE}, \textit{Suzaku}, and \textit{NICER}, with a combined exposure time of $\sim10^6$ seconds. The \chandra\ observations represent the highest-resolution study of \grs\ in its low state. In total, these observations covered $\sim120$ predicted occurences of the short eclipse, many accretion-related dips, and two peculiar drops  in count rate that were off-cycle with the companion star eclipse (one in \textit{Suzaku} data and one in \textit{Chandra} data). 

We also investigate the possibility that the eclipse originates from a separate LMXB. So far the 5.3 ks eclipses have only been observed while \grs\ is in its low state (counting the peculiar dip in the \textit{Suzaku} observation as another example of the  eclipse). There is a $\sim0.75$" difference in the position between the low and high state, which is within \textit{Chandra}'s astrometric uncertainties. The probability of there being two eclipsing LMXBs so close together and offset from the center of the core is low, but it is curious that the X-ray emission (from both resolved and unresolved sources) does peak around \grs. If the long 5.3 ks eclipse does indeed originate from a separate source, then these eclipses would indicate there is a second eclipsing X-ray binary in Terzan 6, which is also of interest.

If there is indeed a third object causing the long eclipse in \grs\, it would be one of the the only known X-ray binaries with a tertiary companion. The system would thus represent an important early step in the study of X-ray triples. High-mass stars are almost always found in binaries, and often in even higher-order combinations. Few X-ray binaries are known to have such architectures though.  The presence of a third object suggests both a unique past and interesting future for the system, as it could affect the orbit of the two inner stars and potentially donate mass. One of the only known analogs to it would be the PSR B1620-26 system, consisting of a Jupiter-sized planet orbiting a pulsar and white dwarf at a distance of $\sim20$ au \citep{1992Natur.355..145W}. \grs\ could represent a version of this system at an earlier point in its evolution, and potentially with an M-dwarf instead of Jupiter sized planet. Another potential analog system would be RX J1744.7-2713, which was recently found to have three candidate companions in wide orbits \citep{Prasow-Émond_2022}.  

In conclusion, while our analysis and models explain some of the unusual features observed, GRS 1747-312 remains an intriguing system for further study. While the true nature of what caused the 5.3 ks eclipse is still uncertain, all of the possible scenarios are highly interesting and some are potentially the first of their kind.

\smallskip
\section{Acknowledgements}

We thank the late, and deeply missed, Tomaso Belloni for his comments and suggestions. RD thanks Max Moe, Ilya Mandel, and
Steinn Sigurdsson for insightful conversations. CP is supported by the Harvard College Research Program. VLK acknowledges support from NASA Contract to the Chandra X-Ray Center NAS8-03060. RD was supported in part by NSF grant AST-2009520. RD was supported in part by a Scholarly Studied Grant from The Smithsonian Institution. RS acknowledges grant number 12073029 from the National Natural Science Foundation of China (NSFC). 

\section{Data Availability}

All data underlying this paper are public, and can be accessed through NASA's High Energy Astrophysics Science Archive Research Center.

\clearpage
\bibliography{sample631}{}

\begin{thebibliography}{}
\expandafter\ifx\csname natexlab\endcsname\relax\def\natexlab#1{#1}\fi
\providecommand{\url}[1]{\href{#1}{#1}}
\providecommand{\dodoi}[1]{doi:~\href{http://doi.org/#1}{\nolinkurl{#1}}}
\providecommand{\doeprint}[1]{\href{http://ascl.net/#1}{\nolinkurl{http://ascl.net/#1}}}
\providecommand{\doarXiv}[1]{\href{https://arxiv.org/abs/#1}{\nolinkurl{https://arxiv.org/abs/#1}}}

\bibitem[{{Arnaud}(1996)}]{arnaud96}
{Arnaud}, K.~A. 1996, in Astronomical Society of the Pacific Conference Series, Vol. 101, Astronomical Data Analysis Software and Systems V, ed. G.~H. {Jacoby} \& J.~{Barnes}, 17

\bibitem[{{Astropy Collaboration} {et~al.}(2022){Astropy Collaboration}, {Price-Whelan}, {Lim}, {Earl}, {Starkman}, {Bradley}, {Shupe}, {Patil}, {Corrales}, {Brasseur}, {N{\"o}the}, {Donath}, {Tollerud}, {Morris}, {Ginsburg}, {Vaher}, {Weaver}, {Tocknell}, {Jamieson}, {van Kerkwijk}, {Robitaille}, {Merry}, {Bachetti}, {G{\"u}nther}, {Aldcroft}, {Alvarado-Montes}, {Archibald}, {B{\'o}di}, {Bapat}, {Barentsen}, {Baz{\'a}n}, {Biswas}, {Boquien}, {Burke}, {Cara}, {Cara}, {Conroy}, {Conseil}, {Craig}, {Cross}, {Cruz}, {D'Eugenio}, {Dencheva}, {Devillepoix}, {Dietrich}, {Eigenbrot}, {Erben}, {Ferreira}, {Foreman-Mackey}, {Fox}, {Freij}, {Garg}, {Geda}, {Glattly}, {Gondhalekar}, {Gordon}, {Grant}, {Greenfield}, {Groener}, {Guest}, {Gurovich}, {Handberg}, {Hart}, {Hatfield-Dodds}, {Homeier}, {Hosseinzadeh}, {Jenness}, {Jones}, {Joseph}, {Kalmbach}, {Karamehmetoglu}, {Ka{\l}uszy{\'n}ski}, {Kelley}, {Kern}, {Kerzendorf}, {Koch}, {Kulumani}, {Lee}, {Ly}, {Ma}, {MacBride}, {Maljaars}, {Muna}, {Murphy}, {Norman},
  {O'Steen}, {Oman}, {Pacifici}, {Pascual}, {Pascual-Granado}, {Patil}, {Perren}, {Pickering}, {Rastogi}, {Roulston}, {Ryan}, {Rykoff}, {Sabater}, {Sakurikar}, {Salgado}, {Sanghi}, {Saunders}, {Savchenko}, {Schwardt}, {Seifert-Eckert}, {Shih}, {Jain}, {Shukla}, {Sick}, {Simpson}, {Singanamalla}, {Singer}, {Singhal}, {Sinha}, {Sip{\H{o}}cz}, {Spitler}, {Stansby}, {Streicher}, {{\v{S}}umak}, {Swinbank}, {Taranu}, {Tewary}, {Tremblay}, {de Val-Borro}, {Van Kooten}, {Vasovi{\'c}}, {Verma}, {de Miranda Cardoso}, {Williams}, {Wilson}, {Winkel}, {Wood-Vasey}, {Xue}, {Yoachim}, {Zhang}, {Zonca}, \& {Astropy Project Contributors}}]{2022ApJ...935..167A}
{Astropy Collaboration}, {Price-Whelan}, A.~M., {Lim}, P.~L., {et~al.} 2022, \apj, 935, 167, \dodoi{10.3847/1538-4357/ac7c74}

\bibitem[{{Bahramian} \& {Degenaar}(2023)}]{2023hxga.book..120B}
{Bahramian}, A., \& {Degenaar}, N. 2023, in Handbook of X-ray and Gamma-ray Astrophysics. Edited by Cosimo Bambi and Andrea Santangelo, 120, \dodoi{10.1007/978-981-16-4544-0_94-1}

\bibitem[{{Barbuy} {et~al.}(1997){Barbuy}, {Ortolani}, \& {Bica}}]{1997A&AS..122..483B}
{Barbuy}, B., {Ortolani}, S., \& {Bica}, E. 1997, \aaps, 122, 483, \dodoi{10.1051/aas:1997148}

\bibitem[{{Borkovits} {et~al.}(2022){Borkovits}, {Mitnyan}, {Rappaport}, {Pribulla}, {Powell}, {Kostov}, {B{\'\i}r{\'o}}, {Cs{\'a}nyi}, {Garai}, {Gary}, {Kaye}, {Kom{\v{z}}{\'\i}k}, {Terentev}, {Omohundro}, {Gagliano}, {Jacobs}, {Kristiansen}, {LaCourse}, {Schwengeler}, {Czavalinga}, {Seli}, {Huang}, {P{\'a}l}, {Vanderburg}, {Rodriguez}, \& {Stevens}}]{2022MNRAS.510.1352B}
{Borkovits}, T., {Mitnyan}, T., {Rappaport}, S.~A., {et~al.} 2022, \mnras, 510, 1352, \dodoi{10.1093/mnras/stab3397}

\bibitem[{{Cash}(1979)}]{cash79}
{Cash}, W. 1979, \apj, 228, 939, \dodoi{10.1086/156922}

\bibitem[{{Chenevez} {et~al.}(2009){Chenevez}, {Kuulkers}, {Beckmann}, {Bird}, {Brandt}, {Domingo}, {Ebisawa}, {Jonker}, {Kretschmar}, {Markwardt}, {Oosterbroek}, {Paizis}, {Risquez}, {Sanchez-Fernandez}, {Shaw}, \& {Wijnands}}]{2009ATel.2235....1C}
{Chenevez}, J., {Kuulkers}, E., {Beckmann}, V., {et~al.} 2009, The Astronomer's Telegram, 2235, 1

\bibitem[{{Church} \& {Ba{\l}uci{\'n}ska-Church}(2004)}]{church04}
{Church}, M.~J., \& {Ba{\l}uci{\'n}ska-Church}, M. 2004, \mnras, 348, 955, \dodoi{10.1111/j.1365-2966.2004.07162.x}

\bibitem[{Di~Stefano(2020)}]{di2020dynamical}
Di~Stefano, R. 2020, Monthly Notices of the Royal Astronomical Society, 491, 495

\bibitem[{{Eubanks} {et~al.}(2021){Eubanks}, {Hein}, {Lingam}, {Hibberd}, {Fries}, {Perakis}, {Kennedy}, {Blase}, \& {Schneider}}]{2021arXiv210303289E}
{Eubanks}, T.~M., {Hein}, A.~M., {Lingam}, M., {et~al.} 2021, arXiv e-prints, arXiv:2103.03289, \dodoi{10.48550/arXiv.2103.03289}

\bibitem[{{Fiocchi} {et~al.}(2007){Fiocchi}, {Bazzano}, {Ubertini}, \& {Zdziarski}}]{fiocchi07}
{Fiocchi}, M., {Bazzano}, A., {Ubertini}, P., \& {Zdziarski}, A.~A. 2007, \apj, 657, 448, \dodoi{10.1086/510573}

\bibitem[{{Fruscione} {et~al.}(2006){Fruscione}, {McDowell}, {Allen}, {Brickhouse}, {Burke}, {Davis}, {Durham}, {Elvis}, {Galle}, {Harris}, {Huenemoerder}, {Houck}, {Ishibashi}, {Karovska}, {Nicastro}, {Noble}, {Nowak}, {Primini}, {Siemiginowska}, {Smith}, \& {Wise}}]{2006SPIE.6270E..1VF}
{Fruscione}, A., {McDowell}, J.~C., {Allen}, G.~E., {et~al.} 2006, in Society of Photo-Optical Instrumentation Engineers (SPIE) Conference Series, Vol. 6270, Society of Photo-Optical Instrumentation Engineers (SPIE) Conference Series, ed. D.~R. {Silva} \& R.~E. {Doxsey}, 62701V, \dodoi{10.1117/12.671760}

\bibitem[{Ibragimov {et~al.}(2011)Ibragimov, Kajava, \& Poutanen}]{Ibragimov_2011}
Ibragimov, A., Kajava, J. J.~E., \& Poutanen, J. 2011, Monthly Notices of the Royal Astronomical Society, 415, 1864, \dodoi{10.1111/j.1365-2966.2011.18836.x}

\bibitem[{{in't Zand} {et~al.}(2003){in't Zand}, {Hulleman}, {Markwardt}, {M{\'e}ndez}, {Kuulkers}, {Cornelisse}, {Heise}, {Strohmayer}, \& {Verbunt}}]{2003A&A...406..233I}
{in't Zand}, J.~J.~M., {Hulleman}, F., {Markwardt}, C.~B., {et~al.} 2003, \aap, 406, 233, \dodoi{10.1051/0004-6361:20030681}

\bibitem[{{Kuulkers} {et~al.}(2003){Kuulkers}, {den Hartog}, {in't Zand}, {Verbunt}, {Harris}, \& {Cocchi}}]{2003A&A...399..663K}
{Kuulkers}, E., {den Hartog}, P.~R., {in't Zand}, J.~J.~M., {et~al.} 2003, \aap, 399, 663, \dodoi{10.1051/0004-6361:20021781}

\bibitem[{{Lellouch} {et~al.}(2022){Lellouch}, {Moreno}, {Bockel{\'e}e-Morvan}, {Biver}, \& {Santos-Sanz}}]{2022A&A...659L...1L}
{Lellouch}, E., {Moreno}, R., {Bockel{\'e}e-Morvan}, D., {Biver}, N., \& {Santos-Sanz}, P. 2022, \aap, 659, L1, \dodoi{10.1051/0004-6361/202243090}

\bibitem[{L{\"u}tzgendorf {et~al.}(2012)L{\"u}tzgendorf, Gualandris, Kissler-Patig, Gebhardt, Baumgardt, Noyola, Kruijssen, Jalali, de~Zeeuw, \& Neumayer}]{lutzgendorf2012high}
L{\"u}tzgendorf, N., Gualandris, A., Kissler-Patig, M., {et~al.} 2012, Astronomy \& Astrophysics, 543, A82

\bibitem[{Mardling \& Aarseth(1999)}]{mardling1999dynamics}
Mardling, R., \& Aarseth, S. 1999, The Dynamics of Small Bodies in the Solar System: A Major Key to Solar Systems Studies, 385

\bibitem[{Meylan {et~al.}(1991)Meylan, Dubath, \& Mayor}]{meylan1991two}
Meylan, G., Dubath, P., \& Mayor, M. 1991, Astrophysical Journal, Part 1 (ISSN 0004-637X), vol. 383, Dec. 20, 1991, p. 587-593. Research supported by SNSF., 383, 587

\bibitem[{{Mihalas} \& {Binney}(1981)}]{1981gask.book.....M}
{Mihalas}, D., \& {Binney}, J. 1981, {Galactic astronomy. Structure and kinematics}

\bibitem[{{Nasa High Energy Astrophysics Science Archive Research Center (Heasarc)}(2014)}]{heasoft14}
{Nasa High Energy Astrophysics Science Archive Research Center (Heasarc)}. 2014, {HEAsoft: Unified Release of FTOOLS and XANADU}, Astrophysics Source Code Library, record ascl:1408.004.
\newblock \doeprint{1408.004}

\bibitem[{{Panurach} {et~al.}(2021){Panurach}, {Strader}, {Bahramian}, {Chomiuk}, {Miller-Jones}, {Heinke}, {Maccarone}, {Shishkovsky}, {Sivakoff}, {Tremou}, {Tudor}, \& {Urquhart}}]{2021ApJ...923...88P}
{Panurach}, T., {Strader}, J., {Bahramian}, A., {et~al.} 2021, \apj, 923, 88, \dodoi{10.3847/1538-4357/ac2c6b}

\bibitem[{{Park} {et~al.}(2006){Park}, {Kashyap}, {Siemiginowska}, {van Dyk}, {Zezas}, {Heinke}, \& {Wargelin}}]{2006ApJ...652..610P}
{Park}, T., {Kashyap}, V.~L., {Siemiginowska}, A., {et~al.} 2006, \apj, 652, 610, \dodoi{10.1086/507406}

\bibitem[{{Pavlinsky} {et~al.}(1994){Pavlinsky}, {Grebenev}, \& {Sunyaev}}]{1994ApJ...425..110P}
{Pavlinsky}, M.~N., {Grebenev}, S.~A., \& {Sunyaev}, R.~A. 1994, \apj, 425, 110, \dodoi{10.1086/173967}

\bibitem[{Prasow-Émond {et~al.}(2022)Prasow-Émond, Hlavacek-Larrondo, Fogarty, Rameau, Guité, Mawet, Gandhi, Rao, Steiner, Artigau, Lafrenière, Fabian, Walton, Weiss, Doyon, 任彬, Rhea, Bégin, Vigneron, \& Naud}]{Prasow-Émond_2022}
Prasow-Émond, M., Hlavacek-Larrondo, J., Fogarty, K., {et~al.} 2022, The Astronomical Journal, 164, 7, \dodoi{10.3847/1538-3881/ac6d57}

\bibitem[{{Predehl} {et~al.}(1991){Predehl}, {Hasinger}, \& {Verbunt}}]{1991A&A...246L..21P}
{Predehl}, P., {Hasinger}, G., \& {Verbunt}, F. 1991, \aap, 246, L21

\bibitem[{{Remillard} {et~al.}(2022){Remillard}, {Loewenstein}, {Steiner}, {Prigozhin}, {LaMarr}, {Enoto}, {Gendreau}, {Arzoumanian}, {Markwardt}, {Basak}, {Stevens}, {Ray}, {Altamirano}, \& {Buisson}}]{2022AJ....163..130R}
{Remillard}, R.~A., {Loewenstein}, M., {Steiner}, J.~F., {et~al.} 2022, \aj, 163, 130, \dodoi{10.3847/1538-3881/ac4ae6}

\bibitem[{{Saji} {et~al.}(2016){Saji}, {Mori}, {Matsumoto}, {Dotani}, {Iwai}, {Maeda}, {Mitsuishi}, {Ozaki}, \& {Tawara}}]{2016PASJ...68S..15S}
{Saji}, S., {Mori}, H., {Matsumoto}, H., {et~al.} 2016, \pasj, 68, S15, \dodoi{10.1093/pasj/psw011}

\bibitem[{{Scargle}(1982)}]{1982ApJ...263..835S}
{Scargle}, J.~D. 1982, \apj, 263, 835, \dodoi{10.1086/160554}

\bibitem[{{Scargle}(1998)}]{1998ApJ...504..405S}
---. 1998, \apj, 504, 405, \dodoi{10.1086/306064}

\bibitem[{{Scargle} {et~al.}(2013{\natexlab{a}}){Scargle}, {Norris}, {Jackson}, \& {Chiang}}]{2013ApJ...764..167S}
{Scargle}, J.~D., {Norris}, J.~P., {Jackson}, B., \& {Chiang}, J. 2013{\natexlab{a}}, \apj, 764, 167, \dodoi{10.1088/0004-637X/764/2/167}

\bibitem[{{Scargle} {et~al.}(2013{\natexlab{b}}){Scargle}, {Norris}, {Jackson}, \& {Chiang}}]{2013arXiv1304.2818S}
---. 2013{\natexlab{b}}, arXiv e-prints, arXiv:1304.2818, \dodoi{10.48550/arXiv.1304.2818}

\bibitem[{{Trager} {et~al.}(1995){Trager}, {King}, \& {Djorgovski}}]{1995AJ....109..218T}
{Trager}, S.~C., {King}, I.~R., \& {Djorgovski}, S. 1995, \aj, 109, 218, \dodoi{10.1086/117268}

\bibitem[{{Vats} {et~al.}(2018){Vats}, {Wijnands}, {Parikh}, {Ootes}, {Degenaar}, \& {Page}}]{vats18}
{Vats}, S., {Wijnands}, R., {Parikh}, A.~S., {et~al.} 2018, \mnras, 477, 2494, \dodoi{10.1093/mnras/sty733}

\bibitem[{{White} \& {Swank}(1982)}]{1982ApJ...253L..61W}
{White}, N.~E., \& {Swank}, J.~H. 1982, \apjl, 253, L61, \dodoi{10.1086/183737}

\bibitem[{{Wolff} {et~al.}(2021){Wolff}, {Guillot}, {Bogdanov}, {Ray}, {Kerr}, {Arzoumanian}, {Gendreau}, {Miller}, {Dittmann}, {Ho}, {Guillemot}, {Cognard}, {Theureau}, \& {Wood}}]{2021ApJ...918L..26W}
{Wolff}, M.~T., {Guillot}, S., {Bogdanov}, S., {et~al.} 2021, \apjl, 918, L26, \dodoi{10.3847/2041-8213/ac158e}

\bibitem[{{Wolszczan} \& {Frail}(1992)}]{1992Natur.355..145W}
{Wolszczan}, A., \& {Frail}, D.~A. 1992, \nat, 355, 145, \dodoi{10.1038/355145a0}

\end{thebibliography}
\bibliographystyle{aasjournal}

\onecolumngrid
\vspace{2cm}

\clearpage

\appendix


\section{Derivation of Ingresses \& Egresses in \chandra\ Light Curves}
\label{chap:ingress_egress}

Unlike the \textit{RXTE} and \textit{NICER} data, \chandra\ count rates are small, and the time resolution is limited by the CCD frame readout time of $\tau_{\rm \rm CCD}\approx$3.24~s.  Therefore, in order to estimate the durations of ingress and egress, we carry out fits to the light curves binned at $33\cdot\tau_{\rm \rm CCD}\approx10^2$~s using a simplified piecewise linear model representing an eclipse shape.  The model comprises a constant baseline rate, with the ingress and egresss described by linearly descending and ascending count rates respectively, flanking a constant minimum rate.  The rate at time $t$,
\begin{eqnarray}
r(t) =& r_{\rm \rm baseline} & ~~~~{\rm for}~ t_{\rm \rm min} \leq t < t_{\rm \rm begin} \nonumber \\
=& r_{\rm \rm dip} + \frac{r_{\rm \rm baseline}-r_{\rm \rm dip}}{t_{\rm \rm begin}-t_{\rm \rm ingres}} \cdot (t-t_{\rm \rm ingress}) & ~~~~{\rm for}~ t_{\rm \rm begin} \leq t \leq t_{\rm \rm ingress} \nonumber \\
=& r_{\rm \rm dip} & ~~~~{\rm for}~ t_{\rm \rm ingress} < t < t_{\rm \rm egress} \nonumber \\
=& r_{\rm \rm baseline} + \frac{r_{\rm \rm dip}-r_{\rm \rm baseline}}{t_{\rm \rm egress}-t_{\rm \rm end}} \cdot (t-t_{\rm \rm end}) & ~~~~{\rm for}~ t_{\rm \rm egress} \leq t \leq t_{\rm \rm end} \nonumber \\
=& r_{\rm \rm baseline} & ~~~~{\rm for}~ t_{\rm \rm end} < t \leq t_{\rm \rm max} \,,
\end{eqnarray}
where $t_{\rm \rm min} < t_{\rm \rm begin} < t_{\rm \rm ingress} < t_{\rm \rm egress} < t_{\rm \rm end} < t_{\rm \rm max}$ are the times when the light curve data begin, the ingress begins, the ingress ends and the maximum eclipse begins, the maximum eclipse ends and the egress begins, the egress ends, and the light curve data ends.  We typically use data extending $\approx$2-3~ks below and above $t_{\rm \rm begin,end}$ to define the baseline; the estimates are robust to the precise choices of these extensions, provided that $t_{\rm \rm min,max}$ exclude periods of obvious variability (e.g., the flare at $+5$~ks in ObsID 23444 or the spike at $+10$~ks in ObsID 23443, see Figure~\ref{fig:Chandra_lc}).  We fit this model, with parameters $\mathbf{\theta}=\{r_{\rm \rm baseline},r_{\rm \rm dip},t_{\rm \rm begin},t_{\rm \rm ingress},t_{\rm \rm egress},t_{\rm \rm end}\}$ to the several of the light curves in Figure~\ref{fig:Chandra_lc} using a Poisson likelihood with a Metropolis Markov Chain Monte Carlo procedure.  Where only an ingress [egress] is visible, we exclude $\{t_{\rm \rm egress},t_{\rm \rm end}\}$ [$\{t_{\rm \rm begin},t_{\rm \rm ingress}\}$] from $\mathbf{\theta}$.  We report the duration of the eclipse as the difference $\Delta\tau=t_{\rm \rm end}-t_{\rm \rm begin}$, and the durations of ingress and egress as $\tau_{\rm \rm ingress}=t_{\rm \rm ingress}-t_{\rm \rm begin}$ and $\tau_{\rm \rm egress}=t_{\rm \rm end}-t_{\rm \rm egress}$ in Table~\ref{tab:dips}.  We obtain $10,000$ draws from the posterior distribution and discard the first $1,000$ as burn-in, and use the rest to compute uncertainties as the 68\% highest posterior density (HPD) intervals, which are appropriate for highly skewed distributions; the ingresses and egresses are effectively unresolved in \chandra\ light curves, but the estimates are consistent with values derived for \textit{RXTE} and \textit{NICER}.


\clearpage

\section{\chandra\ Source Position Analysis Methods}\label{chap:astrometry}

\subsection{Changes in Source Position During vs Outside of the Short Eclipse}

As an initial check for shifts in centroid, we generate separate event lists outside and inside eclipse for observations in the high state, and look for changes in the centroid. For observations outside eclipse, we estimate the centroid by eye using a circular region which borders the edges of the pile-up. For observations inside eclipse we use a $2$" circular region. We find that the centroid location between in eclipse and out of eclipse differed by $(\delta\rm{RA},\delta\rm{DEC})$ of ($.75"\pm.15",.08"\pm.07"$) and ($.2"\pm.16",.05"\pm.19"$) for ObsIDs 23444 and 21218, respectively. 

\begin{figure}[h]
    \centering
    \includegraphics[width=0.48\textwidth]{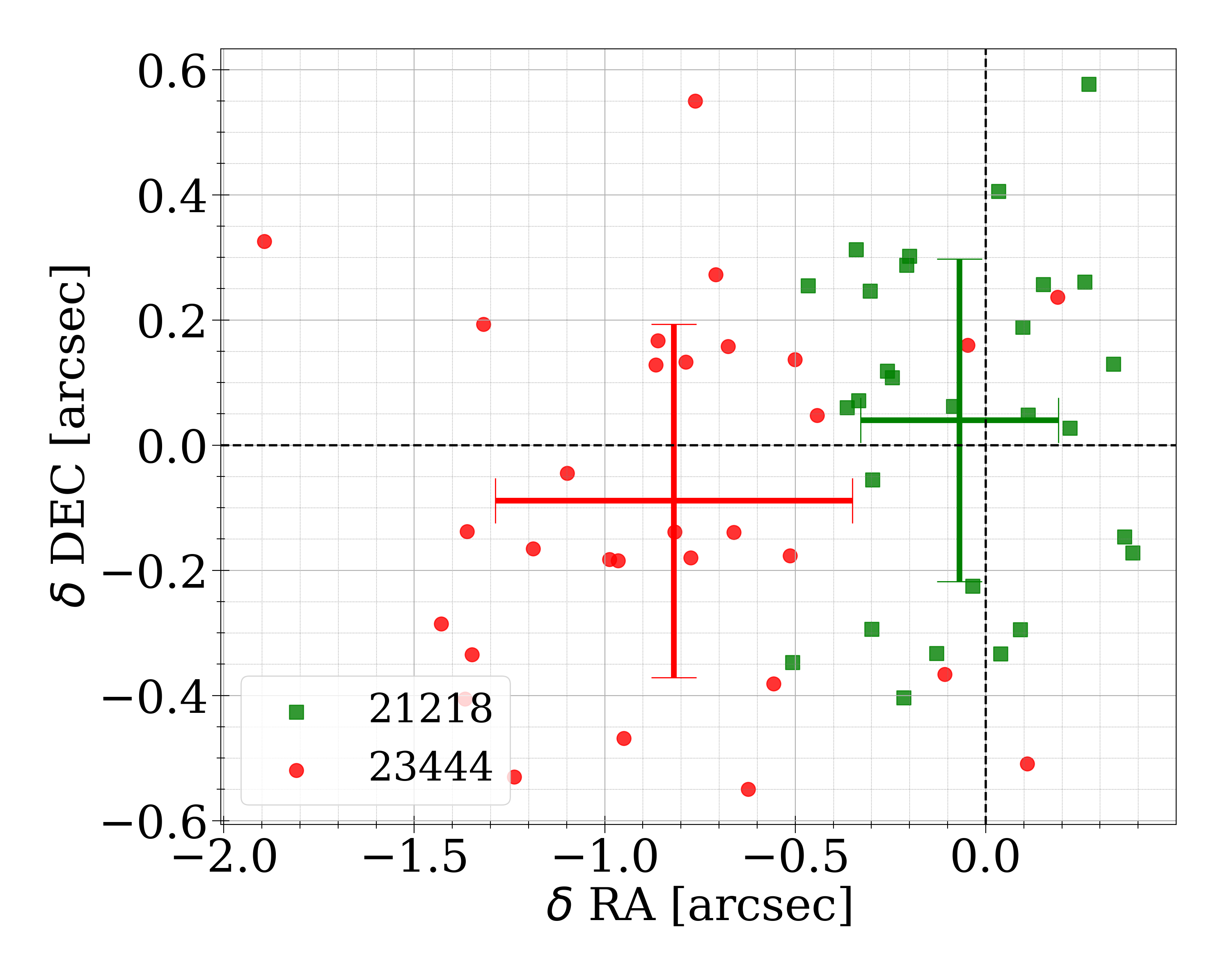}
    \caption{
    Illustrating the variations in the positions of \grs\ in the high state relative to that during the short eclipse.  The data are grouped into the same number of events as are found during the eclipse (294 and 94 for ObsIDs 21218 and 23444 respectively), and the difference between the computed centroids and the centroid position during the eclipse are plotted as dots (green and red for ObsIDs 21218 and 23444 respectively).  The standard deviation of the points along each coordinate axis are shown as line segments, and represent the uncertainty on the centroid locations.  There is no evidence for a significant bias in the centroid during the short eclipse.
    }
    \label{fig:centroids}
\end{figure}

To see if these changes in centroid were the product of expected noise due to low count statistics during eclipse, we apply the same centroid method to multiple time intervals with equal number of events outside of eclipse. We choose 2" regions for the counts inside eclipse, to avoid incorporating too much background. We choose annuli with radii (1”,15”) and (2”,15”) for centroid measurements out of eclipse, for observations 23444 and 21218 respectively. These annuli are centered on the positions listed in Table \ref{tab:obslog}. There are $\sim100$ counts in the 2" centroid centered on the residual emission during the 2.6 ks eclipse in observation 23444, and $\sim300$ for observation 21218. This resulted in $30$ distinct count groups for ObsID 23444 and $28$ for ObsID 21218 (see Figure \ref{fig:centroids}).

We find that there is a spread of a few arcseconds in centroid locations for both RA and DEC. For ObsID 21218, the in-eclipse centroid location falls within the $1\sigma$ error bounds of the centroid location outside eclipse. For ObsID 23444, the in-eclipse centroid location falls outside the $1\sigma$ error bounds, representing a slightly less than $2\sigma$ difference. In neither case are the in-eclipse centroid location outliers. We conclude that there is no strong evidence for a shift in source position during eclipse versus out of eclipse for ObsID 21218, and that there is a $\sim2\sigma$ position difference for ObsID 23444. 

We also correct for astrometry differences between \textit{Chandra} observations by using nearby point-sources. For each of these sources, 5" regions were drawn and centered on the source using \textbf{ds9}'s \texttt{centroid} function. The locations of all the counts in the region were extracted, and those that were $>3\sigma$ away from the centroid were discarded. The new centroid was calculated from the remaining counts. Not all sources were within the field of view of each observation. For the soure representing \grs, 1.5" regions were also used in the low-state observations. For the high state, annuli with radii (1",15") and (2",15") were used in observations 23444 and 21218, respectively, to avoid the pile-up. These were centered on the sources by hand, and the readout streak was interpolated to ensure the annulus was centered on it. 2" regions were used for the counts received during eclipse in the high states.

\begin{figure}[t]
    \centering
     \includegraphics[width=.46\textwidth]{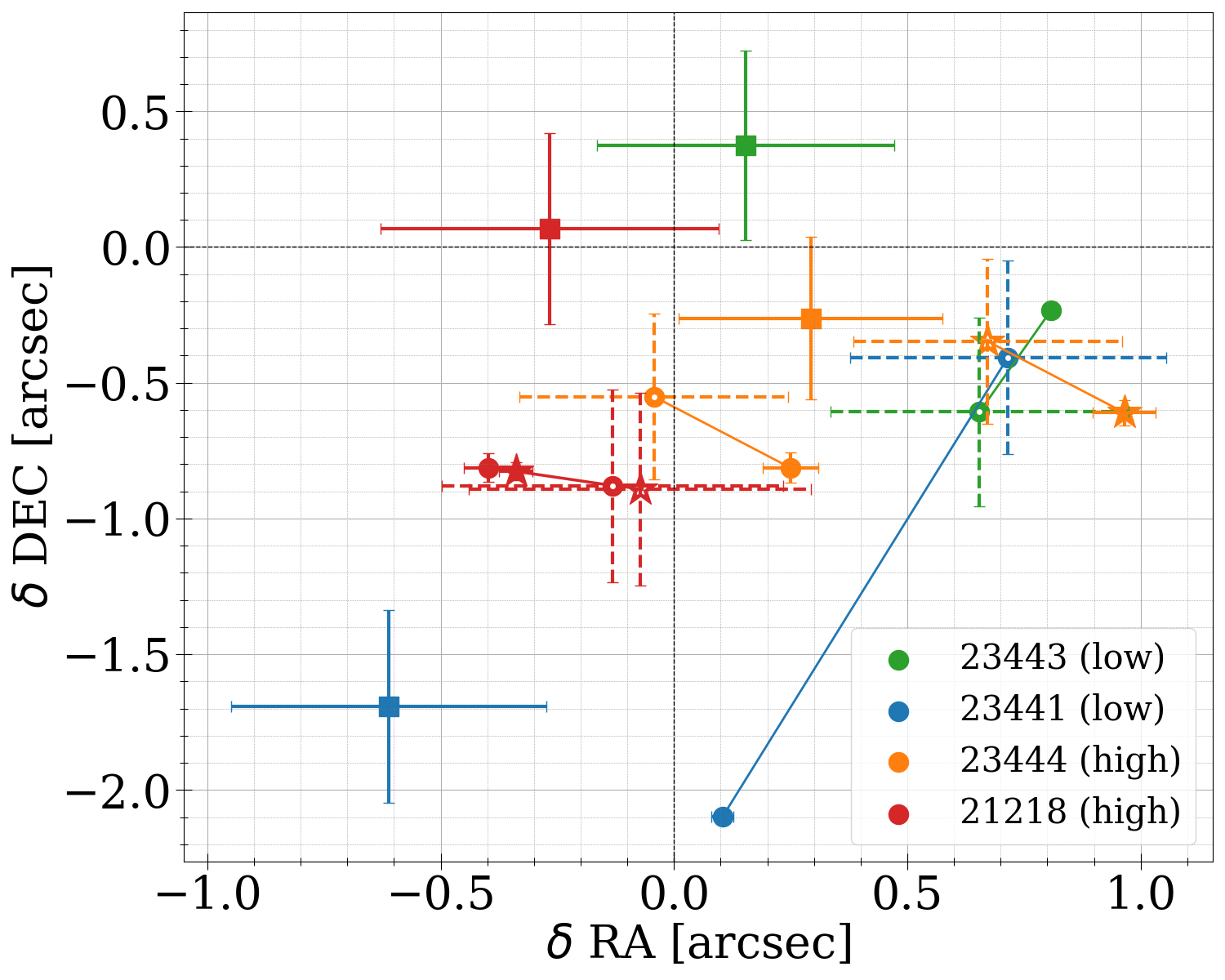}
    \caption{Estimated positional offsets and corrections of \grs\ derived from nearby point sources.  Filled circles represent locations of \grs\ relative to its radio position. Filled boxes are the average offsets determined from other objects in the field. The corresponding correction is then applied to the putative locations of \grs\ (line segments ending in hollow circles with dashed lines indicating error bars). Stars represent the position of \grs\ during the short 2.6 ks eclipse, with hollow stars representing the position after astrometric correction. 
    }
    \label{fig:astrometry}%
\end{figure}

\begin{table}[h]
\caption{Location of Point Sources Used in \chandra\ Astrometry Corrections}
\label{tab:point_sources}
\begin{center}
\begin{tabular}{lccccc}
\toprule
Source & 21218 (high) & 23443 (low) & 23444 (high) & 23441 (low) & Avg Counts \\
\hline
\midrule
\grs\ & (267.69515, -31.27482) & (267.69552, -31.27464) & (267.69536, -31.27482) & (267.69531, -31.27517) & $\sim4500$ \\

1 & (267.66242, -31.34289) & (267.6626, -31.34289) & (267.66274, -31.34306) & (267.66246, -31.34347) & 109.3 \\
2 & (267.64695, -31.30706) & (267.64706, -31.30687) & (267.64711, -31.307) & (267.64685, -31.30745) & 106.0 \\
3 & (267.63601, -31.28597) & (267.63612, -31.28589) & (267.63624, -31.28597) & (267.63588, -31.28636) & 46.5 \\
4 & (267.63087, -31.20289) & (267.63101, -31.20284) & -- & (267.63066, -31.20344) & 137.3 \\
5 & (267.7534, -31.21336) & (267.75365, -31.21298) & (267.75342, -31.21353) & (267.75305, -31.21397) & 76.3 \\
6 & -- & (267.68614, -31.29895) & (267.68627, -31.29915) & -- & 39.5 \\
7 & -- & (267.66586, -31.28995) & (267.66603, -31.29018) & -- & 30.5 \\
8 & -- & (267.75182, -31.27525) & (267.75182, -31.27533) & -- & 33.0 \\
9 & -- & (267.65326, -31.30655) & (267.65352, -31.30671) & (267.65315, -31.30713) & 31.7 \\
10 & (267.6031, -31.31144) & -- & (267.60324, -31.31148) & -- & 103.5 \\
11 & -- & (267.7391, -31.33173) & (267.73916, -31.33187) & (267.73872, -31.33235) & 23.0 \\
12 & (267.65593, -31.23891) & (267.6564, -31.23897) & (267.65627, -31.23894) & (267.65608, -31.23947) & 27.0 \\
\bottomrule
\hline
\multicolumn{6}{l}{$*$: {Average number of counts between all observations the source was present in.}} \\

\end{tabular}

\end{center}
\end{table}


\clearpage

\section{Transit Timing Variations}\label{chap:etvs}

\begin{table}[!htb]
    \begin{minipage}{.5\linewidth}
      \caption{Egress Timing Variations}
      \centering
\begin{tabular}{l|c|c|c|c}

\toprule
MJD & $n$ & $\sigma_{\rm lower}$ & $\sigma_{\rm upper}$ & $O-C \quad \Delta t$ \\
\hline
\midrule
51594.56691 & -916 & 1.9s & 1.8s & -4.6s \\
51600.74667 & -904 & 4.1s & 4.1s & 3.3s \\
51602.80659 & -900 & 3.4s & 3.3s & 3.21s \\
51615.16612 & -876 & 3.7s & 3.7s & 1.02s \\
51768.11526 & -579 & 1.3s & 1.4s & -4.45s \\
52066.28884 & 0 & 3.3s & 3.3s & 0.97s \\
52471.57832 & 787 & 1.5s & 1.5s & -3.46s \\
52719.79881 & 1269 & 2.1s & 2.0s & 5.93s \\
52728.03849 & 1285 & 3.1s & 3.2s & 4.62s \\
53084.40484 & 1977 & 2.3s & 2.3s & 6.7s \\
53090.58461 & 1989 & 2.1s & 2.1s & -1.09s \\
53093.67449 & 1995 & 2.3s & 2.4s & 7.53s \\
53096.24939 & 2000 & 3.6s & 3.7s & 6.2s \\
53099.85425 & 2007 & 5.1s & 5.3s & 9.37s \\
53233.74912 & 2267 & 2.2s & 2.3s & 3.13s \\
53240.44387 & 2280 & 1.7s & 1.7s & 1.62s \\
53247.13861 & 2293 & 3.0s & 3.0s & 1.79s \\
53251.25845 & 2301 & 5.0s & 5.0s & 7.65s \\
53484.02954 & 2753 & 1.8s & 1.8s & 6.07s \\
53487.63440 & 2760 & 1.6s & 1.6s & 7.09s \\
53497.93400 & 2780 & 3.0s & 3.0s & 4.15s \\
53938.75712 & 3636 & 2.9s & 2.8s & 8.85s \\
53947.51178 & 3653 & 3.5s & 3.5s & 8.32s \\
54149.89903 & 4046 & 6.8s & 6.8s & 8.66s \\
54546.94882 & 4817 & 3.2s & 3.1s & 7.82s \\
54553.64357 & 4830 & 1.7s & 1.7s & 11.08s \\
54559.30835 & 4841 & 1.9s & 1.9s & 6.36s \\
54565.48811 & 4853 & 3.1s & 3.1s & 8.07s \\
54886.32083 & 5476 & 3.0s & 2.9s & 16.1s \\
54890.44067 & 5484 & 5.3s & 5.1s & 10.47s \\
\bottomrule
\end{tabular}
    \end{minipage}%
    \begin{minipage}{.5\linewidth}
      \centering
        \caption{Ingress Timing Variations}
\begin{tabular}{l|c|c|c|c}
\toprule
MJD & $n$ & $\sigma_{\rm lower}$ & $\sigma_{\rm upper}$ & $O-C \quad \Delta t$ \\
\hline
\midrule
51448.79742 & -1199 & 1.3s & 1.2s & 0.28s \\
51599.68665 & -906 & 2.4s & 2.3s & -10.07s \\
51601.74657 & -902 & 2.2s & 2.2s & -3.89s \\
51767.05525 & -581 & 1.1s & 1.1s & -2.99s \\
52063.68391 & -5 & 1.9s & 1.9s & -2.66s \\
52066.25882 & 0 & 1.6s & 1.5s & -0.41s \\
52077.07340 & 21 & 1.1s & 1.1s & 0.93s \\
52347.43807 & 546 & 1.1s & 1.0s & -2.74s \\
52720.79879 & 1271 & 1.4s & 1.4s & -4.0s \\
52727.49353 & 1284 & 2.0s & 2.0s & -4.4s \\
53084.88987 & 1978 & 1.9s & 1.9s & -8.32s \\
53092.61458 & 1993 & 1.6s & 1.5s & -5.6s \\
53095.70446 & 1999 & 1.7s & 1.7s & -4.37s \\
53099.30932 & 2006 & 2.7s & 2.7s & -6.36s \\
53101.88422 & 2011 & 2.0s & 2.0s & -4.25s \\
53235.36412 & 2270 & 1.9s & 1.9s & -8.8s \\
53238.86898 & 2277 & 1.8s & 1.8s & -9.05s \\
53247.10867 & 2293 & 2.6s & 2.5s & -7.24s \\
53253.28843 & 2305 & 1.0s & 1.1s & 0.85s \\
53486.57451 & 2758 & 1.0s & 1.0s & -3.48s \\
53490.69436 & 2766 & 0.8s & 0.8s & -4.15s \\
53494.29922 & 2773 & 1.5s & 1.4s & -5.0s \\
53496.35914 & 2777 & 1.8s & 1.9s & -13.24s \\
53938.72723 & 3636 & 1.6s & 1.7s & -5.75s \\
53944.90699 & 3648 & 1.6s & 1.5s & -2.48s \\
54147.80923 & 4042 & 4.4s & 4.5s & -5.63s \\
54416.11397 & 4563 & 1.8s & 1.7s & -4.94s \\
54547.94893 & 4819 & 1.5s & 1.4s & -7.78s \\
54553.61372 & 4830 & 1.5s & 1.5s & -7.33s \\
54559.79348 & 4842 & 2.4s & 2.5s & -10.53s \\
54565.45826 & 4853 & 3.2s & 3.2s & -9.89s \\
54883.20112 & 5470 & 1.1s & 1.2s & -6.04s \\
54884.23108 & 5472 & 1.9s & 2.0s & -5.46s \\
54888.35092 & 5480 & 2.0s & 2.0s & -5.58s \\
54890.92582 & 5485 & 6.2s & 6.1s & -0.46s \\
\bottomrule
\end{tabular}
    \end{minipage} 
\end{table}

\begin{figure}[H]
    \centering
    \includegraphics[width=.6\textwidth]{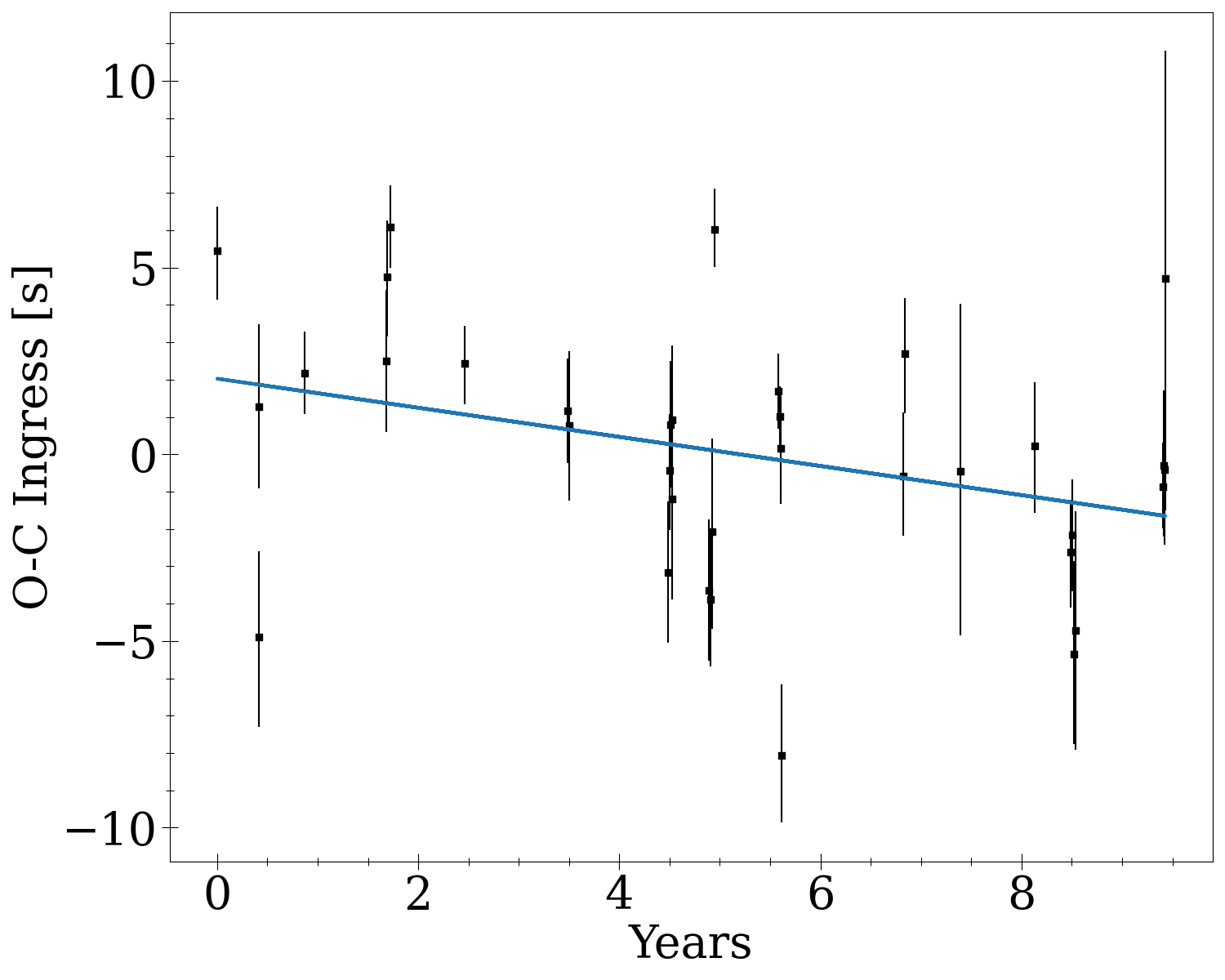}
    \caption{Transit timing variations for ingresses in \textit{RXTE} data.}
    \label{fig:ingresses_RXTE_Fig}
\end{figure}

\clearpage

\section{Methods for Hardness Ratio Analysis}\label{chap:more_HR}

We estimate the significance of the difference in $HR$ during the dip compared to the intervals just before and after the dip using a metric based on the Kullback-Leibler divergence ($D_{\rm KL}(p||q)$). This is a measure of how much one probability density function $q(x)$ overlaps another probability density function $p(x)$, and is calculated as the expectation of the log of the ratio of $q(x)$ to $p(x)$
\begin{equation}
    D_{\rm KL}(p||q) \equiv -E_p\left[\log\frac{q}{p}\right] = -\int~dx~p(x)~\log\frac{q(x)}{p(x)} \,,
    \label{eq:D_KL}
\end{equation}
where the integral is carried out over the domain of coverage of $p(x)$, and represents the entropy of $p(x)$ relative to $q(x)$.  That is, it describes how much excess information is required if one were to describe $q(x)$ in terms of $p(x)$.  When the overlap is maximum, i.e., $q(x)=p(x)$, the ``distance'' between the two distributions, $D_{\rm KL}(p||q=p)=0$, and when there is no overlap, $D_{\rm KL}(p||q)=+\infty$.  Notice that the ``distance'' computed in this manner is directional, i.e., $D_{\rm KL}(p||q) \ne D_{\rm KL}(q||p)$.  However, in all cases $D_{\rm KL}\ge{0}$.  We compute $D_{\rm KL}$ via Equation~\ref{eq:D_KL}
for the distance of the posterior density of $HR$ computed over the duration of the dip $p=p(HR|{\rm dip})$ (the green shaded intervals and curves in Figure~\ref{fig:statistics}~(top,~middle)) to the posterior densities of $HR$ compute over the left (yellow shaded intervals and curves in Figure~\ref{fig:statistics}) and right flanks (green shaded intervals and curves in Figure~\ref{fig:statistics}) around the dip. We do this on a uniform and identical grid over a range of (-2,2). We then determine how deviant the observed $D_{\rm KL}({\rm dip}||{\rm left})$ and $D_{\rm KL}({\rm dip}||{\rm right})$ are relative to statistical fluctuations expected for $p(HR|dip)$, by carrying out 500 bootstraps with replacement for the events within the dip and computing the $p$-value of the chance that the observed $D_{\rm KL}$ could be attributed to a chance fluctuation.  The last row of figures in Figure~\ref{fig:statistics} shows where the estimated $D_{\rm KL}$ fall within the bootstrapped range (see Table~\ref{table:HR_p_values}).

\begin{figure}[H]
    \centering
    \includegraphics[width=.6\textwidth]{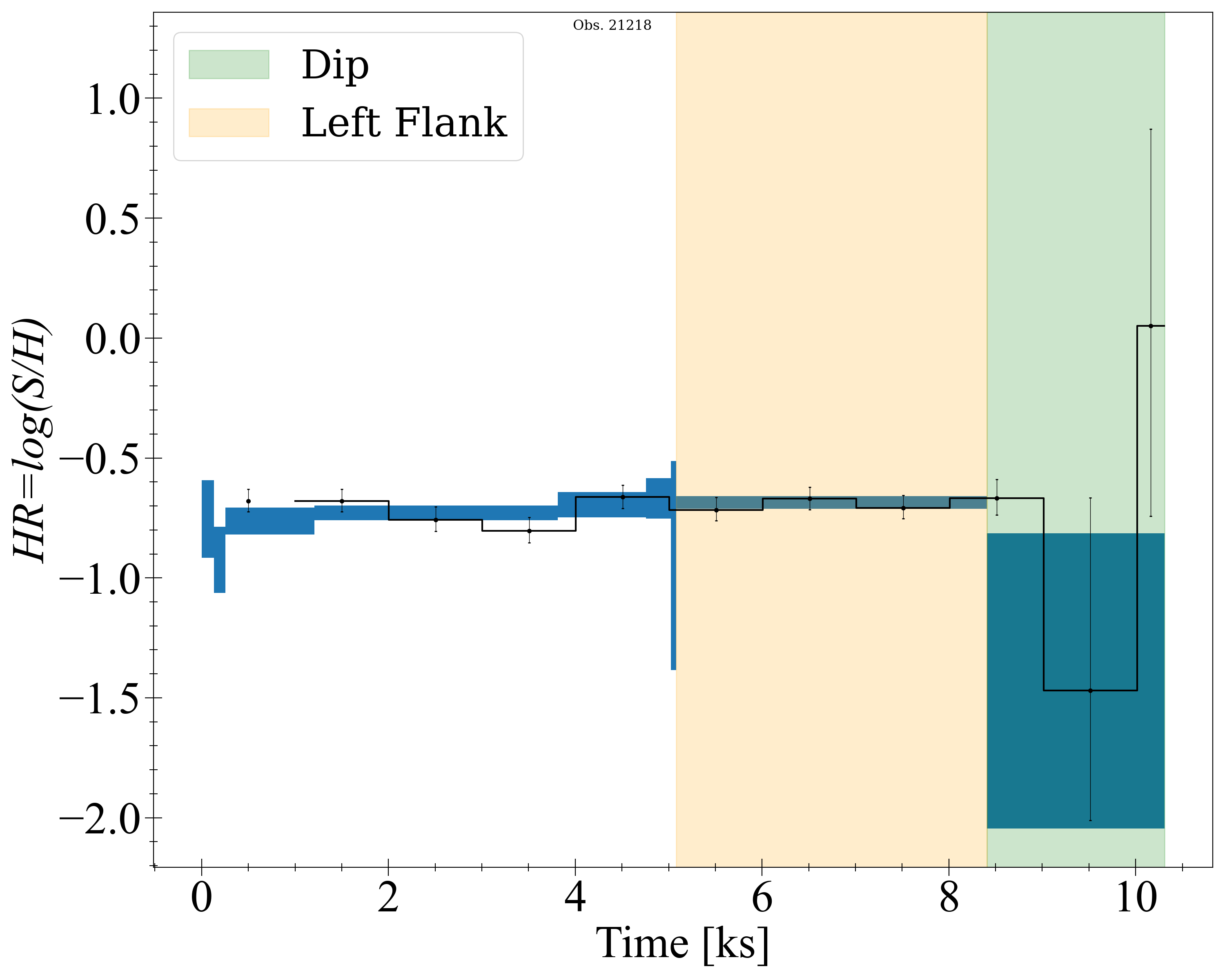}
    \caption{Hardness Ratios of each Bayesian Block for \chandra\ ObsID 21218, with 1000s bin HR plotted on top.}
    \label{fig:21218_hr}
\end{figure}

\clearpage

\section{Model Mathematics} \label{ref:math_section}

\subsection{Geometric Arguments Assuming Circular X-ray Emission Regions}

As stated above, we define $\tau_\mathrm{in,low}$ to be the time during which the flux is lowest; this interval starts when the trailing edge of the XRS has just passed over the leading edge of the eclipser, and ends when the leading edge of the XRS has is just about to pass over the trailing edge of the eclipser.\footnote{Thus, $\tau_{\rm \mathrm{in}}$ is the time interval from the point at which the flux first deviates from baseline until the point at which the flux again reaches up to baseline. $\tau_\mathrm{in,low}$ is the time interval corresponding to the deepest portion of the transit.}
We define $\rho_{\rm \mathrm{in}}$
to be the ratio $R_x/R_2$, and $\beta_{\rm \mathrm{in}}$ to be $b/R_2$, where $b$ is the projected distance of closest approach between the center of the XRS and the center of the eclipser. Both Star~1 and Star~2 are modeled as circular disks. Referring to Figure \ref{fig:diagram2}, we see that we can  define the ${\cal R}_{\rm \mathrm{in}}$ and ${\cal R}_{\rm \mathrm{in,low}}$ in terms of  $R_2$, $\rho_{\rm \mathrm{in}}$, and $\beta_{\rm \mathrm{in}}$.

\begin{equation}
\begin{aligned}
  {\cal R}_{\rm \mathrm{in}} = R_2\, \Bigg[\Big(1 + \rho_{\rm \mathrm{in}}\Big)^2 -\beta_{\rm \mathrm{in}}^2\Bigg]^\frac{1}{2}\\
   {\cal R}_{\rm \mathrm{in,low}} = R_2\, \Bigg[\Big(1 - \rho_{\rm \mathrm{in}}\Big)^2 -\beta_{\rm \mathrm{in}}^2\Bigg]^\frac{1}{2}\\ 
\end{aligned}
\end{equation}

We can now write:
\begin{equation}
   {\cal R}_\mathrm{in, low}^2 =\big(\frac{\tau_{\rm in,low}v_{\rm in}}{2}\big)^2 = R_2^2\, \Bigg[(1-\rho_{\rm \mathrm{in}})^2 -\beta_{\rm \rm in}^2\Bigg] 
\end{equation}

\begin{equation}
   {\cal R}_\mathrm{in}^2 =\big(\frac{\tau_{\rm in}v_{\rm in}}{2}\big)^2 = R_2^2\, \Bigg[(1+\rho_{\rm \mathrm{in}})^2 -\beta_{\rm \mathrm{in}}^2\Bigg] 
\end{equation}

Subtracting Eq~(6) from Eq~(7), we get:
\begin{equation}
   4\, \rho_{\rm \mathrm{in}} = \big(\frac{v_{\rm \mathrm{in}}}{2\, R_2}\big)^2 \big(\tau_{\rm \mathrm{in}}^2-\tau_{\rm \mathrm{in,low}}^2\big)
\end{equation}

We can also derive an explicit formula for $\beta$:
$$
    \beta^2 = 1 - 2\, \kappa\, \rho + \rho^2
$$

where $\kappa$ is defined as

\begin{equation}
    \kappa = \frac{\Big[\Big(\frac{\tau}{\tau_{\rm low}}\Big)^2+1\Big]}{\Big[\Big(\frac{\tau}{\tau_{\rm low}}\Big)^2-1\Big]}
\end{equation}

\clearpage

\section{THIRD BODY MODELS CONTINUED}\label{sec:more_3_body_plots}

\begin{figure}[H]
    \centering
    \includegraphics[width=.48\textwidth]{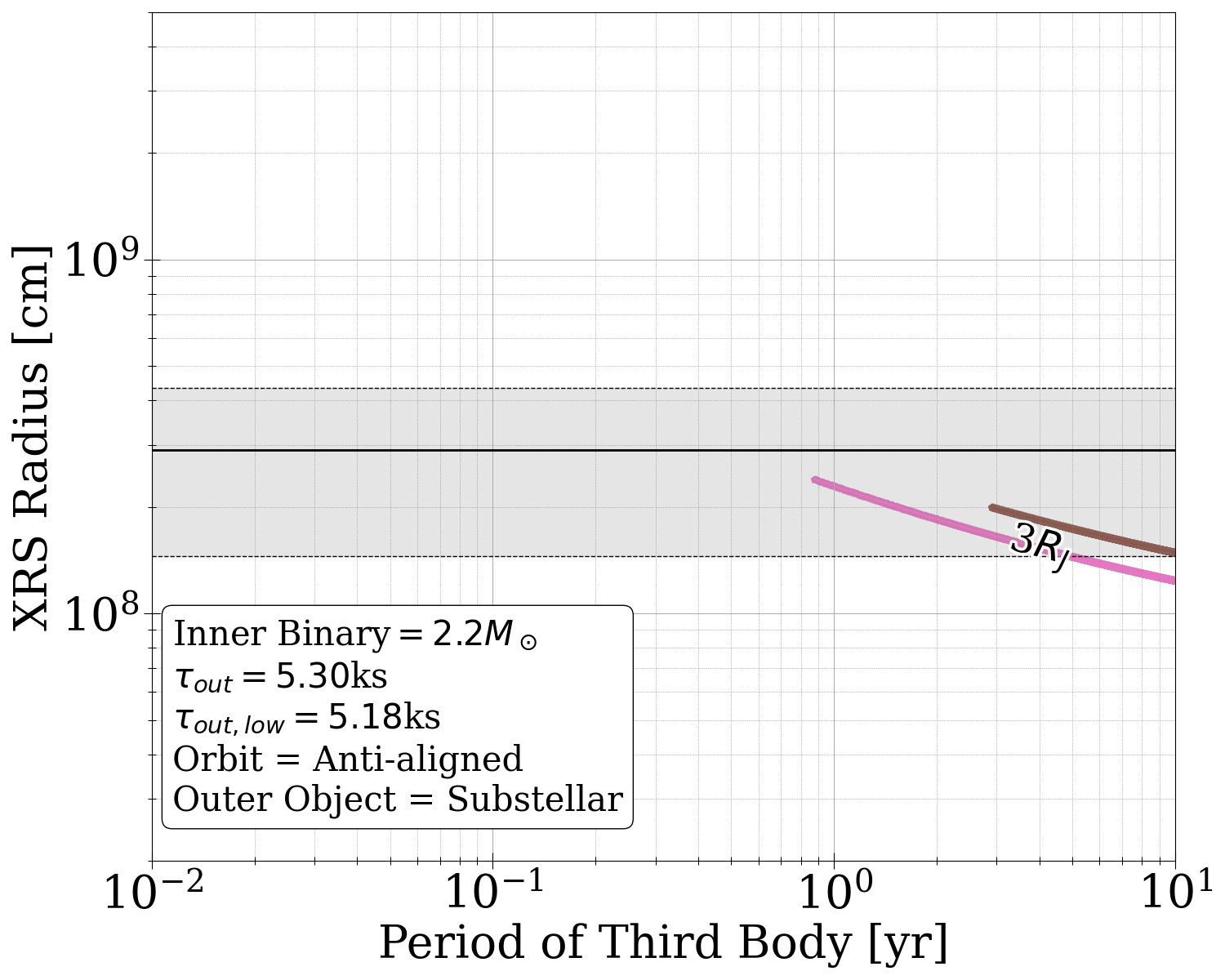}
    \includegraphics[width=.48\textwidth]{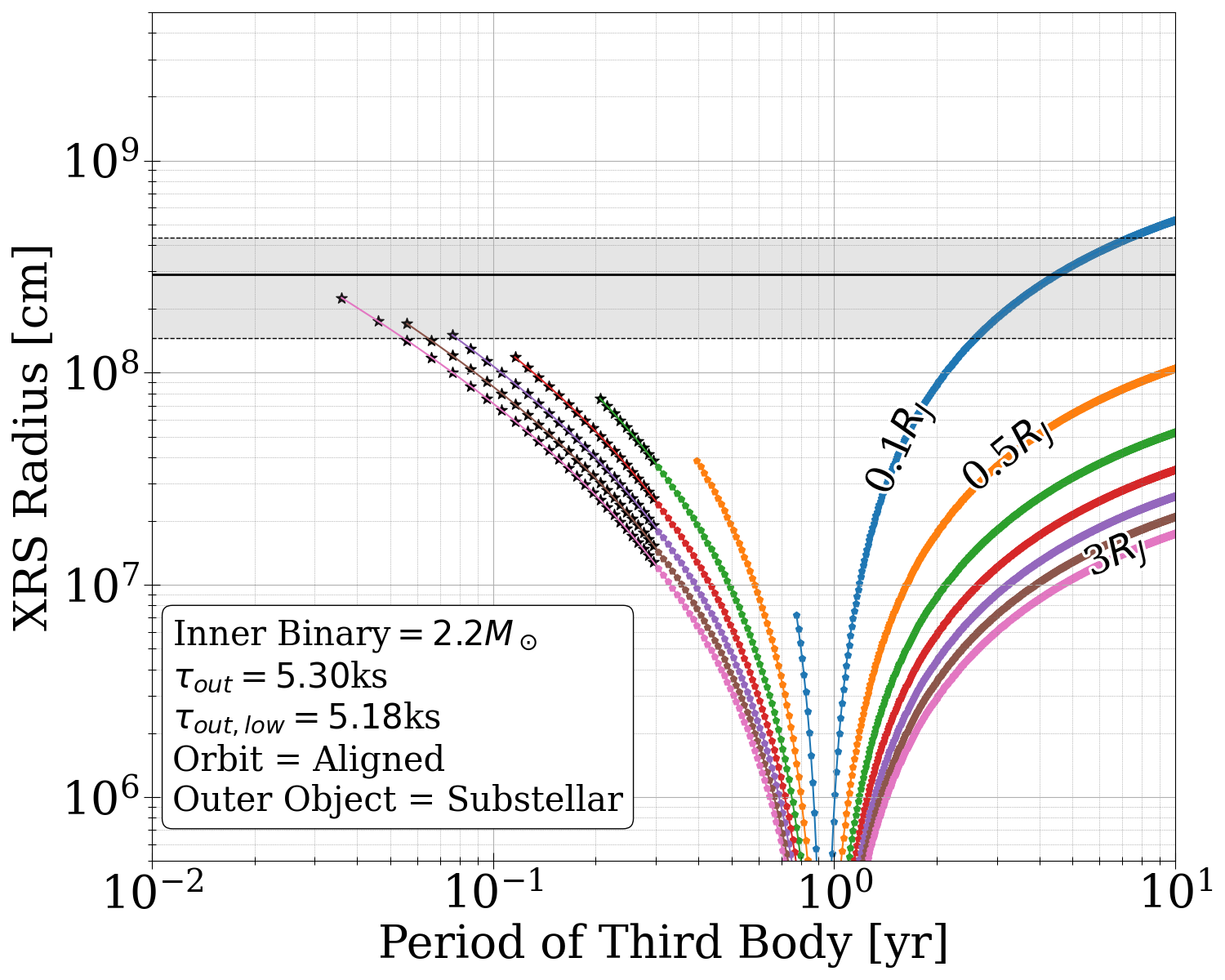}
    \caption{The same plots as in Figure \ref{fig:3_body_models_planets}, but with  $\tau_{\rm out,low}=5.18$ks.}
    \label{fig:smaller_tau_planets}
\end{figure}

\begin{figure}[H]
    \centering
    \includegraphics[width=.48\textwidth]{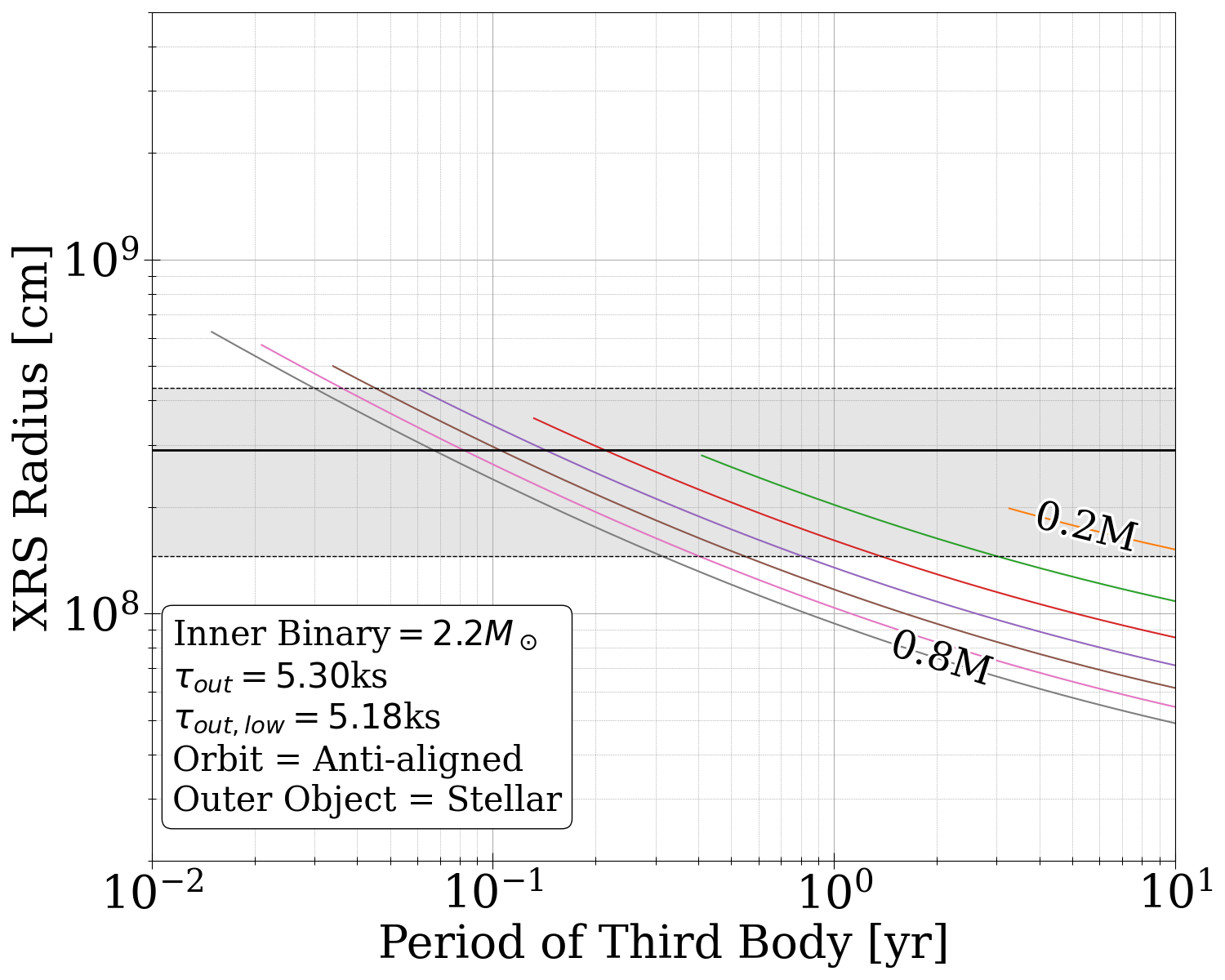}
    \includegraphics[width=.48\textwidth]{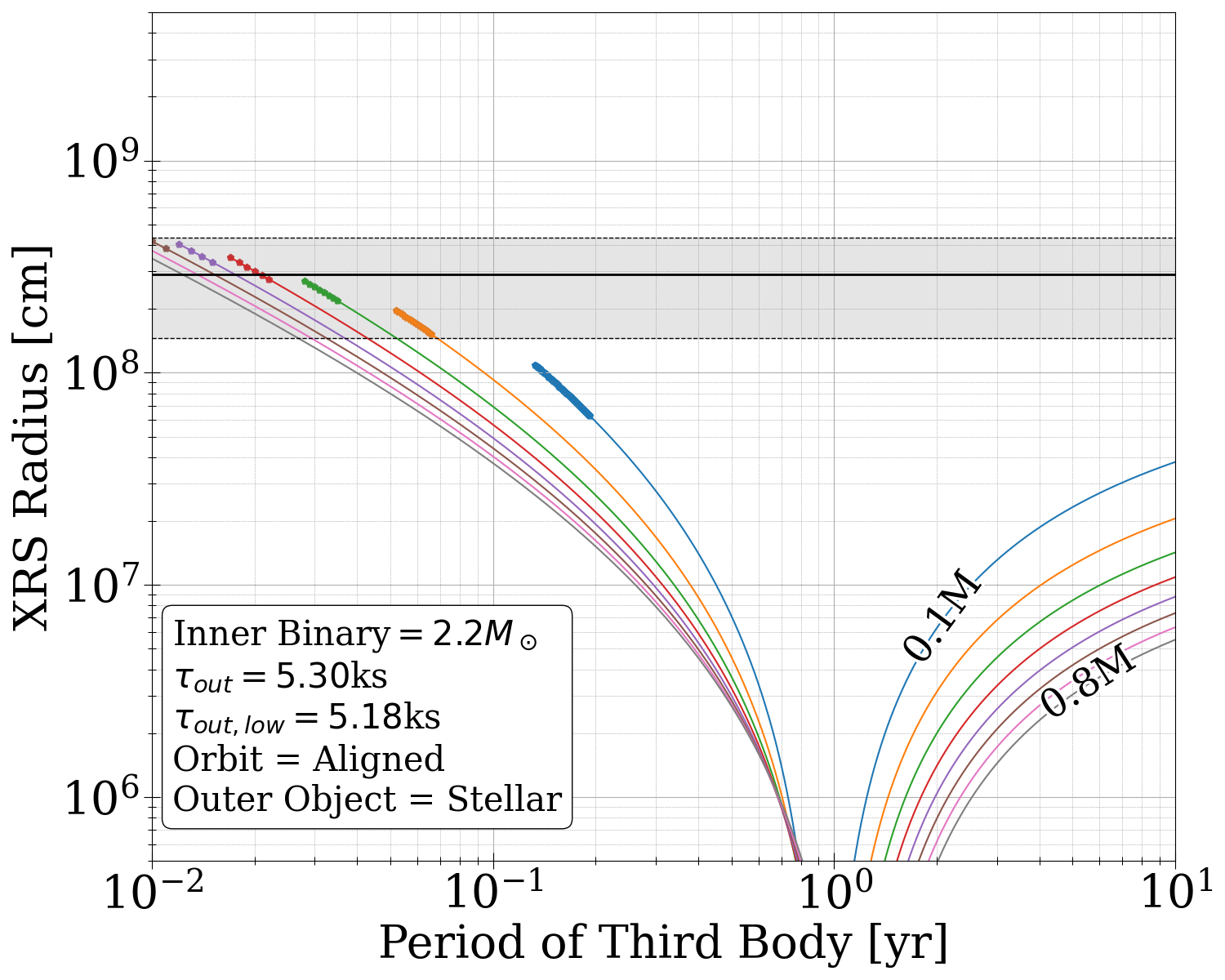}
    \caption{The same plots as in Figure \ref{fig:3_body_models}, but with $\tau_{\rm out,low}=5.18$ks.}
    \label{fig:smaller_tau}
\end{figure}

\newpage

\section{Example MODEL Parameters for the 5.3 ks eclipse}\label{sec:params}

\setlength{\tabcolsep}{3pt}
\begin{table}[!htb]
    \caption{Three Body Models in which $E_1$ fills its Roche lobe ($A_1$ is a $1.4M_\odot$ NS)}
    \begin{minipage}{.5\linewidth}
      $E_1$ ($.8M_\odot$) fills its RL, $E_2$ is MS
      \centering
    \begin{tabular}{ccccc}
    \toprule
    $R_3$ [$R_\odot$] & $P_3$ [d] & $v_{\rm rel}$ [km/s] & $\beta$ &  $R_x$ [e8 cm] \\
    \hline
    \midrule
    0.36 & 10.2 & 93 & 0.26 & 4.93\\
    0.46 & 8.0 & 106 & 0.51 & 2.81 \\
    0.56 & 5.8 & 127 & 0.52 & 3.27\\
    \bottomrule
    \end{tabular}
    \end{minipage}%
    \begin{minipage}{.5\linewidth}
      \centering
    $E_1$ ($.8M_\odot$) fills its RL, $E_2$ ($.07M_\odot$) is a brown dwarf
\begin{tabular}{cccccc}

\toprule
   $R_3$ [$R_J$] & $P_3$ [d] & $v_{\rm rel}$ [km/s] & $\beta$ & $R_x$ [e8 cm] \\
\hline
\midrule
1.5 & 45.0 & 39 & 0.35 & 1.97 \\
2.0 & 35.0 & 46 & 0.56 & 2.08\\
2.5 & 20.0 & 61 & 0.46 &  2.48 \\
\bottomrule
\end{tabular}
    \end{minipage} 

\label{tab:RL_MS}

\end{table}

\begin{table}[!htb]
    \caption{Three Body Models in which $E_2$ fills its Roche lobe ($A_1$ is a $1.4M_\odot$ NS)}
    \begin{minipage}{1\linewidth} 
    \centering
        $E_1$ ($.8M_\odot$) is MS, $E_2$ ($.8M_\odot$) fills its RL

\centering
\label{tab:MS_RL}

\hskip-1.5cm\begin{tabular}{ccccc}

\toprule
    $R_3$ [$R_\odot$] & $P_3$ [d] & $v_{\rm rel}$ [km/s] & $\beta$ &  $R_x$ [e8 cm] \\
\hline
\midrule
2.59 & 2.00 & 204 & 0.96 & 2.99\\
4.08 & 4.00 & 153 & 0.99 & 1.07 \\
5.36 & 6.00 & 130 & 1.00 & 0.58\\
\bottomrule
\end{tabular}
\end{minipage} 

\end{table}

\begin{table}[!htb]
    \caption{Three Body Models in which there are two compact objects ($A_1$ is a $6M_\odot$ BH)}
    \begin{minipage}{.5\linewidth}
      \centering
      $E_1$ ($1.4M_\odot$) is a NS, $E_2$ ($.8M_\odot$)  fills its RL
    \begin{tabular}{ccccc}
    \toprule
    $R_3$ [$R_\odot$] & $P_3$ [d] & $v_{\rm rel}$ [km/s] & $\beta$ &  $R_x$ [e8 cm]\\
    \hline
    \midrule
2.85 & 2.00 & 266 & 0.94 & 5.08 \\
4.50 & 4.00 & 151 & 0.99 & 1.04\\
6.00 & 6.00 & 80 & 1.0 & 0.22\\
    \bottomrule
    \end{tabular}
    \end{minipage}%
    \begin{minipage}{.5\linewidth}
      \centering
      $E_1$ ($.8M_\odot$) fills its RL, $E_2$ ($1.4M_\odot$) is a NS filling $90\%$ of its RL
\begin{tabular}{ccccc}
\toprule
   $R_3$ [$R_\odot$] & Period [d] & $v_{\rm rel}$ [km/s] & $\beta$ & $R_x$ [e8 cm] \\
\hline
\midrule
3.10 & 2.00 & 294 & 0.93 & 5.77\\
4.90 & 4.00 & 196 & 0.99 & 1.62\\
6.40 & 6.00 & 139 & 1.0 & 0.62\\
\bottomrule
\end{tabular}
    \end{minipage} 

        \label{tab:2_CO}

\end{table}

\begin{table}[!htb]

    \caption{Models in which there are two seperate LMXBs}

    \begin{minipage}{.5\linewidth}
    \centering
    Seperate LMXB is a NS ($1.4M_\odot$), $E_2$ ($.8M_\odot$)  fills its RL
    \begin{tabular}{ccccc}
    \toprule
    $R_3$ [$R_\odot$] & $P_3$ [d] & $v_{\rm rel}$ [km/s] & $\beta$ &  $R_x$ [e8 cm] \\
    \hline
    \midrule
1.82 & 1.00 & 277 & 0.82 & 8.67 \\
2.38 & 1.50 & 241 & 0.93 & 5.05 \\
2.88 & 2.00 & 220 & 0.96 & 3.44\\
4.57 & 4.00 & 174 & 0.99 & 1.36 \\
5.99 & 6.00 & 152 & 1.0 & 0.79 \\
    \bottomrule
    \end{tabular}
    \end{minipage}%
    \begin{minipage}{.5\linewidth}
      \centering
      Seperate LMXB is a BH ($6M_\odot$), $E_2$ ($.8M_\odot$)  fills its RL
\begin{tabular}{ccccc}
\toprule
   $R_3$ [$R_\odot$] & Period [d] & $v_{\rm rel}$ [km/s] & $\beta$ & $R_x$ [e8 cm] \\
\hline
\midrule
1.79 & 1.00 & 403 & 0.54 & 18.64\\
2.35 & 1.50 & 352 & 0.83 & 10.85\\
2.84 & 2.00 & 320 & 0.91 & 7.40 \\
4.51 & 4.00 & 254 & 0.98 & 2.94 \\
5.91 & 6.00 & 222 & 0.99 & 1.71 \\
\bottomrule
\end{tabular}
    \end{minipage} 
    \label{tab:2_seperate}
\end{table}

\vspace{1cm}
\FloatBarrier

\end{document}